\documentclass[prc,aps,eqsecnum,floatfix,showpacs,nofootinbib]{revtex4}
\usepackage{dcolumn}
\usepackage{bm}
\usepackage{color}
\usepackage{graphicx}
\usepackage{epsf}
\usepackage{pstricks}
\usepackage{slashed}
\usepackage{amssymb}
\usepackage{hyperref}
\usepackage{array}

\begin{document}

\title{
 Minimally non-local nucleon-nucleon potentials with
chiral two-pion exchange including $\Delta$'s}
\author{M.\ Piarulli$^{\rm a}$,
L. Girlanda$^{\rm b,c}$,
R.\ Schiavilla$^{\,{\rm a,d}}$,
R.\ Navarro P\'{e}rez$^{\,{\rm e}}$,
J.E.\ Amaro$^{\,{\rm e}}$,
and E. Ruiz Arriola$^{\,{\rm e}}$}
\affiliation{
$^{\,{\rm a}}$\mbox{Department of Physics, Old Dominion University, Norfolk, VA 23529, USA}\\
$^{\,{\rm b}}$\mbox{Department of Mathematics and Physics, University of Salento, I-73100 Lecce, Italy}\\
$^{\,{\rm c}}$\mbox{INFN-Lecce, I-73100 Lecce, Italy}\\
$^{\,{\rm d}}$\mbox{Theory Center, Jefferson Lab, Newport News, VA 23606, USA}\\
$^{\,{\rm e}}$\mbox{Departamento de F\'{\i}sica At\'{o}mica, Molecular y Nuclear
and Instituto Carlos I de F\'{\i}sica Te\'{o}rica y Computacional}\\
\mbox{Universidad de Granada, E-18071 Granada, Spain}
}
\date{\today}

\begin{abstract}
We construct a coordinate-space chiral potential, including $\Delta$-isobar
intermediate states in its two-pion-exchange component up to order
$Q^3$ ($Q$ denotes generically the low momentum scale).  The contact interactions
entering at next-to-leading and next-to-next-to-next-to-leading orders ($Q^2$ and
$Q^4$, respectively) are rearranged by Fierz transformations to yield terms at most quadratic in the
relative momentum operator of the two nucleons.  The low-energy constants
multiplying these contact interactions are fitted to the 2013 Granada database,
consisting of 2309 $pp$ and 2982 $np$ data (including, respectively, 148 and
218 normalizations) in the laboratory-energy range 0--300 MeV.  For the total
5291 $pp$ and $np$ data in this range, we obtain a $\chi^2$/datum of roughly 1.3
for a set of three models characterized by long- and short-range cutoffs, $R_{\rm L}$ and
$R_{\rm S}$ respectively, ranging from $(R_{\rm L},R_{\rm S})=(1.2,0.8)$ fm down to
$(0.8,0.6)$ fm.  The long-range (short-range) cutoff regularizes the one- and two-pion
exchange (contact) part of the potential.
\end{abstract} 


\index{}\maketitle

\section{Introduction}
\label{sec:intro}
The nucleon-nucleon ($NN$) interaction is a basic building block in
nuclear physics as it makes it possible to describe nuclear structure
and nuclear reactions.  If the forces were known accurately and
precisely, the nuclear many-body problem would become a large-scale
computation where precision and accuracy are defined in terms of the
preferred numerical method.  However, the lack of direct knowledge of
the forces among constituents at separation distances relevant for
nuclear structure and reactions drastically changes the rules of the
game.  Indeed, the use of a large but {\it finite} body of scattering
data below a given maximal energy to provide constraints on the
interaction transforms the whole setup into a statistical inference
problem, based on the conventional least $\chi^2$-method.  This fact
was recognized already in 1957~\cite{Stapp:1956mz} (see
Ref.~\cite{arndt1966chi} for an early review) and, after many years,
culminated in the admirable Nijmegen partial wave analysis (PWA) of
1993~\cite{Stoks93}, based on the crucial observations that
charge-dependent one-pion-exchange (CD-OPE), tiny but essential
electromagnetic and relativistic effects, and a judicious selection of
the scattering database could actually provide a satisfactory fit with
$\chi^2/{\rm datum} \sim 1$ for a total number of data consisting, as
of 1993, of 1787 $pp$ and 2514 $np$ (normalizations included) at the
$3\, \sigma$ level. These criteria have set the standard for PWA's and
the design of high quality phenomenological potentials~\cite{Stoks94,
  Wiringa95,Rentmeester:1999vw,Machleidt:2000ge,Rentmeester:2003mf,
  Gross08,Navarro13,Perez:2013oba,Navarro14}.  The inference point of
view is mainly phenomenological and requires a balanced interplay
between {\it which} data qualify as constraints and {\it which} models
provide the most likely description of the data. None of these
choices is free of prejudices and they are actually intertwined; a
circumstance that should be kept in mind when assessing the
reliability and predictive power of the theory aiming at a faithful
representation of the input data and their uncertainties. 

The quantum mechanical nature of the PWA with a given cutoff in
energy leads to inverse scattering ambiguities which increase at short
distances (see, for example, Refs.~\cite{Inverse,Baye:2014oea} and
references therein). Remarkably, a universal and
model-independent low-energy interaction arises when unobserved
high energy components above the cutoff are explicitly integrated
out of the Hilbert space preserving the scattering
amplitude~\cite{Bogner:2001gq,Bogner:2003wn}.  While this $V_{\rm
low-k}$ framework is an extremely appealing setup based on
Wilsonian renormalization, to date this universal interaction has
not been determined from data {\it directly} and one has to proceed
via a fitted and bare $NN$ interaction since off-shellness is
required~\cite{Arriola:2014fqa}.  However, inferring a $NN$ interaction
from data, is not the full story, and three-nucleon, and
possibly higher multi-nucleon, interactions are needed to describe
residual contributions to nuclear binding energies~\cite{Carlson2014}.
As is well known, their strength and form are also affected by the
{\it chosen} off-shell behavior of the $NN$ interaction and a
universal $V_{\rm low-k}$ three-nucleon interaction remains to be found.

In an ideal situation all steps in the inference process, including
the scattering data selection itself, should be carried out with the
``true'' theory, which for nuclear physics is quantum chromodynamics
(QCD), the fundamental theory of interacting quarks and gluons.  Assuming, as
we do, that the theory is correct, QCD would just tell us which experiments
are right and which are wrong, or whether the reported uncertainties are
realistic with a given confidence level on the side of the experiment.  At the
same time one would set constraints on the QCD parameters such as the
light quark masses and $\Lambda_{\rm QCD}$, or equivalently the pion
mass $m_\pi$ and the pion weak decay constant $F_\pi$.  While there
has been impressive progress in bringing lattice QCD simulations for light
quarks closer to nuclear physics working conditions (see Refs.~\cite{Hatsuda:2012hw,
Detmold:2015jda} and references therein), we do not yet envisage, at least
not in the near future, the realization of conditions that would allow one
to establish, on QCD grounds, the correctness of the about 8000 currently 
available published $pp$ and $np$ scattering data below pion production
threshold.  Instead, already in the early 90's the phenomenological analysis
carried out by the Nijmegen group made it possible to pin down the pion
masses with a precision of 1 MeV from their PWA of $pp$ and $np$
data~\cite{Klomp:1991vz}.

In practice, we must content ourselves with an approximation scheme to
the true theory in conjunction with a phenomenological approach.  This
specifically means assuming a sufficiently
flexible parametrization of the interaction in terms of the relevant
degrees of freedom which does not overlook some relevant physical
feature.  In what follows it is instructive to briefly review both the
process and criteria taken into account to select a consistent
database as well as the QCD-based theory used to describe it.
Our aim is to make the reader aware of all the fine details which
are needed in order to credibly falsify the theoretical model, QCD grounded or not, against
the data and keep an open mind about the out-coming result.

On the theoretical side, we will assume along with Weinberg~\cite{Weinberg:1990rz}
that there is a chiral effective field theory ($\chi$EFT) capable of systematically
describing the strong interactions among nucleons, $\Delta$-isobars, and
pions, as well as the electroweak interactions of these hadrons with external 
(electroweak) fields.  In the specific case of two nucleons, the requirements
imposed by $\chi$EFT can be incorporated into a non-relativistic quantum
mechanical potential, constructed by a perturbative matching, order
by order in the chiral expansion, between the on-shell scattering amplitude
and the solution of the Schr\"odinger equation (see, for example, the review
paper by Machleidt and Entem~\cite{Entem11}).  Such a theory provides the
most general scheme accommodating all possible interactions compatible
with the relevant symmetries of QCD at low energies, in particular chiral
symmetry.  By its own nature, $\chi$EFT needs to be organized within a given
power counting scheme and the resulting chiral potentials can conveniently
be separated into long- and short-distance contributions, the latter
(short-distance ones) featuring the needed counter-terms for renormalization.
At leading order in the chiral expansion one has the venerable one-pion-exchange
(OPE) potential which, as already mentioned, emerges as a universal
and indispensable long-distance feature for an accurate description of
proton-proton and neutron-proton scattering data~\cite{Stoks93}.
Higher orders in the chiral expansion incorporate the two-pion-exchange
(TPE) potential~\cite{Kaiser:1997mw}, due to leading and sub-leading
$\pi N$ couplings (the sub-leading couplings $c_1$, $c_3$, and $c_4$
can consistently be obtained from low energy $\pi N$ scattering data).
The inclusion of TPE allows one to reduce the short-range cutoff
separating long- and short-distance contributions, which helps in
reducing the impact of details in the unknown short-distance behavior
of the potentials.  Nonetheless, we will note in Sec.~\ref{sec:res} that
uncertainties are dominated by this diffuse separation between short 
and long distances.

There are many practical advantages deriving from a $\chi$EFT that
explicitly includes $\Delta$-isobar degrees of freedom, the most immediate
one being a numerical consistency between the values of the low-energy constants
$c_1$, $c_3$ and $c_4$ inferred from either $\pi N$ or $NN$ scattering.
Such a theory also naturally leads to three-nucleon forces induced by TPE
with excitation of an intermediate $\Delta$ (the Fujita-Miyazawa
three-nucleon force) as well as to two-nucleon electroweak currents
(see for example Ref.~\cite{Pastore2008}).  In addition, there are rather
strong indications from phenomenology that $\Delta$ isobars play
an important role in nuclear structure and reactions.  An illustration of this
are the three-nucleon forces involving excitation of intermediate $\Delta$'s, needed
to reproduce the observed energy spectra and level ordering of low-lying
states in s- and p-shell nuclei or the correct spin-orbit splitting of P-wave
resonances in low-energy $n$-$\alpha$ scattering (for a review, see
Ref.~\cite{Carlson2014}).  Another illustration is the relevance of electroweak
$N$-to-$\Delta$ transition currents in radiative and weak capture
processes involving few-nucleon systems~\cite{Schiavilla92}, specifically the radiative
captures of thermal neutrons on deuteron and $^3$He~\cite{Viviani96,Girlanda10}
or the weak capture of protons on $^3$He (the so-called $hep$ process)~\cite{Marcucci01}.  
It is for these reasons that in the present work we construct a minimally non-local
coordinate space chiral potential, that includes $\Delta$
intermediate states in its TPE component---it is described in detail
in Sec.~\ref{sec:pots}.  Such a coordinate-space representation offers many
computational advantages for {\it ab initio} calculations of nuclear structure
and reactions, in particular for the type of quantum Monte Carlo calculations of s- and
p-shell nuclei very recently reviewed in Ref.~\cite{Carlson2014}.

On the experimental side, there are currently $\sim 8000$ published
$pp$ and $np$ scattering data below pion production threshold
corresponding to 24 different scattering observables, including
differential cross sections, spin asymmetries, and total cross
sections~\cite{Bystricky1,Bystricky2}, see Ref.~\cite{Navarro14}
for updated $pp$ and $np$ abundance plots in the $(E_{\rm lab}, \theta_{\rm cm})$ plane.
However, not all of these data are mutually
compatible and a decision has to be made as to which are more likely to
be correct.  In principle, the $NN$ scattering amplitude can be determined
uniquely, provided a complete set of experiments is given---a rare
situation for the case under consideration. Therefore, a theoretical
{\it  model} is needed to provide a smooth energy dependence which
allows one to interpolate between different energy values, and helps in
deciding on the mutual consistency of {\it nearby} data in $(E_{\rm lab}, \theta_{\rm cm})$ plane.  The PWA
carried out in Granada parametrizes~\cite{Navarro13}\footnote{The Granada
database is located in the HADRONICA website
\url{http://www.ugr.es/~amaro/hadronica/}.}~the interaction, for inter-nucleon
distances $r$ less than 3 fm, in terms of a set equidistant delta-shells
separated by $\Delta r =0.6$ fm (in other words, a coarse-grained
parametrization), while retaining only the OPE component for $r > 3$ fm.
The choice of $\Delta r$ corresponds to the shortest de Broglie
wavelength at about pion production threshold, and consequently
all the data are weighted with their quoted experimental uncertainty.
The result of the analysis has been a $3\,\sigma$ self-consistent
database comprising a total of 6713 $pp$ and $np$ scattering data.
More details on the data analysis specific to our potential are
presented in Sec.~\ref{sec:data}.  One important aspect of the
Granada PWA is the correlation pattern among the fitting parameters,
namely different partial waves are mostly uncorrelated which,
together with the large number of selected data, speaks in favor 
of a lack of bias in the selection process.  Actually the
correlation length which decides on the specific form of the potential
should be smaller than the distance $\Delta r = 0.6$ fm in the
coarse-grained parametrization.

Chiral potentials have been subjected to PWA and confronted to $pp$
and $np$ scattering data up to lab energy of 350 MeV.  Within the
$\chi$EFT framework the Nijmegen group used the TPE potential~\cite{Kaiser:1997mw}
to carry out $pp$~\cite{Rentmeester:1999vw} and $np+pp$~\cite{Rentmeester:2003mf}
analyses determining the chiral constants $c_3$ and $c_4$ from these data
while constraining $c_1$ from $\pi N$ data.  Taking the chiral constants from
$\pi N$ analyses, Entem and Machleidt~\cite{Entem03} used a next-to-next-to-next-to-leading
order (N3LO or $Q^4$, $Q$ generically specifying the low momentum scale) chiral
potential to fit $pp$ and $np$ scattering data up to lab energy of 290 MeV.
The resulting $\chi^2$/datum were 1.1 for 2402 $np$ data and 1.50 for 2057
$pp$ data, and consequently a global $\chi^2$/datum of 1.28.  The chiral
TPE potential~\cite{Kaiser:1997mw} was also used within the coarse
grained framework to determine the chiral constants in Ref.~\cite{Perez:2013oba}
with a global $\chi^2$/datum of 1.07, based on 6713 $pp$ and $np$ scattering data.

Other available chiral potentials~\cite{Epelbaum:2004fk,Entem2015} have not
been confronted to scattering data directly but rather to phase shifts obtained
in the Nijmegen analysis (the recent upgrade~\cite{Epelbaum:2014efa}
of Ref.~\cite{Epelbaum:2004fk} relies on the same procedure, while
in Ref.~\cite{Entem2015} a study of peripheral phase shifts is carried out
with two- and three-pion exchange potentials up to order $Q^5$).  As we will show in Sec.~\ref{sec:res},
there is a substantial difference between fitting scattering data and
fitting phase shifts mainly because of the existing correlations among
the many partial waves and mixing angles.  Actually, a good $\chi^2$-fit
to phase shifts may yield quite a bad $\chi^2$ in a fit to data.  Moreover, the
spread in phase-shift values among different high-quality potentials fitting
the same data reflects the differences in the potential representation and
turns out to be {\it larger} than the estimated statistical errors (compare
Fig.~1 of Ref.~\cite{NavarroPerez:2012vr} with Fig.~3 of Ref.~\cite{Perez:2014jsa}).
The consequences of these larger errors have been discussed in Ref.~\cite{Perez:2013za}. 

The previous comments address the use of chiral potentials to fit
{\it selected} $NN$ scattering databases which have been obtained from
{\it phenomenological} representations of the interactions.  An
obvious question which comes to mind is whether chiral potentials,
being credible and general low energy representations of QCD in the
$NN$ sector, should be used themselves to {\it select} the database.
Within the coarse grained framework the impact of chiral interactions
on the selection of the database has also been studied in Ref.~\cite{Perez:2013oba}.
The result was that a larger number of data were rejected but at the
same time the number of parameters was reduced.  This poses the
interesting question on {\it what} is the meaning of improvement---a
particularly critical issue when the potential itself (chiral or not) must
be tested against the selected data.  Obviously an incorrect model will
appear to be correct if a sufficiently large number of data is discarded.
However, the theory with just delta-shells+OPE is more general than
that with delta-shells+(OPE+TPE), and hence data selection based on
the former is more reliable.  In any case, the results of Ref.~\cite{Perez:2013oba}
show also that the long range part of the next-to-next-to-leading order (N2LO or $Q^3$)
chiral potential can indeed fit the delta-shells+OPE selected data
satisfactorily with a $\chi^2$/datum of 1.07, when the potential is taken
to be valid for inter-nucleon distances ranging from 1.8 fm outwards.

The present paper is organized as follows.  In the next section
we describe the potential, while in Sec.~\ref{sec:data} we provide
a brief discussion of the data fitting.  In Sec.~\ref{sec:res} we report
the $\chi^2$ values obtained in the fits as well as the values for the
low-energy constants that characterize the potential, and show
the calculated phase shifts for the lower partial waves (S,
P, and D waves) and compare them to those from recent PWA's.
There, we also provide tables of the $pp$, $np$
and $nn$ effective range parameters and of deuteron properties,
including a figure of the deuteron S and D waves.
Finally, in Sec.~\ref{sec:conc} we summarize our conclusions.
A number of details are relegated to Appendices~\ref{app:a1}-\ref{app:a5}.
\section{Potentials}
\label{sec:pots}
\begin{center}
\begin{figure}[bth]
\includegraphics[width=6in]{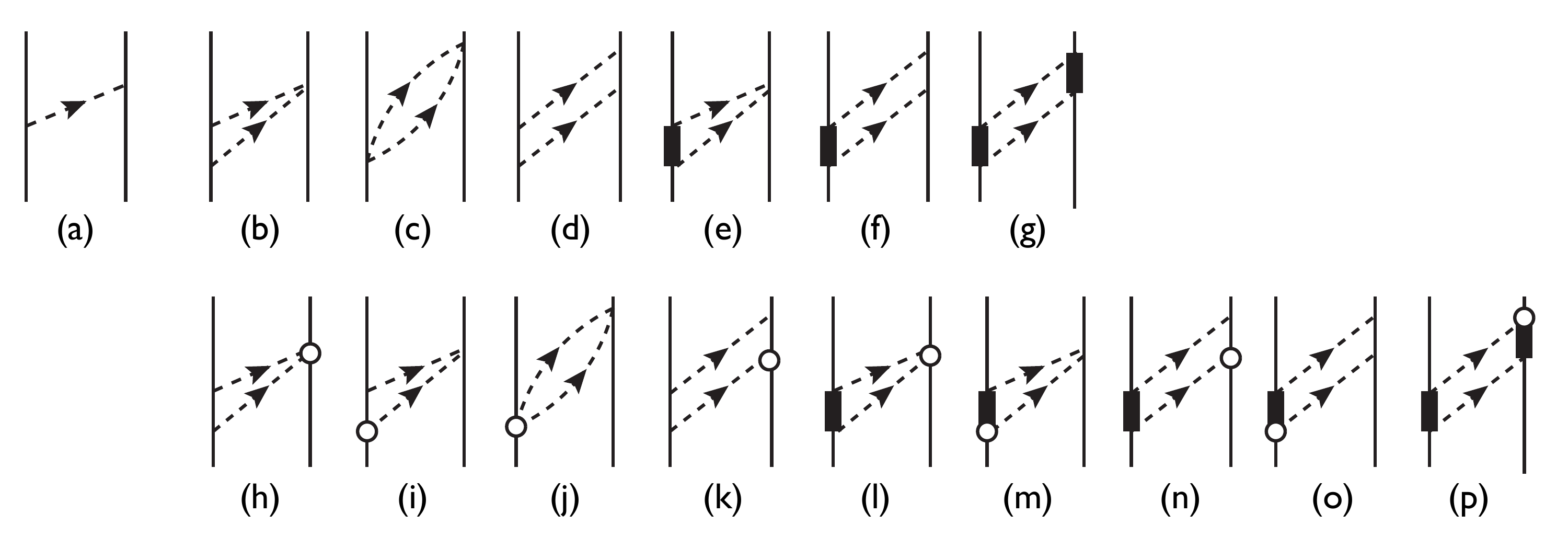}
\caption{OPE and TPE contributions at LO [(a)], NLO [(b)-(g)], and N2LO [(h)-(p)]. Nucleons,
$\Delta$ isobars, and pions are denoted, respectively, by the solid, thick-solid, and
dashed lines; both direct and crossed box contributions are retained in
diagrams (d), (f)-(g), (k), (n)-(p). The open circles denote $\pi N$ and $\pi N\Delta$
couplings from the sub-leading chiral Lagrangians $\mathcal{L}_{\pi N}^{(2)}$~\cite{Fettes00}
and $\mathcal{L}_{\pi N \Delta}^{(2)}$~\cite{Krebs07}. Note that relativistic 
$1/M_N$-corrections ($M_N$ is the nucleon mass) included in $\mathcal{L}_{\pi N}^{(2)}$ Lagrangian are not considered here. In particular
 the contributions of diagrams (i), (k) and (n) are neglected.}
\label{fig:f1}
\end{figure}
\end{center}

The two-nucleon potential includes a strong interaction component derived from
$\chi$EFT up to next-to-next-to-next-to-leading order (N3LO or $Q^4$) and denoted
as $v_{12}$, and an electromagnetic interaction component, including up to terms
quadratic in the fine structure constant $\alpha$ (first and second order Coulomb,
Darwin-Foldy, vacuum polarization, and magnetic moment interactions), and denoted
as $v_{12}^{\rm EM}$.  The $v_{12}^{\rm EM}$ component is the same as that adopted
in the Argonne $v_{18}$ (AV18) potential~\cite{Wiringa95}.  The component induced by
the strong interaction is separated into long- and short-range parts, labeled, respectively,
$v_{12}^{\rm L}$ and $v_{12}^{\rm S}$.  The $v_{12}^{\rm L}$ part includes the one
pion-exchange (OPE) and two pion-exchange (TPE) contributions, illustrated in
Fig.~\ref{fig:f1}: panel (a) represents the OPE contribution at leading order (LO or $Q^0$);
panels (b)-(g) represent the TPE contributions at next-to leading order (NLO or $Q^2$)
without and with $\Delta$-isobars in the intermediate states; lastly, panels (h)-(p) represent
sub-leading TPE contributions at next-to-next-to leading order (N2LO or $Q^3$).  The NLO
and N2LO loop corrections contain ultraviolet divergencies, which are isolated in dimensional
regularization and then reabsorbed into contact interactions by renormalization of the
associated low energy constants (LEC's)~\cite{Epelbaum98,Pastore09}.  Additional loop
corrections at NLO and N2LO only lead to renormalization of OPE and contact
interactions~\cite{Epelbaum98,Viviani14}, and will not be discussed any further here.

\begin{table}[bth]
\caption{\label{tab:table}%
Values of (fixed) low energy constants (LEC's): $g_A$ and $h_A=3\,g_A/\sqrt{2}$ are
adimensional, $F_\pi=2\, f_\pi$ is in MeV, and the remaining LEC's are in GeV$^{-1}$.}
\begin{ruledtabular}
\begin{tabular}{cccccccccc}
$g_A$  & $h_A$ & $F_{\pi}$ & $c_1$ & $c_2$ & $c_3$ & $c_4$ & $b_3+b_8$ \\
\hline
$1.29$ &$2.74$ & $184.80$ &$-0.57$ &$-0.25$ &$-0.79$ & $1.33$& $1.40$ \\
\end{tabular}
\end{ruledtabular}
\end{table}
The LO, NLO, and N2LO terms are well known, and explicit expressions for them
can be found in Refs.~\cite{Epelbaum98,Pastore09,Kaiser1998,Epelbaum04,Krebs07}.
The LO and NLO terms depend on the the pion decay amplitude $F_{\pi}$, and
the nucleon and $N$-to-$\Delta$ axial coupling constants, respectively $g_A$ and
$h_A=3\,g_A/\sqrt{2}$ (this value for $h_A$ is from the large $N_c$ expansion
or strong-coupling model~\cite{Green76},
and is in good agreement with the value inferred from the empirical $\Delta$-width).
The sub-leading N2LO terms also depend on the LEC's $c_1$, $c_2$, $c_3$, and
$c_4$ and the combination of LEC's $(b_3+b_8)$, respectively from the second order
$\pi N$ and $\pi N\Delta$ chiral Lagrangians $\mathcal{L}_{\pi N}^{(2)}$~\cite{Fettes00}
and $\mathcal{L}_{\pi N \Delta}^{(2)}$~\cite{Krebs07}. 
The values of these LEC's, as determined by fits to $\pi N$ scattering data~\cite{Krebs07},
and of the masses and other physical constants adopted
in the present study are listed in Tables~\ref{tab:table} and~\ref{tab:table1}. 
\begin{table}[bth]
\caption{\label{tab:table1}%
Values of charged and neutral pion masses, proton and neutron masses, $\Delta$-nucleon
mass difference, and electron mass (all in MeV), and of the (adimensional) fine structure
constant $\alpha$.  Note that $\hbar c$ is taken as 197.32697 MeV$\,$fm.}
\begin{ruledtabular}
\begin{tabular}{cccccccc}
$m_{\pi_0}$ &$m_{\pi_{\pm}}$ & $M_n$ &$M_p$ &$\Delta M$& $m_e$ & $\alpha^{-1}$ \\
\hline
$134.9766$& $139.5702$ & $939.56524$  & $938.27192$&$293.1$ & $0.510999$ &  $137.03599$ \\
\end{tabular}
\end{ruledtabular}
\end{table}

In the static limit, the momentum-space LO, NLO, N2LO terms are functions
of the momentum transfer ${\bf k}$; hereafter, we define ${\bf k}={\bf p}^\prime-{\bf p}$
and ${\bf K}=({\bf p}^\prime+{\bf p})/2$, where ${\bf p}$ and ${\bf p}^\prime$ are the initial
and final relative momenta of the two nucleons.  Coordinate-space expressions
for the TPE terms are obtained by using the spectral function representation~\cite{Epelbaum04},
however with no spectral cutoff~\footnote{This detail is important, since the lack
of a spectral cutoff ensures the correct analytical properties of
the partial wave scattering amplitude in the complex $p_{\rm cm}$
plane, namely the proper branch-cut structure of the TPE
potential with the opening of the left cut at $p_{\rm cm}= \pm i\,
m_\pi$.  Moreover, it produces the correct asymptotic behavior of
the potential avoiding mid-range distortions.  We refer to
Refs.~\cite{Valderrama:2008kj,PavonValderrama:2010fb} for a
discussion of these issues.  As a matter of fact, the N3LO-$\slashed{\Delta}$
upgrade in Ref.~\cite{Epelbaum:2014efa} improves over
the work in Ref.~\cite{Epelbaum:2004fk} by removing the spectral cutoff.},
\begin{equation}
 v_{\rm L}^{l,{\rm TPE}} (r)\!=\!\frac{1}{2\pi^2r}\int_{2 m_{\pi}}^{\infty}d\mu\,\mu\, e^{-\mu r}f^l(\mu)\,
 {\rm Im}[\, \widetilde{v}_{\rm L}^{\, l,{\rm TPE}}(0^+-i\mu)\,]\, ,
\end{equation}
in terms of the left-cut discontinuity at $k=0^+ -i\, \mu$.  Here
$f^c(\mu)=f^{\tau}(\mu) =1$, $f^{\sigma}(\mu)=f^{\sigma\tau}(\mu)=2/3$
and $f^{t}(\mu)=f^{t\tau}(\mu)=-(3+3\mu r+\mu^2 r^2)/(3\,r^2)$, and the functions
$\widetilde{v}_{\rm L}^{\,l,{\rm TPE}}(k)$ are the momentum-space TPE components of the
potential at NLO and N2LO, 
\begin{equation}
\widetilde{v}_{12}^{\,{\rm L, TPE}}=
\sum_{l=1}^{6} \widetilde{v}_{\rm L}^{\,l,{\rm TPE}} (k) \, \widetilde{O}^{\, l}_{12} \ , 
\end{equation}
with $\widetilde{O}^{l=1,\dots,6}_{12}=[{\bf 1}\,,\, {\bm \sigma}_1\cdot {\bm \sigma}_2\, , 
\,{\bm \sigma}_1\cdot {\bf k}\,\,{\bm \sigma}_2\cdot {\bf k}] \otimes[{\bf 1}\, ,
\, {\bm \tau}_1\cdot {\bm \tau}_2]$
denoted as $c, \tau, \sigma, \sigma\tau, t , t\tau$.  Those corresponding 
to diagrams (b)-(d) and (h)-(k) in Fig.~\ref{fig:f1} are known in closed form
(see, for example, Ref.~\cite{Epelbaum04}) and are listed in Appendix~\ref{app:a1}
for completeness; the remaining ones corresponding to diagrams
(e)-(g) and (l)-(p) have been derived in terms of a parametric integral, and they too are
given in Appendix~\ref{app:a1}.  The radial functions $v_{\rm L}^l(r)$ are singular
at the origin (they behave as $1/r^n$ with $n$ taking on values up to $n=6$,
see Refs.~\cite{Valderrama:2008kj,PavonValderrama:2010fb} for analytical expressions), and
each is regularized by a cutoff of the form
\begin{equation}
\label{eq:ctff}
 C_{R_{\rm L}}(r)=1-\frac{1}{(r/R_{\rm L})^6 \,  e^{(r-R_{\rm L})/a_L} +1} \ , 
\end{equation}
where in the present work three values for the radius $R_L$
are considered $R_L=(0.8,1.0,1.2)$ fm with the diffuseness $a_L$
fixed at $a_L=R_L/2$ in each case.  The potential
 $v_{12}^{\rm L}$, including the well known OPE components
 at LO regularized by the cutoff in Eq.~(\ref{eq:ctff}), then reads in coordinate space
\begin{equation}
\label{eq:vlr}
v_{12}^{\rm L}= \left[\sum_{l=1}^{6} v_{\rm L}^l (r) \, O^l_{12}\right]
+v_{\rm L}^{\sigma T}(r)  \, O^{\sigma T}_{12} + v_{\rm L}^{t T}(r)  \, O^{t T}_{12} \ ,
\end{equation}
where 
\begin{equation}
O^{l=1,...,6}_{12}=\left[{\bf 1}\, ,\, {\bm \sigma}_1\cdot {\bm \sigma}_2\, , \,S_{12}\right]
\otimes\left[{\bf 1}\, ,\, {\bm \tau}_1\cdot {\bm \tau}_2\right] \ ,
\end{equation}
$O^{\sigma T}_{12}={\bm \sigma}_1\cdot {\bm \sigma}_2\,T_{12}$, and
$O^{tT}_{12}=S_{12}\,T_{12}$, and $T_{12}=3\,\tau_{1z}\tau_{2z}-{\bm
  \tau}_1\cdot {\bm \tau}_2$ is the isotensor operator.  The terms
proportional to $T_{12}$ account for the charge-independence breaking
induced by the difference between the neutral and charged pion masses
in the OPE.  However, this difference is ignored in the NLO and N2LO
loop corrections which have been evaluated with
$m_{\pi}=\left(2\,m_{\pi^+}+m_{\pi^0}\right)/3$.  Additional (and
small) isospin symmetry breaking terms arising from OPE~\cite{Friar04}
and TPE~\cite{Epelbaum05} and from OPE and one-photon
exchange~\cite{Kolck98,Kaiser06} have also been neglected.

The potential $v_{12}^{\rm S}$ includes charge-independent (CI) contact interactions
at LO, NLO and N3LO, and charge-dependent (CD) ones at LO and NLO,
in momentum-space $v_{12}^{\rm S}({\bf k}, {\bf K})=v_{12}^{\rm S, CI}({\bf k}, {\bf K})
+v_{12}^{\rm S, CD}({\bf k}, {\bf K})$ with
\begin{widetext}
\begin{eqnarray}
\!\!\!\!v_{12}^{\rm S, CI}({\bf k}, {\bf K})\!\!&=&\!\!\left(C_S+C_1\,k^2+D_1\,k^4\right)
+\left(C_2\,k^2+D_2\,k^4\right){\bm \tau}_1\cdot {\bm \tau}_2
+\left(C_T+C_3\,k^2+D_3\,k^4\right){\bm \sigma}_1\cdot {\bm \sigma}_2\nonumber\\
\!\!\!\!&&+\left(C_4\,k^2+D_4\,k^4\right){\bm \sigma}_1\cdot {\bm \sigma}_2\,{\bm \tau}_1\cdot {\bm \tau}_2
+\left(C_5+D_5\,k^2\right) S_{12}({\bf k}) +\left(C_6+D_6\,k^2\right) S_{12}({\bf k})
\,{\bm \tau}_1\cdot {\bm \tau}_2\nonumber\\
\!\!\!\!&&+i\left(C_7+D_7\,k^2\right){\bf S}\cdot \left({\bf K} \times {\bf k}\right)
+i\,D_8\,k^2\,{\bf S}\cdot \left({\bf K}\, \times {\bf k}\right){\bm \tau}_1\cdot {\bm \tau}_2
+D_{9}\left[{\bf S}\cdot \left({\bf K} \times {\bf k}\right)\right]^2+D_{10}\left({\bf K} \times {\bf k}\right)^2\nonumber\\
\!\!\!\!&&+D_{11}\left({\bf K} \times {\bf k}\right)^2{\bm \sigma}_1\cdot {\bm \sigma}_2
+D_{12}\,k^2K^2+D_{13}\,k^2K^2{\bm \sigma}_1\cdot {\bm \sigma}_2
+D_{14}\,K^2\,  S_{12}({\bf k})\nonumber\\
\!\!\!\!&&+D_{15}\,K^2\, S_{12}({\bf k})
\,{\bm \tau}_1\cdot {\bm \tau}_2
\label{eq:sci}\ , \\
\!\!\!\! v_{12}^{\rm S,CD}({\bf k}, {\bf K})\!\!&=&\!\!\left[C_0^{\rm IT}
+C_1^{\rm IT}\,k^2+C_2^{\rm IT}\,k^2\,{\bm \sigma}_1\cdot {\bm \sigma}_2+C_3^{\rm IT}\,S_{12}({\bf k})
+i\,C_4^{\rm IT}{\bf S}\cdot \left({\bf K} \times {\bf k}\right)\right]T_{12}\nonumber\\
\!\!\!\!&&+\left[C_0^{\rm IV}+C_1^{\rm IV}\,k^2+C_2^{\rm IV}\,k^2\,{\bm \sigma}_1\cdot {\bm \sigma}_2
+C_3^{\rm IV}\,S_{12}({\bf k})
+i\,C_4^{\rm IV}{\bf S}\cdot \left({\bf K} \times {\bf k}\right)\right](\tau_{1z}+\tau_{2z})\ ,
\label{eq:scib}
\end{eqnarray}
\end{widetext}
where $S_{12}({\bf k})= 3\,{\bm\sigma}_1 \cdot {\bf k}\,\, {\bm\sigma}_2 \cdot {\bf k}-
k^2\,{\bm\sigma}_1\cdot{\bm \sigma}_2$, $C_S$ and $C_T$ are the LO LEC's in standard notation, while
$C_{i=1,\dots, 7}$ and $D_{i=1,\dots, 15}$ are generally linear
combinations of those in the ``standard'' set, as defined, for example,
in Ref.~\cite{Entem11}.  In the NLO and N3LO contact interactions terms
proportional to $K^2$ and $K^4$, which would lead to $p^2$ and $p^4$
operators in coordinate space (${\bf p}\longrightarrow-i{\bm \nabla}$ is the
relative momentum operator), have been removed by a Fierz rearrangement,
for example
\begin{equation}
K^{m} \longrightarrow -\frac{1+{\bm \tau}_1\cdot{\bm \tau}_2}{2}
\,\frac{1+{\bm \sigma}_1\cdot{\bm \sigma}_2}{2}\,
 \frac{k^{m}}{2^{m}} 
\end{equation}
with $m=2$ or 4.  Of course, mixed terms of the type $k^2\, K^2$ or
${\bf K}\times {\bf k}$ cannot be Fierz-transformed away.  In the
potential $v_{12}^{\rm S, CD}({\bf k}, {\bf K})$ only terms up to
NLO, involving charge-independence breaking (proportional to $T_{12}$)
and charge-symmetry breaking (proportional to $\tau_{1z}+\tau_{2z}$),
are accounted for.  The associated LEC's, while providing some additional
flexibility in the data fitting discussed below (especially
$C_0^{\rm IV}$ in reproducing the singlet $nn$ scattering length),
are not well constrained.

A couple of comments are now in order.  The first is that
strict adherence to power counting would require inclusion of
additional one-loop as well as two-loop TPE and three-pion exchange
contributions at order $Q^4$.  These contributions have been
neglected, since they are known to be small (see, for example,
Ref.~\cite{Entem11}).  Furthermore it is the $D_i$ LEC's at $Q^4$
that are critical for a good reproduction of phase shifts in lower
partial waves, particularly D-waves, and a good fit to the $NN$
database~\cite{Entem11} in the 0--300 MeV range of energies
considered in the present study.

The second comment is in reference to isospin symmetry breaking.
We have not included explicitly contributions from OPE and one-photon
exchange~\cite{Kolck98,Kaiser06}.  As noted in Ref.~\cite{Perez:2013oba},
this $\pi$-$\gamma$ interaction is small and ambiguous, and requires
regularization at short distances.  So its main effect can be effectively
shifted into a counter-term.  While this can be improved, we will see below
our final fitting results do not seem to require these long-range isospin
breaking effects.

The potential $v_{12}^{\rm S}({\bf k}, {\bf K})$ is regularized via
a Gaussian cutoff depending only on the momentum transfer $k$,
\begin{equation}
 \widetilde{C}_{R_{\rm S}}(k)\!=\!  e^{-  R^2_{\rm S} k^2/4}\longrightarrow
 C_{R_{\rm S}}(r)=\frac{1}{\pi^{3/2}R_{\rm S}^3} e^{-(r/R_{\rm S})^2} ,
\end{equation}
which leads to a coordinate-space representation only mildly
non-local, containing at most terms quadratic in the relative
momentum operator.  It reads (see Appendix~\ref{app:a2})
\begin{eqnarray}
\label{eq:vr}
v_{12}^{\rm S}&=&\left[\sum_{l=1}^{19} v_{\rm S}^l (r) \, O^l_{12}\right]+\{\,v_{\rm S}^p(r)
+v_{\rm S}^{p\sigma}(r)\,{\bm \sigma}_1\cdot {\bm \sigma}_2\nonumber\\
&&+v_{\rm S}^{pt}(r)\,S_{12}+
v_{\rm S}^{pt\tau}(r)\,S_{12}\,{\bm \tau}_1\cdot {\bm \tau}_2\,\, ,\,\,{\bf p}^2\,\}\ ,
\end{eqnarray}
where $O^{l=1,\dots,6}_{12}$ have been defined above,
\begin{equation}
O^{l=7,\dots,11}_{12}={\bf L}\cdot{\bf S}\,,\,
 {\bf L}\cdot{\bf S}\,{\bm \tau}_1\cdot {\bm \tau}_2\, ,\, ({\bf L}\cdot{\bf S})^2\, ,\, {\bf L}^2\, ,\, 
 {\bf L}^2\, {\bm \sigma}_1\cdot {\bm \sigma}_2 \ ,
 \end{equation}
referred to as $b$, $b\tau$, $bb$, $q$, $q\sigma$, and 
\begin{equation}
O^{l=12,\dots,19}_{12}=\left[{\bf 1}\, ,\, {\bm \sigma}_1\cdot {\bm \sigma}_2\, , \,S_{12}\, ,\, 
 {\bf L}\cdot{\bf S}\right] \otimes\left[T_{12}\, ,\, \tau_1^z+\tau_2^z\right] \ ,
 \end{equation}
referred to as $T$, $\tau z$, $\sigma T$, $\sigma \tau z$, $tT$, $t \tau z$, $bT$, $b\tau z$.
The four additional terms, denoted as $p$, $p\sigma$, $pt$, and $pt\tau$, in the
anti-commutator of Eq.~(\ref{eq:vr}) are ${\bf p}^2$-dependent.
We consider, in combination with $R_L=(0.8,1.0,1.2)$ fm,
$R_{\rm s}=(0.6,0.7,0.8)$ fm, corresponding to typical momentum-space
cutoffs $\Lambda_{\rm S}=2/R_{\rm S}$ from about 660 MeV down to 500 MeV.
While the use of a Gaussian cutoff mixes up orders in the power counting---for example,
the LO contact interactions proportional to $C_S$
and $C_T$ in Eq.~(\ref{eq:sci}) generate contributions at NLO and N3LO---such
a choice nevertheless leads to smooth functions for the potential components
$v_{\rm S}^l(r)$ and the resulting deuteron waves.  Sharper cutoffs,
like those $\propto {\rm exp}\left[-(r/R)^n\right]$ with $n=4$,
as suggested in Ref~\cite{Gezerlis14}, or $n=6$, as in one of the earlier
versions of the present model, generate wiggles in the deuteron
waves at $r \sim R$ (as well as mixing
of power-counting orders).

\section{Data analysis}
\label{sec:data}
Setting aside electromagnetic (EM) contributions (Coulomb and higher order ones) for the
time being, the invariant on-shell scattering amplitude $M$ for the $NN$
system can be expressed in terms of five independent complex functions---the Wolfenstein
parametrization---as
\begin{equation}
\label{eq:eq3}
M({\bf p}^\prime,{\bf p})=a +m\, {\bm \sigma}_1\cdot \hat{\bf n}\, {\bm \sigma}_2\cdot \hat{\bf n}
+(g-h)\,{\bm \sigma}_1\cdot \hat{\bf m}\, {\bm \sigma}_2\cdot \hat{\bf m}
+(g+h)\,{\bm \sigma}_1\cdot \hat{\bf l}\, {\bm \sigma}_2\cdot \hat{\bf l}
+c \left({\bm \sigma}_1+{\bm \sigma}_2\right)\cdot \hat{\bf n} \ ,
\end{equation}
where $\hat{\bf l}$, $\hat{\bf m}$, $\hat{\bf n}$ are three
orthonormal vectors along the directions of
${\bf p }^\prime+{\bf p}$, ${\bf p}^\prime-{\bf p}$, and
${\bf p} \times {\bf p}^\prime$, and ${\bf p}^\prime$, ${\bf p}$
are the final and initial relative momenta, respectively.   The functions
$a,m,g,h$, and $c$ are taken to depend on the energy in the laboratory (lab) frame
and the scattering angle $\theta$ in the center-of-mass (cm) frame.
Any scattering observable can be constructed out of these amplitudes~\cite{Bystricky1,Bystricky2}.

The $NN$ amplitude is diagonal in pair spin $S$, and pair isospin and isospin projection
$TM_T$, and is expanded in partial waves as
\begin{eqnarray}
\label{eq:eq1}
M^{S,TM_T}_{M_S^\prime \, M_S}(E,\theta)&=& \sqrt{4\pi} \sum_{JLL^\prime}{i}^{L-L^\prime}\,
\sqrt{2L+1}\, \frac{1-(-)^{L+S+T}}{2}\,
\langle L^\prime (M_S-M^\prime_S), S M^\prime_S\mid J M_S\rangle \nonumber \\
&&\langle L 0, SM_S\mid J M_S\rangle \,Y_{L^\prime}^{M_S-M_S^\prime}(\theta,0)\,
\frac{S^{JS,TM_T}_{L^\prime  L}(p) - \delta_{L^\prime  L}}{{i}\, p} \ ,
\label{eq:am}
\end{eqnarray}
where $L$ and $J$ denote respectively the orbital and total angular momenta,
the $\langle\, \dots \, \rangle$ are Clebsch-Gordan coefficients, the $Y_{L}^{M_L}(\theta,\phi)$
are spherical harmonics, the $\delta_{L^\prime L}$ are Kronecker deltas, and the
$S^{JS,TM_T}_{L^\prime L}$ are $S$-matrix elements.
Denoting phase shifts as $\delta^{JS,TM_T}_{L^\prime L}$, the $S$-matrix
is simply given by
\begin{equation}
S^{JS}_{JJ} = e^{2i\delta^{JS}} \ ,
\end{equation}
in single channels with $L=L^\prime=J$, and by
\begin{equation}
S^{J} = \left[
\begin{array}{c c}
 e^{ 2i\delta^{J}_-}\, \cos{2 \epsilon_{J}} & 
  i \,e^{ i(\delta^{J}_- + \delta^{J}_+)} \sin{2 \epsilon_{J}}\\
 i\, e^{i(\delta^{J}_{-}+\delta^{J}_{+})} \sin{2 \epsilon_J} & e^{2i\delta^{J}_{+}} \cos{2 \epsilon_J}
\end{array}
\right],
\end{equation}
in coupled channels with $S=1$ and $L,L^\prime=J\mp 1$ ($\epsilon_J$ is
the mixing angle).  Hereafter, for notational simplicity we drop from
the phase shifts unnecessary subscripts as well as the superscripts $TM_T$,
with $T=1$ and $M_T=1,0,-1$ for respectively $pp$, $np$, and $nn$.  The $S$-matrix
elements and phase shifts are obtained from solutions of the Schr\"odinger
equation with suitable boundary conditions, as discussed Appendix~\ref{app:a3}.
In terms of the amplitudes $M^{S}_{M_S^\prime \,M_S}$, the functions $a,m,g,h$,
and $c$ then read
\begin{eqnarray}
a&=&\left( M^1_{11} +M^1_{00}+M^0_{00}+M^1_{-1-1}\right)\!/4\ , \\
 c &=& i\left(M^1_{10}-M^1_{01}+M^1_{0 -1}-M^1_{-10}\right)/(4\sqrt{2})  \ ,\\
m &=&\left(-M^1_{1-1} +M^1_{00}-M^0_{00}-M^1_{-11}\right)/4 \ ,\\
g&=& \left(M^1_{11}+M^1_{1-1}+M^1_{-11}+M^1_{-1-1}-2\, M^0_{00} \right)/8\ ,\\
h &=&{\rm cos}\,\theta\left( M^1_{11}-M^1_{1-1}-M^1_{-11}+M^1_{-1-1}-2\, M^1_{00} \right)/8\nonumber \\
&&+\sqrt{2}\, {\rm sin}\,\theta\left(M^1_{10} +M^1_{01} -M^1_{0 -1}-M^1_{-10}\right)/8 \ ,
\end{eqnarray}
and this can be further simplified by noting that $M^1_{0-1}=-M^1_{01}$,
$M^1_{1-1}=M^1_{-11}$,  $M^1_{-10}=-M^1_{10}$, and  $M^1_{11}=M^1_{-1-1}$.

When EM interactions are included,
the full scattering amplitudes $M$ are conveniently separated
into a part due to nuclear interactions and another one stemming 
from EM interactions,
\begin{equation}
 M = M_{\rm EM} + M_{\rm N}\ .
\end{equation}
The $pp$ EM amplitudes contain Coulomb with leading relativistic
corrections, vacuum polarization, and magnetic moments contributions,
whereas the $np$ ones contain magnetic moment contributions only (see
Ref.~\cite{Navarro13} for a compendium of formulas and references
to the original papers; for completeness, however, the determination of the $pp$ phase
shifts relative to EM functions and of the $pp$ effective range expansion
is summarized in Appendix~\ref{app:a4}).
Due to the finite range of the $NN$ force, the nuclear part
of the scattering amplitudes, $M_{\rm N}$, converges with a maximum
total angular momentum of $J=15$.  In contrast, EM scattering
amplitudes, $M_{\rm EM}$, require a summation of about thousand
partial waves due to the long range and tensor character of
the dipolar magnetic interactions.  While these corrections are
numerically tiny, they are nevertheless indispensable for an accurate description of
the data~\cite{Stoks:1990us}.

We use the database developed in Granada and specified in detail in
Ref.~\cite{Navarro13}, where a selection of the large collection of $np$
and $pp$ scattering data taken from 1950 till 2013 was made.  The
adopted criterium was to represent the $NN$ interaction with a general
and flexible parametrization, based on a minimal set of theoretical assumptions so as
to avoid any systematic bias in the selection process.  The aim of the
method, first suggested by Gross and Stadler~\cite{Gross08}, was to
obtain a $3\,\sigma$ self-consistent database.  This entails removing
$3\,\sigma$ outliers and re-fitting iteratively until convergence.  The procedure
results in a database with important statistical features~\cite{Navarro14}
and therefore amenable to statistical analysis, and leads to the identification of a
consistent subset among the large body of 6713 $np$ and
$pp$ experimental cross sections and polarization observables~\footnote
%
%
{This implies that experiments where the
errors are overestimated or underestimated by the experimentalists
may be rejected, not by the model itself, but by the incompatibility
with the rest of the copious data proven to be faithfully
represented by the model.  An extensive discussion of these issues is
presented in Refs.~\cite{Navarro13,Navarro14}.
%
%
}. 
In the present study, in particular, we
are concerned with a subset of this $3\, \sigma$-self-consistent
database, namely data below 300 MeV lab energy.  This database is
organized in the following way: there are $N$ sets of data, each one
corresponding to a different experiment.  Each data set contains
measurements at fixed $E_{\rm lab}$ and different scattering angles
$\theta$. However a  few observables are measured at different
$E_{\rm  lab}$ and fixed $\theta$, like, for example, total cross sections
since their measurement does not involve the scattering angle
($\theta=0$).  An experiment may have a specified systematic error
(normalized data), no systematic error (absolute data), or an
arbitrarily large systematic error (floated data).

We briefly describe the fitting procedure.  The total figure of merit is
defined as the usual $\chi^2$ function
\begin{equation}
 \chi^2= \sum_{t=1}^N \chi_t^2 \, , 
\label{eq:chi2}
\end{equation}
where $\chi_t^2$ refers to the corresponding contribution from each
data set, which we explain next.  In all cases, the $\chi_t^2$ for a
data set is given by
\begin{equation}
 \chi^2_t=\sum_{i=1}^{n}\frac{\left(o_i/Z_t-t_i\right)^2}{\left(\delta o_i/Z_t\right)^2}+\frac{(1-1/Z_t)^2}{(\delta_{\rm sys}/Z_t)^2}\ ,
\label{eq:chi2a}
\end{equation}
where $o_i$ and $t_i$ are the measured and calculated values
of the observable at point $i$, $\delta o_i$ and $\delta_{\rm sys}$
are the statistical and systematic errors, respectively, and $Z_t$ is
a scaling factor chosen to minimize the $\chi^2_t$,
\begin{equation}
 Z_t=\left(\sum_{i}^{n}\frac{o_i t_i}{\delta o_i^2}+\frac{1}{\delta_{\rm sys}^2}\right)\Bigg/
\left(\sum_{i}^{n}\frac{ t_i^2}{\delta o_i^2}+\frac{1}{\delta_{\rm sys}^2}\right)\ .
\label{eq:norm}
\end{equation}
The last term in Eq.~(\ref{eq:chi2a}) is denoted $\chi^2_{\rm sys}$.
For absolute data $Z=1$ and $\chi^2_{\rm sys}=0$, while for floated data
use of Eq.~(\ref{eq:norm}) is made with $\delta_{\rm sys}=\infty$ so that
$\chi^2_{\rm sys}=0$.  Normalized data have in most cases $Z\neq 1$
such that $\chi^2_{\rm sys}\neq 1$ and the normalization is counted as
an extra data point~\footnote
%
%
{This actually introduces some model
dependence, since normalization of experimental data is in the eyes
of the beholder, that is different models fitting the {\it same} data,
may yield strictly speaking {\it different} values of $Z$ although
not statistically significant differences in the values; what
changes from potential to potential are the correlations between the
normalization of data and the energy dependence.
}.
%
%
For some normalized data the systematic error can give a rather large
$\chi^2_{\rm sys}$ due to an underestimation of $\delta_{\rm sys}$.  In order
to account for this, we float data that have $\chi^2_{\rm sys}> 9$ and no extra
normalization data is counted.
This is in line with the criterion used to build the $pp$ and $np$
database.  Finally, the total $\chi^2$ is the sum of all the
$\chi^2_t$ for each $pp$ and $np$ data set.

The minimization of the objective function $\chi^2$ with respect to the
LEC's in Eqs.~(\ref{eq:sci}) and~(\ref{eq:scib}) is carried out 
with the Practical Optimization Using no Derivatives (for
Squares), POUNDerS~\cite{POUNDerS}.  This derivative-free algorithm is
designed for minimizing sums of squares and uses interpolation
techniques to construct residuals at each point.  In the optimization
procedure, we fit first phase shifts and then refine the fit by
minimizing the $\chi^2$ obtained from a direct comparison with the
database.  In fact, sizable changes in the total $\chi^2$ are
found when passing from phase shifts to observables, so this
refining is absolutely necessary to claim reasonable fits to
data.  This is a general feature which is often found, and reflects
the different weights in the $\chi^2$ contributions of the two
different fitting schemes.  Indeed, the initial guiding fit to phase shifts
chooses a prescribed energy grid arbitrarily, which {\it does not}
correspond directly to measured energies, nor necessarily samples
faithfully the original information provided by the experimental data.
Moreover, there are different PWA's which
describe the same data but yield different phase shifts with 
significantly larger discrepancies than reflected by the inferred statistical
uncertainties~\cite{Perez:2013oba,Navarro13,Navarro14}.

\section{Results}
\label{sec:res}
We report results for the potentials $v_{12}+v_{12}^{\rm EM}$ corresponding to three
different choices of cutoffs $(R_{\rm L},R_{\rm S})$: model a with $(1.2,0.8)$ fm,
model b with $(1.0,0.7)$ fm, and model c with $(0.8,0.6)$ fm. 
\begin{table*}[bth]
\caption{\label{tab:table4}Total $\chi^{2}$ for model a with $(R_{\rm L},R_{\rm S})=(1.2,0.8)$ fm,
model b with $(1.0,0.7)$ fm, and model c $(0.8,0.6)$ fm, and the AV18;
$N_{pp}$ ($N_{np}$) denotes the number of $pp$ ($np$) data, including observables and normalizations.}
\begin{ruledtabular}
\begin{tabular}{crrrrrrrr|rrrrrrrr}
  \multicolumn{1}{c}{} &  & & & &  \multicolumn{2}{c}{$\chi^{2}(pp)$}
                       &  & & & && &\multicolumn{2}{c}{$\chi^{2}(np)$}  \\
\textrm{Lab Energy (MeV)} & 
$N_{pp}^{\rm a}$ &$N_{pp}^{\rm b}$& $N_{pp}^{\rm c}$&$N_{pp}^{18}$&$v_{12}^{\rm a}$  & $v_{12}^{\rm b}$  & $v_{12}^{\rm c}$&  $v_{18}$ & $N_{np}^{\rm a}$ & $N_{np}^{\rm b}$ &$N_{np}^{\rm c}$& $N_{np}^{18}$&$v_{12}^{\rm a}$&  $v_{12}^{\rm b}$ &  $v_{12}^{\rm c}$& $v_{18}$\\
\tableline
          0--300  &2262 &2260 &2258 & 2269 &3353 &3345  &3430  &4191 &2957 &2954 &2949 &2961 &3548 &3523 &3636 &3391
\end{tabular}
\end{ruledtabular}
\end{table*}
Models a,b, and c were fitted to the Granada database of $pp$ and $np$
cross sections, polarization observables, and normalizations up to lab energies of 300 MeV,
to the $pp$, $np$, and $nn$ singlet scattering lengths, and to the deuteron
binding energy.   We list the number of $pp$ and $np$ data (including normalizations)
and corresponding total $\chi^2$ for the three models in Table~\ref{tab:table4},
where we also report for comparison the $\chi^2$ corresponding to the AV18~\cite{Wiringa95}
(of course, without a refit of it) and the same database.  The total number of data
points changes slightly for each of the various models because of fluctuations in the
number of normalizations included in the database according to the criterion
discussed at the end of the previous section.  In the range (0--300) MeV,
the $\chi^2(pp)$/datum and $\chi^2(np)$/datum are about 1.48, 1.48, 1.52
and 1.20, 1.19, 1.23 for models a, b, and c, respectively; the corresponding
global $\chi^2(pp+np)$/datum are 1.33, 1.33, 1.37.  For the AV18, the
$\chi^2(pp)$/datum, $\chi^2(np)$/datum, and global $\chi^2(pp+np)$/datum
are 1.84, 1.14, and 1.46, respectively.  Note that the global $\chi^2$ values
above have been evaluated by taking into account the number of fitting parameters
characterizing these models (34 in the case of models a, b, and c). 
Errors for $pp$ data are significantly smaller than for $np$, thus explaining
the consistently higher $\chi^2(pp)$/datum.  The quality of the fits deteriorates
slightly as the $(R_{\rm L},R_{\rm S})$ cutoffs are reduced from the values
(1.2,0.8) fm of model a down to (0.8,0.6) fm of model c.

The fitted values of the LEC's in Eqs.~(\ref{eq:sci}) and~(\ref{eq:scib})
corresponding to models a, b, and c are listed in Table~\ref{tab:table5}.
The values for the $\pi N$ LEC's in the OPE and TPE terms of these models
have already been given in Tables~\ref{tab:table} and~\ref{tab:table1}.
It is interesting to examine the extent to which these LEC's satisfy
the requirement of naturalness.  To this end, following Machleidt and Entem~\cite{Entem11},
we note that this criterion would imply that the LEC's of the charge-independent
part $v^{\rm S, CI}_{12}$ of the contact potential have the following magnitudes
\begin{equation}
\mid\! C_{S,T}\!\mid\, \sim \frac{1}{f_\pi^2} \simeq 4.6 \,\, {\rm fm}^2\ , \qquad
\mid\! C_i\!\mid\, \sim \frac{1}{\Lambda_\chi^2 \,f_\pi^2} \simeq  0.18\,\, {\rm fm}^4\ , \qquad
\mid\! D_i\!\mid\,  \sim \frac{1}{\Lambda_\chi^4 \,f_\pi^2} \simeq 0.0070\,\, {\rm fm}^6 \ ,
\end{equation}
where $f_\pi=92.4$ MeV and $\Lambda_\chi=1$ GeV.
\begin{table*}[bth]
\caption{\label{tab:table5}Fitted values of the LEC's corresponding to potential models a, b, and c.
The notation $(\pm \,n)$ means $10^{\pm n}$.}
\begin{ruledtabular}
\begin{tabular}{lddd}
\textrm{LECs}&
\multicolumn{1}{c}{Model a}&
\multicolumn{1}{c}{Model b}&
\multicolumn{1}{c}{Model c}\\
\colrule
  $C_S$ (fm$^2$) 			& 0.2003672(+1)	&0.8841864(+1)	&0.2588776(+2) \\
 $C_T$ (fm$^2$) 			&-0.1660743(+1)	&-0.4168038(+1)	&-0.9160861(+1)\\
\hline
 $C_1$ (fm$^4$)			&-0.1759574		&-0.9367926(-1)	&-0.4455626(-3)\\
 $C_2$ (fm$^4$)			&-0.2029026	 	&-0.2520756		&-0.3082608 \\
 $C_3$ (fm$^4$)			&-0.1856897		&-0.2589016		&-0.3222661 \\
 $C_4$ (fm$^4$)			&-0.5745498(-1)	&-0.2453381(-1)	&0.3773411(-1)\\
 $C_5$ (fm$^4$)			&-0.8813877(-1)	&-0.4685034(-1)	&-0.5156581(-2)\\
 $C_6$ (fm$^4$)			&-0.5857848(-1)	&-0.2804770(-1)	&-0.2762013(-1)\\
 $C_7$ (fm$^4$)			&-0.1140923		&0.7338611		&0.7568732\\
\hline
 $D_1$ (fm$^6$)			&-0.9498379(-1)	&-0.6986704(-1)	&-0.2565252(-1)\\
 $D_2$ (fm$^6$)			&-0.7149729(-2)	&0.1681828(-3)	&0.4909682(-2)\\
 $D_3$ (fm$^6$)			&-0.6502509(-2)	&-0.6355876(-2)	&-0.1721433(-1) \\
 $D_4$ (fm$^6$)			&-0.3217370(-2)	&-0.1153354(-2)	&0.2592172(-2) \\
 $D_5$ (fm$^6$)			&0.2692050(-2)	&0.2258031(-2)	&0.2101464(-2)\\
 $D_6$ (fm$^6$)			&-0.6654712(-2)	&-0.2757790(-2)	&-0.4252508(-2)\\
 $D_7$ (fm$^6$)			&-0.2318069(-1)	&0.1451856(-1)	&0.4247406(-1)\\
 $D_8$ (fm$^6$)			&-0.2899833(-1)	&-0.2897869(-1)	&-0.1122591(-1)\\
 $D_9$ (fm$^6$)			&0.2634392(-2)	&0.3909073(-1)	&0.4966263(-1)\\
 $D_{10}$ (fm$^6$)		&-0.1787025		&-0.2061108		&-0.1628166\\
 $D_{11}$ (fm$^6$)		&0.1758785(-1)	&0.3667628(-2)	&-0.2316157(-1)\\
 $D_{12}$ (fm$^6$)		&0.1126531		&0.1023936		&0.5361795(-1) \\
 $D_{13}$ (fm$^6$)		&-0.1649902(-1)	&-0.9890485(-2)	&0.1744601(-2)\\
 $D_{14}$ (fm$^6$)		&0.1989863(-2)	&0.3066270(-2)	&0.7219031(-2)\\
 $D_{15}$ (fm$^6$)		&0.4540768(-2)	&0.2426771(-2)	&0.2979197(-2)\\
\hline\hline
 $C_0^{\rm IV}$ (fm$^2$)	&-0.8730299(-1)	&-0.1162192		&0.6195324\\
 $C_0^{\rm IT}$ (fm$^2$)	&0.5804662(-1)	&0.6669167(-1)	&0.7020630(-1)\\
\hline
 $C_1^{\rm IV}$ (fm$^4$)	&0.6961072(-1)	&0.5088496(-1)	&0.2174468(-1)\\
 $C_2^{\rm IV}$ (fm$^4$)	&0.3507986(-1)	&0.2288370(-1)	&-0.8112580(-2)\\
 $C_3^{\rm IV}$ (fm$^4$)	&0.3862077(-1)	&-0.7707131(-2)	&-0.6115902(-1)\\
 $C_4^{\rm IV}$ (fm$^4$)	&-0.7617836		&-0.1581137(+1)	&-0.1533212(+1) \\ 
 $C_1^{\rm IT}$ (fm$^4$)	&-0.2382471(-1)	&-0.2373048(-1)	&0.7623486(-2)\\
 $C_2^{\rm IT}$ (fm$^4$)	&-0.1325513(-1)	&-0.1013726(-1)	&0.1205547(-2)\\
 $C_3^{\rm IT}$ (fm$^4$)	&-0.1399371(-1)	&-0.1098114(-3)	&0.2109716(-1)\\
 $C_4^{\rm IT}$ (fm$^4$)	&0.2582607		&0.5180368		&0.4955952
\end{tabular}
\end{ruledtabular}
\end{table*}
A glance at Table~\ref{tab:table5} indicates that the LEC's are
generally natural, but for the following exceptions: $C_{S,T}$ in model c,
$C_7$ in models b and c, and $D_1$, $D_{10}$, and $D_{12}$ in al three
models considered.  As already noted, however, the use of a (momentum-space)
Gaussian cutoff mixes orders in the power expansion, since
\begin{equation}
 e^{-R_{\rm S}^2\, k^2/4} = 1 -\frac{R_{\rm S}^2\, k^2}{4} + \frac{R_{\rm S}^4\, k^4}{32}+\dots
\end{equation}
and, as an example, the spin-isospin independent central component
of $v^{\rm S, CI}_{12}$, after inclusion of this cutoff, is modified as
\begin{equation}
C_S+\left(C_1-\frac{R_{\rm S}^2}{4}\, C_S \right) k^2+\left(D_1-\frac{R_{\rm S}^2}{4}\, C_1
+\frac{R_{\rm S}^4}{32}\, C_S \right) k^4 +\dots \ ,
\end{equation} 
suggesting that some of the LEC's multiplying terms linear and quadratic
in $k^2$ may not be natural after all.

In order to estimate the size of the (nominally) LO ($Q^0$) and NLO ($Q^2$)
LEC's associated with the charge-dependent part $v^{\rm S, CD}_{12}$ of the contact
potential, we note that the terms proportional to $C^{\rm IV}_0$
and $C^{\rm IT}_0$ in Eq.~(\ref{eq:scib}) should scale respectively as
$\epsilon \, m^2_\pi$ and $\Delta m_\pi^2$, where $\epsilon$ is related to the $u$-$d$
quark mass difference---we assume that $\epsilon \sim e =\sqrt{4\pi\alpha}$, $e$ being the electric charge
and $\alpha$ the fine structure constant---and
$\Delta m_\pi^2$ is the squared-mass difference between
the charged and neutral pions.  Consequently, one would expect for the LO LEC's
\begin{equation}
\mid \! C^{\rm IV}_0\!\mid \sim \frac{\sqrt{4\pi \alpha}}{\Lambda_\chi^2}\simeq 0.012\,\, {\rm fm}^2\ , \qquad
\mid\!C^{\rm IT}_0\!\mid\, \sim \frac{\Delta m_\pi^2}{m_\pi^2} \, \frac{1}{f_\pi^2}\simeq 0.15\,\, {\rm fm}^2 \ ,
\end{equation}
and for the NLO LEC's
\begin{equation}
\mid\!C^{\rm IV}_i\!\mid \, \sim  \frac{\sqrt{4\pi \alpha}}{\Lambda_\chi^4}\sim 0.0005 \,\, {\rm fm}^4 \ ,\qquad
 \mid\!C^{\rm IT}_i\!\mid\, \sim \frac{\Delta m_\pi^2}{m_\pi^2} \, \frac{1}{\Lambda_\chi^2 f_\pi^2} \simeq 0.0058 \,\, {\rm fm}^4 \ .
\end{equation}
These expectations are not borne out by the actual values reported in Table~\ref{tab:table5}.
Particularly striking are the very large values
obtained for the LEC's $C_4^{\rm IV}$ and $C_4^{\rm IT}$ associated
with the spin-orbit term.

\begin{center}
\begin{figure*}[bth]
\includegraphics[width=7in]{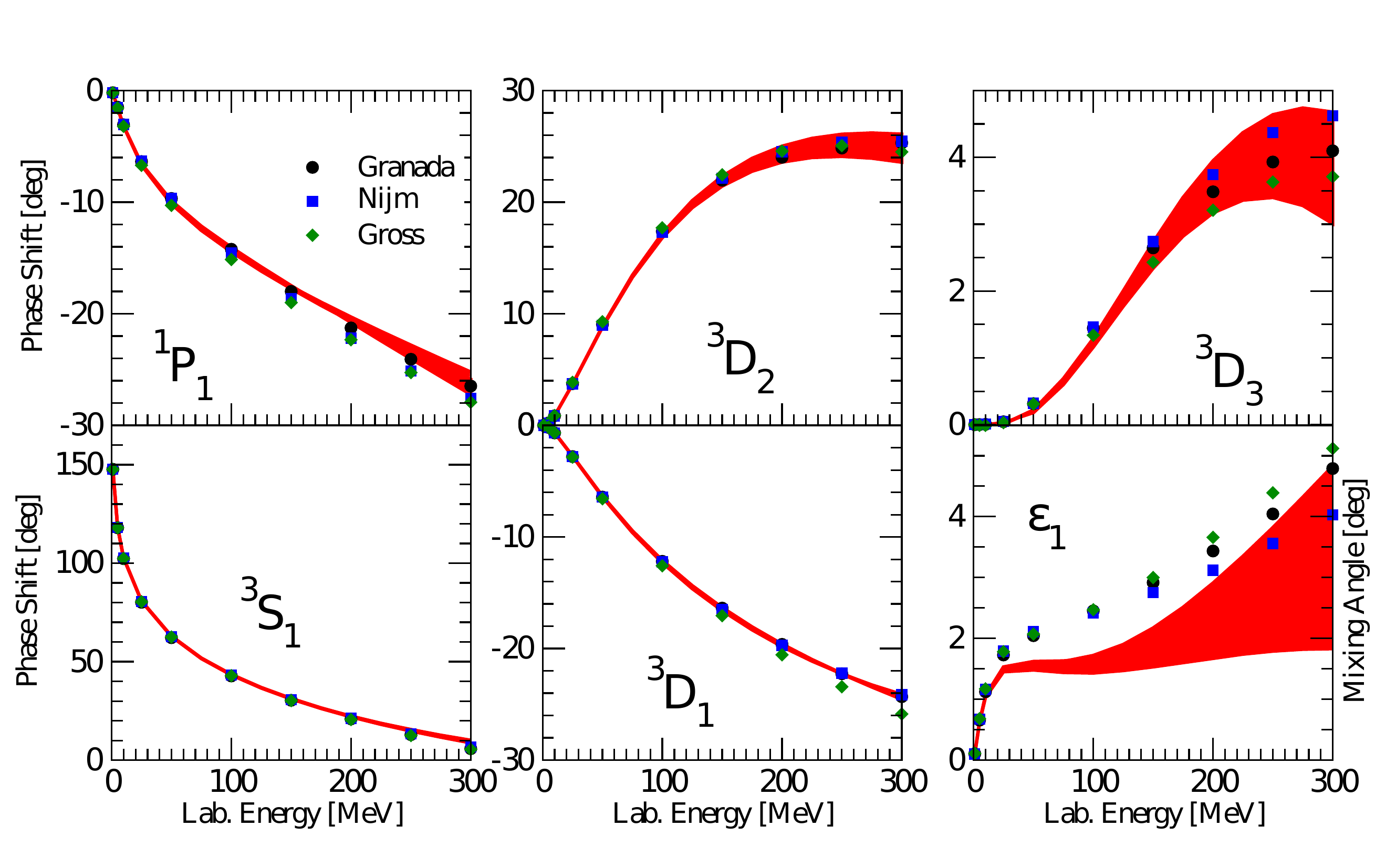}
\caption{(Color online) S-wave, P-wave, and D-wave phase shifts in the $np$ $T$=0 channel, obtained
in the Nijmegen~\cite{Stoks93,Stoks94}, Gross and Stadler~\cite{Gross08}, and
Navarro P\'erez {\it et al.}~\cite{Navarro13} partial-wave
analyses, are compared to those of models a, b, and c, indicated by the band.
For the mixing angle $\epsilon_1$ (phase shift $^3$D$_3$)
the lower limit of the band corresponds to model a (model b) and the
upper limit to model c (model c).}
\label{fig:f2}
\end{figure*}
\end{center}

\begin{center}
\begin{figure*}[bth]
\includegraphics[width=7in]{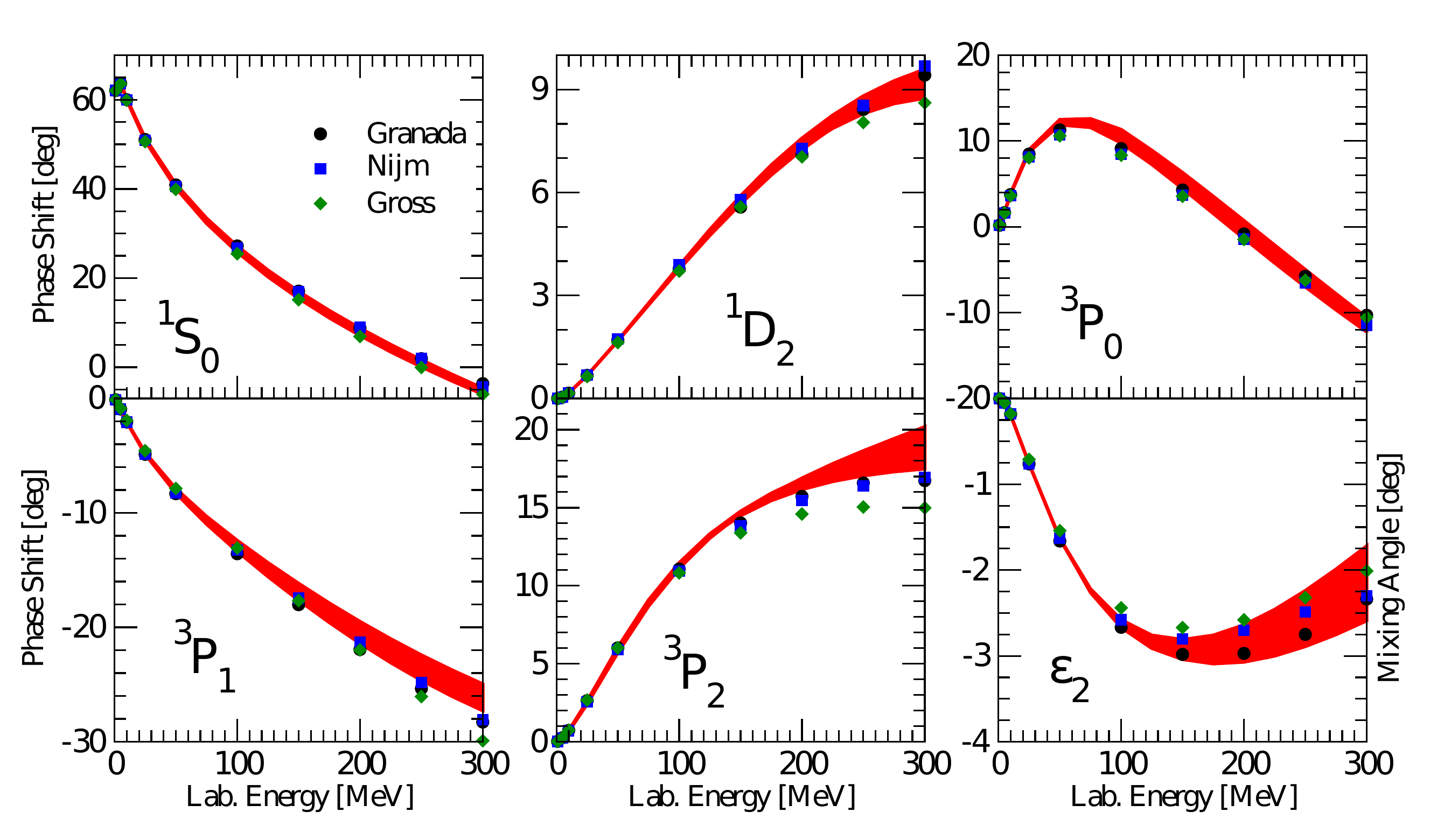}
\caption{(Color online) Same as in Fig.~\protect\ref{fig:f2}, but for the
S-wave, P-wave, and D-wave phase shifts in the $np$ $T$=1 channel.
For the mixing angle $\epsilon_2$
the lower limit of the band corresponds to model c and the
upper limit to model b.}
\label{fig:f3}
\end{figure*}
\end{center}

\begin{center}
\begin{figure*}[bth]
\includegraphics[width=7in]{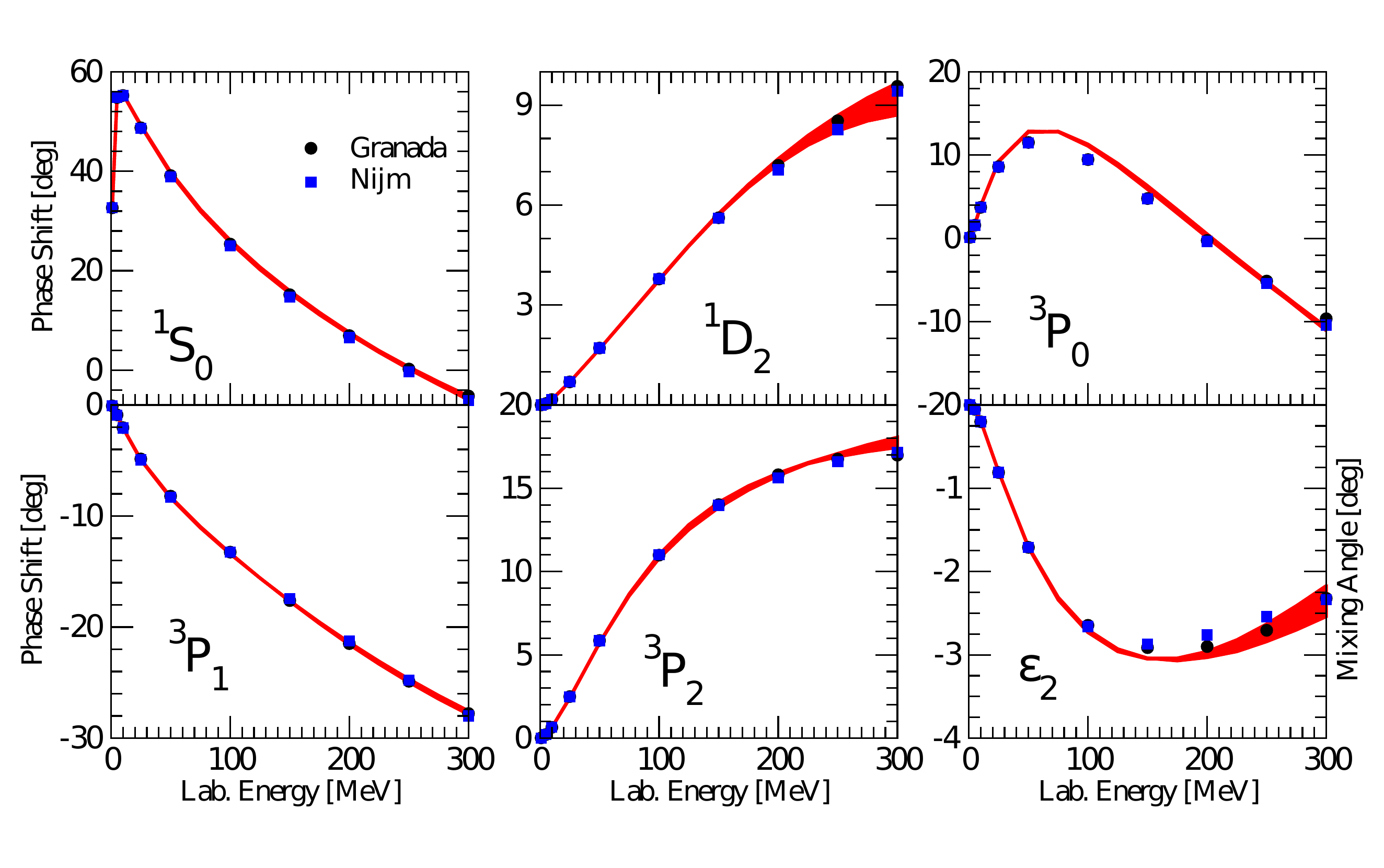}
\caption{(Color online) S-wave, P-wave, and D-wave phase shifts in the $pp$ $T$=1 channel,
obtained in the Nijmegen and Navarro P\'erez {\it et al.} partial-wave
analyses, are compared to those of models a, b, and c, indicated by the band. }
\label{fig:f4}
\end{figure*}
\end{center}

The S-wave, P-wave, and D-wave phase shits for $np$ (in $T=0$ and $T=1$) and
$pp$ are displayed in Figs.~\ref{fig:f2}--\ref{fig:f4} up to 300 MeV lab energies.
The phases calculated with the full models a, b, and c including strong and
electromagnetic interactions are represented by the band.  The $np$ phases
are relative to spherical Bessel functions, while the $pp$ phases are with respect
to electromagnetic functions (see Appendix~\ref{app:a4}).  The cutoff sensitivity,
as represented by the width of the shaded band, is very weak for $pp$, and
generally remains modest for $np$, except for the $T=0$ $^3$D$_3$ phase
and $\epsilon_1$ mixing angle, particularly for energies larger than 150 MeV.
The calculated phases are compared to those obtained in partial-wave analyses (PWA's)
by the Nijmegen~\cite{Stoks93,Stoks94}, Granada~\cite{Navarro13}, and
Gross-Stadler~\cite{Gross08} groups.  Note that the recent Gross and Stadler's
PWA was limited to $np$ data only.  We also should point out that, since the
Nijmegen's PWA of the early nineties which was based on about 1780 $pp$
and 2514 $np$ data in the lab energy range 0--350 MeV, the $NN$ elastic
scattering database has increased very significantly.  Indeed, in the same energy
range the 2013 Granada database contains a total of 2972 $pp$ and 4737 $np$ data.
Especially for the higher partial waves in the $np$ sector and at the larger energies
there are appreciable differences between these various PWA's.  It is also
interesting to observe that these differences are most significant for the $T=0$
$^3$D$_3$ phase and $\epsilon_1$ mixing angle, and therefore correlate
with the cutoff sensitivity displayed in these cases by models a, b, and c.

The low-energy scattering parameters are listed in Table~\ref{tab:table2}, where they are compared
to experimental results.  The singlet and triplet $np$, and singlet $pp$ and $nn$, scattering lengths
are calculated with and without the inclusion of electromagnetic interactions.  Without the latter,
the effective range function is simply given by $F(k^2)=k\, \cot \delta=-1/a+r \, k^2/2$ up to terms
linear in $k^2$.  In the presence of electromagnetic interactions, a more complicated
effective range function must be used; it is reported in Appendix~\ref{app:a4}, along with
the relevant references.  The latest determinations of the empirical values for the singlet
scattering lengths and effective ranges, obtained by retaining only strong interactions (hence the
superscript N), are~\cite{Miller90,Machleidt01,Gonzalez06,Chen08} (as reported in Ref.~\cite{Entem11}):
\begin{eqnarray}
^1a^{\rm N}_{pp} &=& -17.3 \pm 0.4\,\, {\rm fm}\ , \qquad ^1r^{\rm N}_{pp} = 2.85\pm 0.04\,\, {\rm fm}\ , \\
^1a^{\rm N}_{np} &=& -23.74 \pm 0.02\,\, {\rm fm}\ , \qquad ^1r^{\rm N}_{np} = 2.77\pm 0.05\,\, {\rm fm}\ ,\\
^1a^{\rm N}_{nn} &=& -18.95 \pm 0.4\,\, {\rm fm}\ , \qquad ^1r^{\rm N}_{nn} = 2.75\pm 0.11\,\, {\rm fm}\ , 
 \end{eqnarray}
which imply that charge symmetry and charge independence are broken respectively by
\begin{equation}
\Delta a_{\rm CSB}=a_{pp}^{\rm N}-a_{nn}^{\rm N} = 1.65 \pm 0.60 \,\, {\rm fm} \ , \qquad
\Delta r_{\rm CSB}=r_{pp}^{\rm N}-r_{nn}^{\rm N} =0.10 \pm 0.12 \,\, {\rm fm} \ ,
\end{equation}
and
\begin{equation}
\Delta a_{\rm CIB}=(a_{pp}^{\rm N}+a_{nn}^{\rm N})/2- a_{np}^{\rm N}= 5.6 \pm 0.6 \,\, {\rm fm} \ , \qquad
\Delta r_{\rm CIB}=(r_{pp}^{\rm N}+r_{nn}^{\rm N})/2- r_{np}^{\rm N} =0.03 \pm 0.13 \,\, {\rm fm} \ .
\end{equation}
The more significant values for $\Delta a_{\rm CSB}$ and $\Delta a_{\rm CIB}$ can be compared to those
inferred from Table~\ref{tab:table2}:
$(\Delta a_{\rm CSB},\Delta a_{\rm CIB})=(2.13,\,5.11)$ fm  for model a, (2.34, 5.12) fm for model b, and
(1.90, 5.08) fm for model c.

\begin{center}
\begin{figure}[bth]
\includegraphics[width=7in]{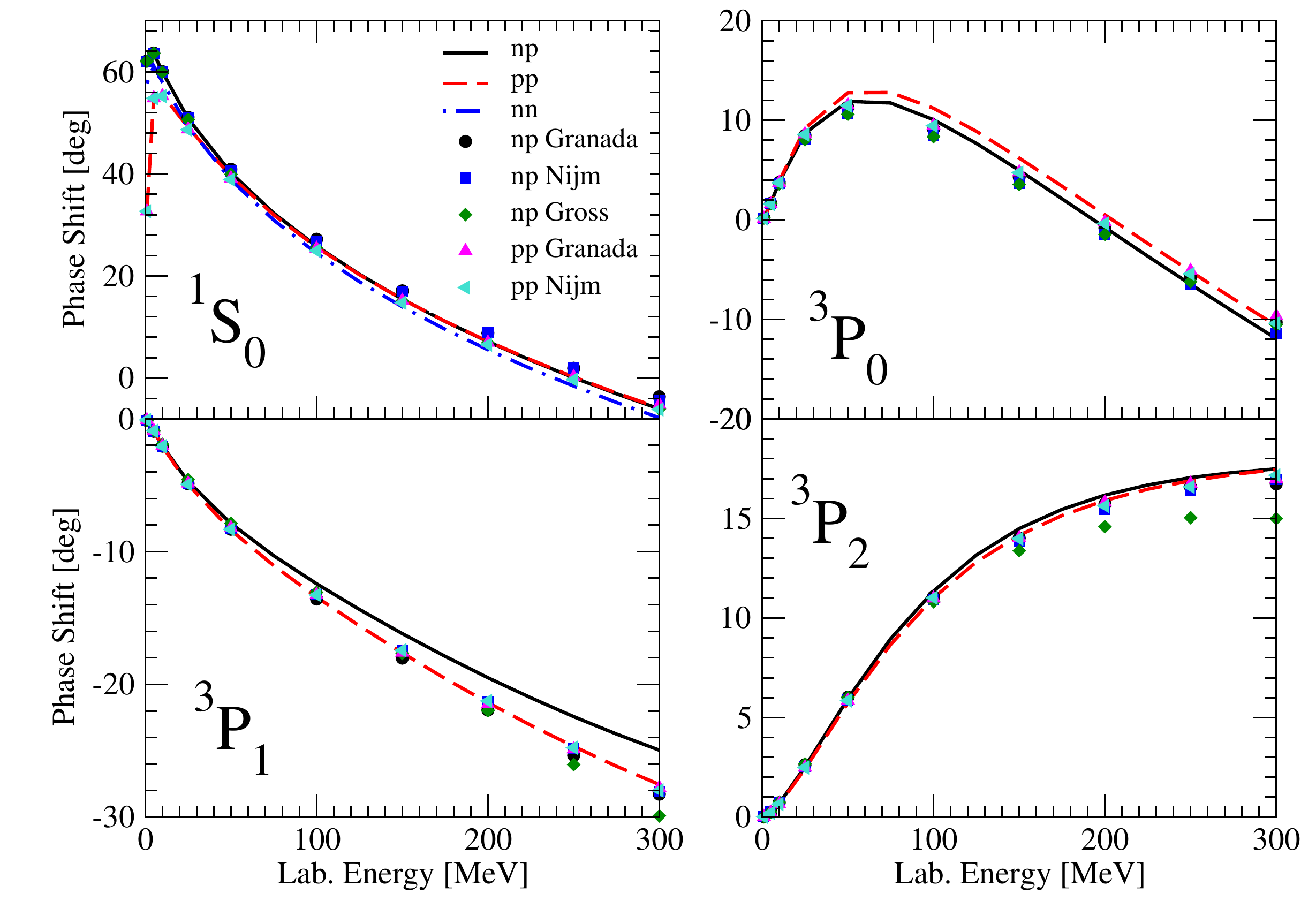}
\caption{(Color online) The $pp$, $np$, and $nn$ $^1$S$_0$ and the $pp$ 
and $np$ $^3$P$_0$, $^3$P$_1$, and $^3$P$_2$ phase shifts obtained with potential model b,
including the full electromagnetic component.}
\label{fig:f1s0em}
\end{figure}
\end{center}

\begin{center}
\begin{figure}[bth]
\includegraphics[width=7in]{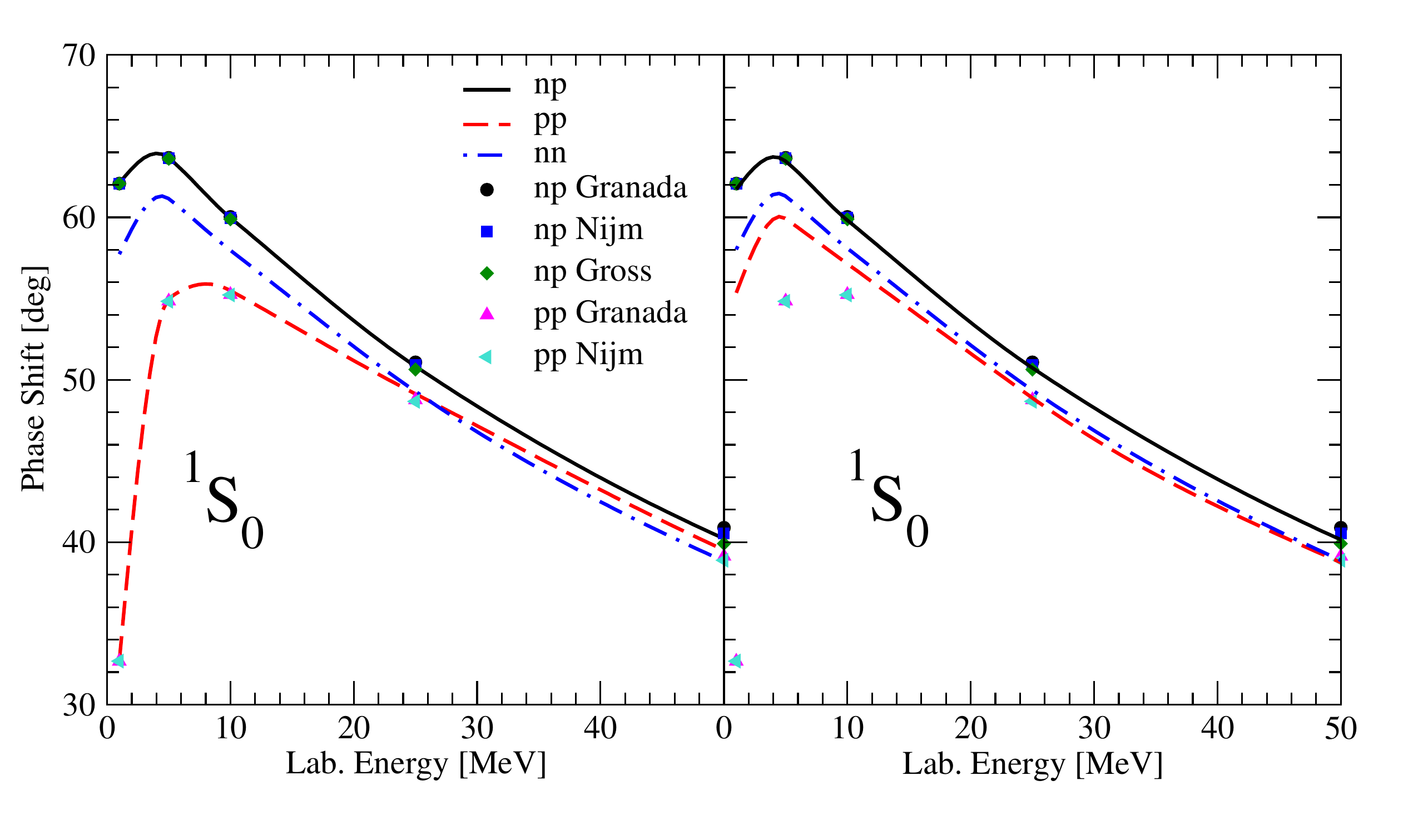}
\caption{(Color online) The $pp$, $np$, and $nn$ $^1$S$_0$ up to lab energy of 50 MeV including (panel left)
and ignoring (panel right) the full electromagnetic component of potential model b. }
\label{fig:f1s0emred}
\end{figure}
\end{center}

In the left upper panel of Fig.~\ref{fig:f1s0em} we show the $^1$S$_0$ phase shifts
for $pp$, $np$ and $nn$ calculated with and without the inclusion of electromagnetic
interactions (only model b is considered).  There is excellent agreement between these
phases and those obtained in the the Granada, Gross and Stadler, and Nijmegen PWA's,
when electromagnetic effects are fully accounted for.  Particularly at low energies (see
Fig.~\ref{fig:f1s0emred}), the latter provide most of the splitting between the $pp$ and
$np$ phases, with remaining differences originating from isospin symmetry breaking
due to the OPE term in $v^{\rm L}_{12}$ and the central terms in $v^{\rm S, CD}_{12}$,
proportional to the LEC's $C_i^{\rm IT}$ and $C_i^{\rm IV}$ with $i=0$--2.
In the absence of electromagnetic interactions, the splitting between the $pp$ and
$nn$ $^1$S$_0$ phases is induced by the charge-symmetry breaking terms of
$v^{\rm S, CD}_{12}$ proportional to the LEC's $C_i^{\rm IV}$ with $i=0$--2; it is
smaller than that between $pp$ and $np$ $^1$S$_0$ phases.

The effects of isospin symmetry breaking are also seen in the $pp$ and
$np$ $^3$P$_J$ phases with $J=0,1,2$ in the upper right and lower
panels of Fig.~\ref{fig:f1s0em}, especially at the higher energies.  The
calculated phases, which correspond again to model b, include electromagnetic
effects, but the latter are negligible beyond 100 MeV.  The splitting between
the $pp$ and $np$ $^3$P$_J$ phases is mostly due to the isotensor and
isovector terms of $v^{\rm S,CD}_{12}$, in particular those proportional to
the LEC's $C_i^{IV}$ and $C_i^{IT}$ with $i=3$ and 4 associated respectively
with the tensor and spin-orbit components of $v^{\rm S,CD}_{12}$---we
have already remarked on the unnaturally large values obtained for
$C_4^{IV}$ and $C_4^{IT}$ in the fits.  There is no evidence on the basis
of the Granada and Nijmegen PWA's for such a large splitting, and so
the latter is likely to be an artifact of the parametrization adopted for $v^{\rm S,CD}_{12}$.
\begin{table*}[bth]
\caption{\label{tab:table2}
The singlet and triplet $np$, and singlet $pp$ and $nn$, scattering lengths and
effective ranges corresponding to the three potential models with
$(R_{\rm L},R_{\rm S})$=(1.2,0.8) fm (model a), (1.0,0.7) fm (model b), and
(0.8,0.6) fm (model c).}
\begin{ruledtabular}
\begin{tabular}{lddddddd}
\textrm{}&
\multicolumn{1}{r}{\textrm{Experiment}}&
\multicolumn{1}{c}{\textrm{$v_{12}^a$}}&
\multicolumn{1}{c}{\textrm{w/o $v_{12}^{\rm EM}$}}&
\multicolumn{1}{c}{\textrm{$v_{12}^b$}}&
\multicolumn{1}{c}{\textrm{w/o $v_{12}^{\rm EM}$}}&
\multicolumn{1}{c}{\textrm{$v_{12}^c$}}&
\multicolumn{1}{c}{\textrm{w/o $v_{12}^{\rm EM}$}}\\
\colrule
$^{1}a_{pp}$ &-7.8063(26)		&-7.766 	 	&-17.014 	&-7.766	&-16.956		&-7.763	&-17.137\\
				  &-7.8016(29)		&	 	 	&			&		&			&&\\
$^{1}r_{pp}$	& 2.794(14)   		& 2.742		&2.818		&2.743	&2.820		&2.730	&2.802\\
 				& 2.773(14)   		& 			&			&		&			&		&\\
$^{1}a_{nn}$	&-18.90(40)		&-18.867 	&-19.148		&-19.025	&-19.301		&-18.719	&-19.039\\
$^{1}r_{nn}$	& 2.75(11)  		& 2.831		&2.827		&2.799	&2.795		&2.738	&2.732\\
$^{1}a_{np}$	&-23.740(20) 		&-23.752		&-23.196		&-23.755	&-23.248		&-23.745 &-23.167\\
$^{1}r_{np}$	& 2.77(5)			&2.665		&2.670		&2.672	&2.677		&2.638	&2.644\\
$^{3}a_{np}$	& 5.419(7)  		& 5.408		&5.391		&5.404	&5.389		&5.412	&5.396\\
$^{3}r_{np}$	& 1.753(8)   		&1.741 		&1.740		&1.737	&1.734		&1.740	&1.745\\
\end{tabular}
\end{ruledtabular}
\end{table*}

\begin{center}
\begin{figure*}[bth]
\includegraphics[width=7in]{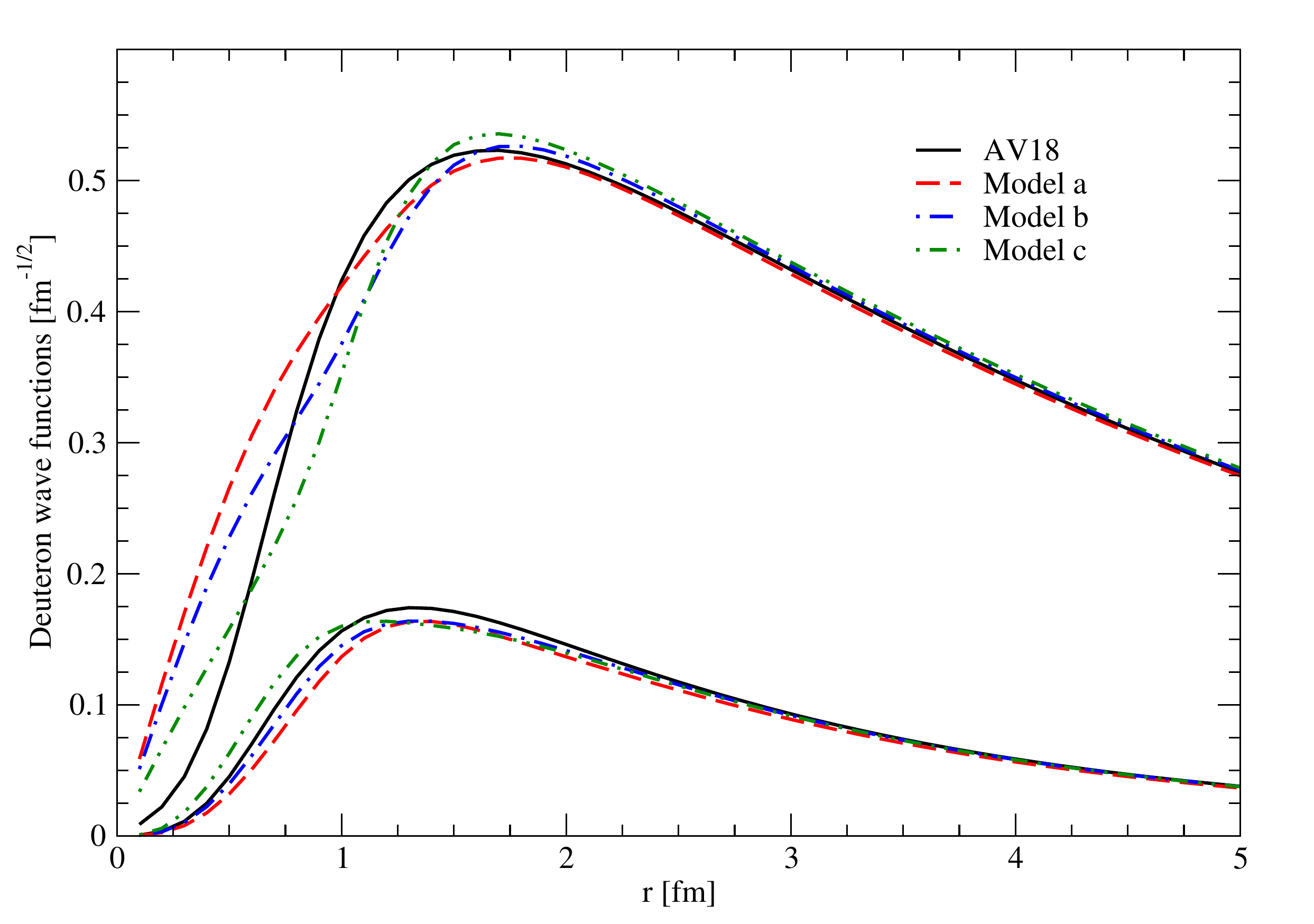}
\caption{(Color online) The $S$-wave and $D$-wave components of the deuteron wave function corresponding
to models a (dashed lines), b (dotted-dashed lines) and c (dotted-dashed-dotted lines) 
are compared with those corresponding to the AV18 (solid lines).}
\label{fig:f5}
\end{figure*}
\end{center}

The static deuteron properties are shown in Table~\ref{tab:table3} and compared
to experimental values~\cite{Vandl82,Ericson83,Rodning90,Huber98,Bishop79}.
The binding energy $E_d$ is fitted exactly and includes the contributions (about
20 keV) of electromagnetic interactions, among which the largest is that due to the
magnetic moment term.  The asymptotic S-state normalization, $A_{\rm S}$,
and the D/S ratio, $\eta$, are both $\sim 2$ standard deviations from experiment
for all models considered.  The deuteron (matter) radius, $r_d$, is exactly reproduced
with model b, but is under-predicted (over-predicted) by about 1.4\% (0.7\%)
with model a (model c).  It is should be noted that this observable has negligible
contributions due to two-body electromagnetic operators~\cite{Piarulli13}.  The
magnetic moment, $\mu_d$, and quadrupole moment, $Q_d$, experimental values
are underestimated by all three models, but these observables are known to have significant
corrections from (isoscalar) two-body terms in nuclear electromagnetic charge
and current~\cite{Piarulli13}.  Their inclusion would bring the calculated values
considerably closer to, if not in agreement with, experiment.
Finally, the S- and D-wave components of the deuteron wave function are
displayed in Fig.~\ref{fig:f5}, where they are compared to those of the Argonne
$v_{18}$ (AV18) model.  There is significant cutoff dependence as $(R_{\rm L},R_{\rm S})$
are reduced from the values (1.2, 0.8) fm of model a down to (0.8, 0.6) fm of
model c.  For $r \lesssim 1$ fm, the S-wave becomes smaller (is pushed out),
while the D-wave becomes larger (is pushed in) in going from model a to model c.
The D-state percentage increases correspondingly (see Table~\ref{tab:table3}).

We note in closing that in Appendix~\ref{app:a5} we provide figures of the various
components of potential models a, b, and  c (their charge-independent parts only) as well as tables
of numerical values for the $pp$ and $np$ S, P, D, F, and G phase shifts obtained
with model b. 
\setlength{\tabcolsep}{3pt}
\begin{table}[bth]
\caption{\label{tab:table3}%
Same as in Table~\protect\ref{tab:table2} but for the deuteron static properties; experimental
values are form Refs.~\cite{Vandl82,Ericson83,Rodning90,Huber98,Bishop79}.}
\begin{ruledtabular}
\begin{tabular}{lllll}
\textrm{}&
{\textrm{Experiment}}&
{\textrm{$v_{12}^{\rm a}$}}&
{\textrm{$v_{12}^{\rm b}$}}&
{\textrm{$v_{12}^{\rm c}$}}\\
\colrule
$E_{d}$ (MeV)		&2.224575(9)	        &2.224575	&2.224574 	&2.224575		\\
$A_{\rm S}$(fm$^{-1/2}$) & 0.8781(44)    &0.8777          &0.8904       &0.8964                \\
$\eta$				&0.0256(4)		&0.0245		&0.0248		&0.0246			\\
$r_{d}$ (fm)			&1.97535(85)		&1.948		&1.975		&1.989					\\
$\mu_{d}$ ($\mu_{0}$) &0.857406(1)	        &0.852 		&0.850		&0.848						\\
$Q_{d}$	 (fm$^2$)		&0.2859(3)		&0.257		&0.268		&0.269				\\
$P_{d}$	(\%)			         &                       &4.94		&5.29 		&5.55			\\
\end{tabular}
\end{ruledtabular}
\end{table}
\section{Conclusions}
\label{sec:conc}
In the present study, we have constructed a coordinate-space nucleon-nucleon
potential with an electromagnetic interaction component including first and second
order Coulomb, Darwin-Foldy, vacuum polarization, and magnetic moment terms,
and a strong interaction component characterized by long- and short-range parts.
The long-range part includes OPE and TPE terms up to N2LO, derived in the static
limit from leading and sub-leading $\pi N$ and $\pi N\Delta$ chiral Lagrangians.
Its strength is fully determined by the nucleon and nucleon-to-$\Delta$ axial
coupling constants $g_A$ and $h_A$, the pion decay amplitude $F_\pi$, and the
sub-leading LEC's $c_1$, $c_2$, $c_3$, $c_4$, and $b_3+b_8$, constrained by
reproducing $\pi N$ scattering data (the values adopted for all these couplings are
listed in Table~\ref{tab:table}).  In coordinate space, this long-range part is represented
by charge-independent central, spin, and tensor components without and with
the isospin dependence ${\bm \tau}_1\cdot {\bm \tau}_2$ (the so-called $v_6$
operator structure), and by charge-dependence-breaking central and tensor
components induced by OPE and proportional to the isotensor operator $T_{12}$.

The short-range part is described by charge-independent contact interactions
specified by a total of 24 LEC's (2 at LO, 7 at NLO, and 15 at N3LO) and
by charge-dependent ones characterized by 10 LEC's (2 at LO and 8 at NLO),
5 of which multiply charge-symmetry breaking terms proportional to $\tau_{1z}+\tau_{2z}$
and the remaining 5 multiply charge-dependence breaking terms proportional
to $T_{12}$.  In the NLO and N3LO contact interactions, Fierz transformations
have been used in order to rearrange terms that in coordinate space would otherwise
lead to powers of ${\bf p}$---the relative momentum operator---higher than two.
The resulting charge-independent (coordinate-space) potential contains, in addition
to the $v_6$ operator structure, spin-orbit, ${\bf L}^2$, quadratic-spin-orbit,
and ${\bf p}^2$ components, while the charge-dependent one retains
central, tensor, and spin-orbit components.

The 34 LEC's in the short-range potential have been constrained
by fitting 5291 $pp$ and $np$ scattering data (including normalizations)
up to 300 MeV lab energies, as assembled in the Granada database,
and the $pp$, $np$, and $nn$ scattering lengths, and the deuteron binding
energy.   The global $\chi^2(pp+np)$/datum is 1.33 for the three different models
we have investigated, each specified by a pair of (coordinate-space) cutoffs,
respectively, $R_{\rm L}$ and $R_{\rm S}$ for the long- and short-range
parts: $(R_{\rm L},R_{\rm S})=(1.2,0.8)$ fm for model a, $(1.0,0.7)$ fm
for model b, and $(0.8,0.6)$ fm for model c.  These cutoffs are close to the
$1/(2\, m_\pi)\sim 0.7$ fm TPE range.  The values of the LEC's corresponding
to the three models are given in Table~\ref{tab:table5}.  They are generally
of natural size, but for a few exceptions, most notably the LEC's $C_4^{\rm IV}$
and $C_4^{\rm IT}$ multiplying the charge-dependent spin-orbit terms,
which lead to relatively large splitting between the $pp$ and $np$
$^3$P$_0$ and $^3$P$_1$ phase shifts---a splitting that is not
consistent with that obtained in both the Nijmegen and Granada  PWA's.
It should also be noted that the degree of unnaturalness increases
as the short-distance cutoffs are reduced.

Our results suggest that discrepancies between the phases calculated
here and those from available PWA's in some of the partial waves,
such as the $\epsilon_1$ mixing angle, could hardly be resolved by
carrying out the database selection using the present interaction.
We should also note that the renowned Entem and Machleidt N3LO
fit up to $E_{\rm lab}=290$ MeV provides a $\chi^2$/datum
of 1.1 for 2402 $np$ data and 1.5 for 2057 $pp$, and hence a global
$\chi^2$/datum of 1.3.  In our case, we describe 2161 (2764) scattering
data and 148 (218) normalizations for $pp$ ($np$), which means that
the average contribution to the $\chi^2$ from each additional datum is
homogeneous and of order one out of about 800 extra data.  So, our fit is
as good as the one of Entem and Machleidt with these additional data.

According to our findings the largest uncertainty in the chiral
theory when fitting up to a maximum lab energy of 300 MeV is
provided by the cutoff dependence.  Under these circumstances it
makes little sense to analyze further uncertainties, but it is nonetheless
surprising that precisely the model implementing many QCD motivated
theoretical constraints should end up magnifying the uncertainty to a
larger extent than the spread historically found in all so far
successful PWA's to $pp$ and $np$ scattering data.  On the other hand, the reliability
of the long distance chiral interaction does not depend on how the
short distance unknown interaction is organized.  This has been proven
by the first chiral potential fits by the Nijmegen group from their
$pp$~\cite{Rentmeester:1999vw} and $np+pp$~\cite{Rentmeester:2003mf}
analyses and more recently verified with increased statistics by the Granada
group~\cite{Perez:2013oba}.  This leaves open the possibility that
better fits than those found here should be possible by properly altering
the short distance structure. This point has recently been discussed
in Ref.~\cite{Perez:2014bua}.

Of course,  this cutoff uncertainty could be greatly
reduced if the fitting energy range were to be lowered so
as to ensure that differences between fitted data and fitting
theory fulfill the normality requirement and, at the same time, statistical
uncertainties remain at the same level as cutoff uncertainties.
Following the recent suggestion~\cite{Perez:2014bua}, we find that
this happens with the current form of the potential when
$E_{\rm lab} \le 125$ MeV.  In a companion paper we will analyze
the statistical properties of the present fit and how there is a
trade-off of different uncertainty sources.

We conclude by observing that, apart from the ${\bf p}^2$-dependent terms,
the potential constructed here has the same operator structure of the AV18,
and is of slightly better quality than the AV18 (the AV18 global $\chi^2(pp+np)$/datum
on the same database up to 300 MeV lab energies is 1.46).  It should be fairly straightforward to
incorporate it in the few-nucleon calculations based on hyperspherical-harmonics
expansion techniques favored by the Pisa group~\cite{Kievsky08}, or in the quantum
Monte Carlo ones preferred by the ANL/ASU/JLab/LANL collaboration~\cite{Carlson2014}.
The Fortran computer program generating the potential will be made available
upon request.
\acknowledgments
We like to thank J.\ Sarich and S.M.\ Wild in the Mathematics
and Computer Science Division at Argonne National Laboratory
for advise on the implementation of POUNDerS in the $\chi^2$-minimization
programs.  Conversations with F.\ Gross, J.W.\ Van Orden, and
R.B.\ Wiringa at various stages of this project are gratefully acknowledged.
Finally, we also like to thank D.\ Lonardoni and A.\ Lovato for help
on the parallelization of the minimization programs.
The work of R.S. is supported by the U.S. Department of Energy, Office
of Nuclear Science, under contract DE-AC05-06OR23177.  The work
of R.N.P., J.E.A., and E.R.A. is supported by the Spanish DGI (grant
FIS2011-24149) and Junta de Andaluc{\'{\i}a} (grant FQM225).
R.N.P. is also supported by a Mexican CONACYT grant.
The calculations were made possible by grants of computing time
from the National Energy Research Supercomputer Center (NERSC).
\appendix
\section{Coordinate-space representation of the potential $v^{\rm L}_{12}$}
\label{app:a1}
The LO (OPE) terms corresponding to diagram (a) in Fig.~\ref{fig:f1} are given by
\begin{eqnarray}
v^{\rm LO}_{\sigma\tau}(r)&=&\frac{Y_0(r)+2\, Y_+(r)}{3}\ ,\\
v^{\rm LO}_{t\tau}(r)&=&\frac{T_0(r)+2\, T_+(r)}{3} \ ,
\end{eqnarray}
where
\begin{eqnarray}
Y_\alpha(r)&=& \frac{g_A^2}{12\, \pi}\,  \frac{m^3_{\pi_\alpha}}{F_{\pi}^2} \,
\frac{e^{-x_\alpha}}{x_\alpha}\ ,\\
T_\alpha(r)&=&Y_\alpha(r)\left( 1+\frac{3}{x_\alpha}+\frac{3}{x^2_\alpha}\right)\ ,
\end{eqnarray}
and $x_\alpha=m_{\pi_\alpha} r$.  The NLO terms corresponding to diagrams (b)-(d) read~\cite{Kaiser:1997mw}
\begin{widetext}
\begin{eqnarray}
v^{\rm NLO}_\tau (r;\slashed{\Delta})&=&\frac{1}
{8 \pi^3  r^4}\frac{m_\pi}{F_{\pi}^4}\, \bigg[\, x
\left[1+10 g_A^2-g_A^4 (23+4x^2)\right]K_0(2x)\nonumber\\
&&+\left[1+2 g_A^2 (5+2x^2)-g_A^4 (23+12x^2)\right]K_1(2x)\bigg] \ ,\\
v^{\rm NLO}_\sigma (r;\slashed{\Delta})&=&\frac{1}
{2 \pi^3  r^4}\frac{g_A^4}{F_{\pi}^4}\,m_{\pi} \bigg[3x\,K_0(2x)+
 (3+2x^2)K_1(2x)\bigg]\ , \\
v^{\rm NLO}_t (r;\slashed{\Delta})&=&-\frac{1}
{8 \pi^3 r^4}\frac{g_A^4}{F_{\pi}^4 }\,m_{\pi} \bigg[12x\,K_0(2x)+
 (15+4x^2)K_1(2x)\bigg] \ ,
\end{eqnarray}
\end{widetext}
where $x=m_\pi r$ ($m_\pi$ is the average pion mass)
and  $K_n$ are modified Bessel functions of the second kind.
The NLO terms corresponding to diagrams (e)-(f) with a single $\Delta$ intermediate state
are given by
 \begin{eqnarray}
v^{\rm NLO}_{c} (r;\Delta)&=&-\frac{1}
{6 \pi^2  r^5\,y}\frac{g_A^2 h_A^2}{F_{\pi}^4}e^{-2x}\left(6+12x+10x^2+4x^3+x^4\right)\ , \\
 v^{\rm NLO}_{\tau} (r;\Delta)&=&-\frac{1}
{216 \pi^3  r^5}\frac{h_A^2}{F_{\pi}^4}\Bigg[\int_{0}^{\infty}d\mu \frac{\mu^2}
{\sqrt{\mu^2+4x^2}}e^{-\sqrt{\mu^2+4x^2}}(12x^2+5\mu^2+12y^2)\nonumber\\
&&-12y\int_{0}^{\infty}d\mu \frac{\mu}
{\sqrt{\mu^2+4x^2}}e^{-\sqrt{\mu^2+4x^2}}(2x^2+\mu^2+2y^2)\arctan{\frac{\mu}{2y}}\Bigg]\nonumber\\
&&-\frac{1}{216 \pi^3  r^5}\frac{g_A^2 h_A^2}{F_{\pi}^4}\Bigg[-\int_{0}^{\infty}d\mu \frac{\mu^2}
{\sqrt{\mu^2+4x^2}}e^{-\sqrt{\mu^2+4x^2}}(24x^2+11\mu^2+12y^2)\nonumber\\
&&+\frac{6}{y}\int_{0}^{\infty}d\mu \frac{\mu}
{\sqrt{\mu^2+4x^2}}e^{-\sqrt{\mu^2+4x^2}}(2x^2+\mu^2+2y^2)^2\arctan{\frac{\mu}{2y}}\Bigg]\ ,\\
v^{\rm NLO}_{\sigma}(r;\Delta)&=&-\frac{1}
{72 \pi^3  r^5}\frac{g_A^2 h_A^2}{F_{\pi}^4}\Bigg[2\,\int_{0}^{\infty}d\mu \frac{\mu^2}
{\sqrt{\mu^2+4x^2}}e^{-\sqrt{\mu^2+4x^2}}(\mu^2+4x^2)\nonumber \\
&&-\frac{1}{y}\int_{0}^{\infty}d\mu \frac{\mu}
{\sqrt{\mu^2+4x^2}}e^{-\sqrt{\mu^2+4x^2}}(\mu^2+4x^2)(\mu^2+4y^2)\arctan{\frac{\mu}{2y}}\Bigg]\ ,\\
v^{\rm NLO}_{\sigma\tau} (r;\Delta)&=&\frac{1}
{54 \pi^2  r^5\,y}\frac{g_A^2 h_A^2}{F_{\pi}^4}e^{-2x}\left(1+x \right)\left(3+3x+x^2\right)\ , \\
v^{\rm NLO}_{t}(r;\Delta)&=&\frac{1}
{144 \pi^3  r^5}\frac{g_A^2 h_A^2}{F_{\pi}^4}\Bigg[2\,\int_{0}^{\infty}d\mu \frac{\mu^2}
{\sqrt{\mu^2+4x^2}}e^{-\sqrt{\mu^2+4x^2}}(3+3\sqrt{\mu^2+4x^2}+\mu^2+4x^2)\nonumber \\
&&-\frac{1}{y}\int_{0}^{\infty}d\mu\frac{\mu}
{\sqrt{\mu^2+4x^2}}e^{-\sqrt{\mu^2+4x^2}}(\mu^2+4y^2)(3+3\sqrt{\mu^2+4x^2}
+\mu^2+4x^2)\arctan{\frac{\mu}{2y}}\Bigg]\ , \\
v^{\rm NLO}_{t\tau} (r;\Delta)&=&-\frac{1}
{54 \pi^2  r^5\,y}\frac{g_A^2 h_A^2}{F_{\pi}^4}e^{-2x}\left(1+x \right)\left(3+3x+2x^2\right)\ ,
\end{eqnarray}
where $y= \Delta M r$ ($\Delta M$ is the $\Delta$-nucleon mass difference)
and the parametric integral over $\mu$ is carried out numerically.  The
NLO terms corresponding
to diagram (g) with $2\,\Delta$ intermediate states are
\begin{eqnarray}
 v^{\rm NLO}_{c} (r;2\Delta)&=&-\frac{1}
{108 \pi^3  r^5}\frac{h_A^4}{F_{\pi}^4}\Bigg[\int_{0}^{\infty}d\mu\frac{\mu^2}
{\sqrt{\mu^2+4x^2}}e^{-\sqrt{\mu^2+4x^2}}
\left[4y^2+2\frac{(2x^2+\mu^2+2y^2)^2}{(\mu^2+4y^2)}\right]\nonumber\\
&&+\frac{1}{y}\int_{0}^{\infty}d\mu\frac{\mu}
{\sqrt{\mu^2+4x^2}}e^{-\sqrt{\mu^2+4x^2}}
(2x^2+\mu^2+2y^2)(2x^2+\mu^2-6y^2)\arctan{\frac{\mu}{2y}}\Bigg]\ ,\\
v^{\rm NLO}_{\tau} (r;2\Delta)&=&-\frac{1}
{1944 \pi^3  r^5}\frac{h_A^4}{F_{\pi}^4}\Bigg[\int_{0}^{\infty}d\mu\frac{\mu^2}
{\sqrt{\mu^2+4x^2}}e^{-\sqrt{\mu^2+4x^2}}
\left[(24x^2+11\mu^2+24y^2)+6\frac{(2x^2+\mu^2+2y^2)^2}{(\mu^2+4y^2)}\right]\nonumber\\
&&-\frac{3}{y}\int_{0}^{\infty}d\mu\frac{\mu}
{\sqrt{\mu^2+4x^2}}e^{-\sqrt{\mu^2+4x^2}}
(2x^2+\mu^2+2y^2)(2x^2+\mu^2+10y^2)\arctan{\frac{\mu}{2y}}\Bigg]\ ,\\
v^{\rm NLO}_{\sigma} (r;2\Delta)&=&-\frac{1}
{1296 \pi^3  r^5}\frac{h_A^4}{F_{\pi}^4}\Bigg[-6 \int_{0}^{\infty}d\mu\frac{\mu^2}
{\sqrt{\mu^2+4x^2}}e^{-\sqrt{\mu^2+4x^2}}
(\mu^2+4x^2)\nonumber\\
&&+\frac{1}{y}\int_{0}^{\infty}d\mu\frac{\mu}
{\sqrt{\mu^2+4x^2}}e^{-\sqrt{\mu^2+4x^2}}(\mu^2+4x^2)
(\mu^2+12y^2)\arctan{\frac{\mu}{2y}}\Bigg]\ ,\\
v^{\rm NLO}_{\sigma\tau} (r;2\Delta)&=&-\frac{1}
{7776 \pi^3  r^5}\frac{h_A^4}{F_{\pi}^4}\Bigg[-2 \int_{0}^{\infty}d\mu\frac{\mu^2}
{\sqrt{\mu^2+4x^2}}e^{-\sqrt{\mu^2+4x^2}}
(\mu^2+4x^2)\nonumber\\
&&+\frac{1}{y}\int_{0}^{\infty}d\mu\frac{\mu}
{\sqrt{\mu^2+4x^2}}e^{-\sqrt{\mu^2+4x^2}}(\mu^2+4x^2)
(-\mu^2+4y^2)\arctan{\frac{\mu}{2y}}\Bigg]\ ,\\
v^{\rm NLO}_{t} (r;2\Delta)&=&\frac{1}
{2592 \pi^3  r^5}\frac{h_A^4}{F_{\pi}^4}\Bigg[-6 \int_{0}^{\infty}d\mu\frac{\mu^2}
{\sqrt{\mu^2+4x^2}}e^{-\sqrt{\mu^2+4x^2}}
(3+3\sqrt{\mu^2+4x^2}+\mu^2+4x^2)\nonumber\\
&&+\frac{1}{y}\int_{0}^{\infty}d\mu\frac{\mu}
{\sqrt{\mu^2+4x^2}}e^{-\sqrt{\mu^2+4x^2}}(3+3\sqrt{\mu^2+4x^2}+\mu^2+4x^2)
(\mu^2+12y^2)\arctan{\frac{\mu}{2y}}\Bigg]\ ,\\
v^{\rm NLO}_{t\tau} (r;2\Delta)&=&\frac{1}
{15552 \pi^3  r^5}\frac{h_A^4}{F_{\pi}^4}\Bigg[-2 \int_{0}^{\infty}d\mu\frac{\mu^2}
{\sqrt{\mu^2+4x^2}}e^{-\sqrt{\mu^2+4x^2}}
(3+3\sqrt{\mu^2+4x^2}+\mu^2+4x^2)\nonumber\\
&&+\frac{1}{y}\int_{0}^{\infty}d\mu\frac{\mu}
{\sqrt{\mu^2+4x^2}}e^{-\sqrt{\mu^2+4x^2}}(3+3\sqrt{\mu^2+4x^2}+\mu^2+4x^2)
(-\mu^2+4y^2)\arctan{\frac{\mu}{2y}}\Bigg]\ .
\end{eqnarray}

Moving on to the loop corrections at N2LO, the terms corresponding to diagrams (h)-(k)
are given by
\begin{eqnarray}
 v^{\rm N2LO}_c (r;\slashed{\Delta})&=&\frac{3}{2\,\pi^2 r^6}\frac{g_A^2}{F_{\pi}^4}e^{-2x}
\left[2c_1x^2(1+x)^2+c_3(6+12x+10x^2+4x^3+x^4)\right]\ , \\
v^{\rm N2LO}_{\sigma\tau} (r;\slashed{\Delta})&=&\frac{1}{3\,\pi^2 r^6}\frac{g_A^2}{F_{\pi}^4}c_4 e^{-2x}
\left(1+x\right)\left(3+3x+2x^2\right)\ , \\
v^{\rm N2LO}_{t\tau} (r;\slashed{\Delta})&=&-\frac{1}{3\,\pi^2 r^6}\frac{g_A^2}{F_{\pi}^4}c_4 e^{-2x}
\left(1+x\right)\left(3+3x+x^2\right)\ , 
\end{eqnarray}
while those corresponding to diagrams (l)-(o) are given by
\begin{eqnarray}
 v^{\rm N2LO}_{c} (r;\Delta)&=&\frac{1}
{18 \pi^3  r^6}\frac{h_A^2\,y}{F_{\pi}^4}\Bigg[\int_{0}^{\infty}d\mu\frac{\mu^2}
{\sqrt{\mu^2+4x^2}}e^{-\sqrt{\mu^2+4x^2}}
[-24c_1x^2+c_2(5\mu^2+12x^2+12y^2)-6c_3(\mu^2+2x^2)] \nonumber\\
&&+\frac{6}{y}\int_{0}^{\infty}\!\!d\mu\frac{\mu}
{\sqrt{\mu^2+4x^2}}e^{-\sqrt{\mu^2+4x^2}}(\mu^2\!+\!2x^2\!+\!2y^2) 
 [4c_1x^2-2c_2y^2+c_3(\mu^2+2x^2)]\,\arctan{\frac{\mu}{2y}}\Bigg]\ ,\\
v^{\rm N2LO}_{\tau} (r;\Delta)&=&-\frac{1}
{54 \pi^3  r^6}\frac{(b_3+b_8)\,h_A\,y}{F_{\pi}^4}\Bigg[+\int_{0}^{\infty}d\mu\frac{\mu^2}
{\sqrt{\mu^2+4x^2}}e^{-\sqrt{\mu^2+4x^2}}(5\mu^2+12x^2+12y^2)\nonumber\\
&&-12\,y\int_{0}^{\infty}d\mu\frac{\mu}
{\sqrt{\mu^2+4x^2}}e^{-\sqrt{\mu^2+4x^2}}(\mu^2+2x^2+2y^2)\,\arctan{\frac{\mu}{2y}}\Bigg]\nonumber\\
&&-\frac{1}{54 \pi^3  r^6}\frac{(b_3+b_8)\,h_A\,g_A^2\,y}{F_{\pi}^4}\Bigg[-\int_{0}^{\infty}d\mu\frac{\mu^2}
{\sqrt{\mu^2+4x^2}}e^{-\sqrt{\mu^2+4x^2}}(11\mu^2+24x^2+12y^2)\nonumber\\
&&+\frac{6}{y}\int_{0}^{\infty}d\mu\frac{\mu}{\sqrt{\mu^2+4x^2}}e^{-\sqrt{\mu^2+4x^2}}
\left(\mu^2+2x^2+2y^2\right)^2\arctan{\frac{\mu}{2y}}\Bigg]\ ,\\
v^{\rm N2LO}_{\sigma} (r;\Delta)&=&-\frac{1}
{18 \pi^3  r^6}\frac{(b_3+b_8)\,h_A\,g_A^2\,y}{F_{\pi}^4}\Bigg[2\int_{0}^{\infty}d\mu\frac{\mu^2}
{\sqrt{\mu^2+4x^2}}e^{-\sqrt{\mu^2+4x^2}}(\mu^2+4x^2)\nonumber\\
&&-\frac{1}{y}\int_{0}^{\infty}d\mu\frac{\mu}{\sqrt{\mu^2+4x^2}}e^{-\sqrt{\mu^2+4x^2}}
(\mu^2+4x^2)(\mu^2+4y^2)\arctan{\frac{\mu}{2y}}\Bigg]\ ,\\
v^{\rm N2LO}_{\sigma\tau} (r;\Delta)&=&-\frac{1}
{108 \pi^3  r^6}\frac{c_4\,h_A^2\,y}{F_{\pi}^4}\Bigg[2\,\int_{0}^{\infty}d\mu\frac{\mu^2}
{\sqrt{\mu^2+4x^2}}e^{-\sqrt{\mu^2+4x^2}}(\mu^2+4x^2)\nonumber\\
&&-\frac{1}{y}\int_{0}^{\infty}d\mu\frac{\mu}
{\sqrt{\mu^2+4x^2}}e^{-\sqrt{\mu^2+4x^2}}(\mu^2+4x^2)(\mu^2+4y^2)\,\arctan{\frac{\mu}{2y}}\Bigg]\ ,\\
v^{\rm N2LO}_{t} (r;\Delta)&=&\frac{1}
{36 \pi^3  r^6}\frac{(b_3+b_8)\,h_A\,g_A^2\,y}{F_{\pi}^4}\Bigg[2\int_{0}^{\infty}d\mu\frac{\mu^2}
{\sqrt{\mu^2+4x^2}}e^{-\sqrt{\mu^2+4x^2}}(3+3\sqrt{\mu^2+4x^2}+\mu^2+4x^2)\nonumber\\
&&-\frac{1}{y}\int_{0}^{\infty}d\mu\frac{\mu}{\sqrt{\mu^2+4x^2}}e^{-\sqrt{\mu^2+4x^2}}
(3+3\sqrt{\mu^2+4x^2}+\mu^2+4x^2)(\mu^2+4y^2)\arctan{\frac{\mu}{2y}}\Bigg]\ ,\\
v^{\rm N2LO}_{t\tau} (r;\Delta)&=&\frac{1}
{216 \pi^3  r^6}\frac{c_4\,h_A^2\,y}{F_{\pi}^4}\Bigg[2\,\int_{0}^{\infty}d\mu\frac{\mu^2}
{\sqrt{\mu^2+4x^2}}e^{-\sqrt{\mu^2+4x^2}}(3+3\sqrt{\mu^2+4x^2}+\mu^2+4x^2)\nonumber\\
&&-\frac{1}{y}\int_{0}^{\infty}d\mu\frac{\mu}
{\sqrt{\mu^2+4x^2}}e^{-\sqrt{\mu^2+4x^2}}(3+3\sqrt{\mu^2+4x^2}+\mu^2+4x^2)
(\mu^2+4y^2)\,\arctan{\frac{\mu}{2y}}\Bigg]\  .
\end{eqnarray}
Lastly, the contributions corresponding to diagram (p) read
\begin{eqnarray}
 v^{\rm N2LO}_{c} (r;2\Delta)&=&-\frac{2}
{81 \pi^3  r^6}\frac{(b_3+b_8)\,h_A^3\,y}{F_{\pi}^4}\Bigg[
\int_{0}^{\infty}d\mu\frac{\mu^2}{\sqrt{\mu^2+4x^2}}e^{-\sqrt{\mu^2+4x^2}}
[6\frac{(\mu^2+2x^2+2y^2)^2}{\mu^2+4y^2}+11\mu^2+24x^2+12y^2]\nonumber\\
&&-\frac{3}{y}\int_{0}^{\infty}d\mu\frac{\mu}
{\sqrt{\mu^2+4x^2}}e^{-\sqrt{\mu^2+4x^2}}(\mu^2+2x^2+10y^2)(\mu^2+2x^2+2y^2)
\arctan{\frac{\mu}{2y}}\Bigg]\ ,  \\
 v^{\rm N2LO}_{\tau} (r;2\Delta)&=&-\frac{1}
{243 \pi^3  r^6}\frac{(b_3+b_8)\,h_A^3\,y}{F_{\pi}^4}\Bigg[
\int_{0}^{\infty}d\mu\frac{\mu^2}{\sqrt{\mu^2+4x^2}}e^{-\sqrt{\mu^2+4x^2}}
[6\frac{(\mu^2+2x^2+2y^2)^2}{\mu^2+4y^2}+11\mu^2+24x^2+12y^2]\nonumber\\
&&-\frac{3}{y}\int_{0}^{\infty}d\mu\frac{\mu}
{\sqrt{\mu^2+4x^2}}e^{-\sqrt{\mu^2+4x^2}}(\mu^2+2x^2+10y^2)(\mu^2+2x^2+2y^2)
\arctan{\frac{\mu}{2y}}\Bigg]\ , \\
v^{\rm N2LO}_{\sigma} (r;2\Delta)&=&-\frac{1}
{162 \pi^3  r^6}\frac{(b_3+b_8)\,h_A^3\,y}{F_{\pi}^4}\Bigg[
-6\int_{0}^{\infty}d\mu\frac{\mu^2}{\sqrt{\mu^2+4x^2}}e^{-\sqrt{\mu^2+4x^2}}
(\mu^2+4x^2)\nonumber\\
&&+\frac{1}{y}\int_{0}^{\infty}d\mu\frac{\mu}
{\sqrt{\mu^2+4x^2}}e^{-\sqrt{\mu^2+4x^2}}(\mu^2+4x^2)(\mu^2+12y^2)
\arctan{\frac{\mu}{2y}}\Bigg]\ ,\\
v^{\rm N2LO}_{\sigma\tau} (r;2\Delta)&=&-\frac{1}
{972 \pi^3  r^6}\frac{(b_3+b_8)\,h_A^3\,y}{F_{\pi}^4}\Bigg[
-6\int_{0}^{\infty}d\mu\frac{\mu^2}{\sqrt{\mu^2+4x^2}}e^{-\sqrt{\mu^2+4x^2}}
(\mu^2+4x^2)\nonumber\\
&&+\frac{1}{y}\int_{0}^{\infty}d\mu\frac{\mu}
{\sqrt{\mu^2+4x^2}}e^{-\sqrt{\mu^2+4x^2}}(\mu^2+4x^2)(\mu^2+12y^2)
\arctan{\frac{\mu}{2y}}\Bigg]\ ,\\
v^{\rm N2LO}_{t} (r;2\Delta)&=&\frac{1}
{324 \pi^3  r^6}\frac{(b_3+b_8)\,h_A^3\,y}{F_{\pi}^4}\Bigg[
-6\int_{0}^{\infty}d\mu\frac{\mu^2}{\sqrt{\mu^2+4x^2}}e^{-\sqrt{\mu^2+4x^2}}
(3+3\sqrt{\mu^2+4x^2}+\mu^2+4x^2)\nonumber\\
&&+\frac{1}{y}\int_{0}^{\infty}d\mu\frac{\mu}
{\sqrt{\mu^2+4x^2}}e^{-\sqrt{\mu^2+4x^2}}(3+3\sqrt{\mu^2+4x^2}+\mu^2+4x^2)(\mu^2+12y^2)
\arctan{\frac{\mu}{2y}}\Bigg]\ , \\
v^{\rm N2LO}_{t\tau} (r;2\Delta)&=&\frac{1}
{1944 \pi^3  r^6}\frac{(b_3+b_8)\,h_A^3\,y}{F_{\pi}^4}\Bigg[
-6\int_{0}^{\infty}d\mu\frac{\mu^2}{\sqrt{\mu^2+4x^2}}e^{-\sqrt{\mu^2+4x^2}}
(3+3\sqrt{\mu^2+4x^2}+\mu^2+4x^2)\nonumber\\
&&+\frac{1}{y}\int_{0}^{\infty}d\mu\frac{\mu}
{\sqrt{\mu^2+4x^2}}e^{-\sqrt{\mu^2+4x^2}}(3+3\sqrt{\mu^2+4x^2}+\mu^2+4x^2)(\mu^2+12y^2)
\arctan{\frac{\mu}{2y}}\Bigg]\ .
\end{eqnarray}

The radial functions of the charge-independent part of the potential $v^{\rm L}_{12}$
in Eq.~(\ref{eq:vlr}) are defined as
\begin{eqnarray}
v^{c}_{\rm L}(r)&=&v^{\rm NLO}_{c}(r;\Delta)
+v^{\rm NLO}_{c}(r;2\Delta)+v^{\rm N2LO}_{c}(r;\slashed{\Delta})+v^{\rm N2LO}_{c}(r;\Delta)
+v^{\rm N2LO}_{c}(r;2\Delta)\ , \\
v^{\tau}_{\rm L}(r)&=&v^{\rm NLO}_\tau (r;\slashed{\Delta})
+v^{\rm NLO}_{\tau}(r;\Delta)+v^{\rm NLO}_{\tau}(r;2\Delta)+v^{\rm N2LO}_{\tau}(r;\Delta)
+v^{\rm N2LO}_{\tau}(r;2\Delta) \ , \\
v^{\sigma}_{\rm L}(r)&=&v^{\rm NLO}_\sigma (r;\slashed{\Delta})
+v^{\rm NLO}_{\sigma}(r;\Delta)+v^{\rm NLO}_{\sigma}(r;2\Delta)+v^{\rm N2LO}_{\sigma}(r;\Delta)
+v^{\rm N2LO}_{\sigma}(r;2\Delta) \ , \\
v^{\sigma\tau}_{\rm L}(r)&=&v^{\rm LO}_{\sigma\tau}(r)+v^{\rm NLO}_{\sigma\tau}(r;\Delta)
+v^{\rm NLO}_{\sigma\tau}(r;2\Delta)+v^{\rm N2LO}_{\sigma\tau}(r;\slashed{\Delta})
+v^{\rm N2LO}_{\sigma\tau}(r;\Delta)\nonumber\\
&&+v^{\rm N2LO}_{\sigma\tau}(r;2\Delta)\ ,\\
v^{t}_{\rm L}(r)&=&v^{\rm NLO}_t (r;\slashed{\Delta})
+v^{\rm NLO}_{t}(r;\Delta)+v^{\rm NLO}_{t}(r;2\Delta)+v^{\rm N2LO}_{t}(r;\Delta)
+v^{\rm N2LO}_{t}(r;2\Delta) \ , \\
v^{t\tau}_{\rm L}(r)&=&v^{\rm LO}_{t\tau}(r)+
v^{\rm NLO}_{t\tau}(r;\Delta)
+v^{\rm NLO}_{t\tau}(r;2\Delta)+v^{\rm N2LO}_{t\tau}(r;\slashed{\Delta})
+v^{\rm N2LO}_{t\tau}(r;\Delta)\nonumber\\
&&+v^{\rm N2LO}_{t\tau}(r;2\Delta)\ ,
\end{eqnarray}
while those of its charge-dependent part are defined as
\begin{eqnarray}
v^{\sigma T}_{\rm L}(r)&=&\frac{Y_0(r)- Y_+(r)}{3}\ ,\\
v^{t T}_{\rm L}(r)&=&\frac{T_0(r)- T_+(r)}{3} \ .
\end{eqnarray}
Each is multiplied by the cutoff $C_{R_{\rm L}}(r)$,
\begin{equation}
v^l_{\rm L}(r) \longrightarrow C_{R_{\rm L}}(r) \, v^l_{\rm L}(r) \ ,
\end{equation}
with $l=c,\tau,\sigma,\sigma\tau,t,t\tau,\sigma T,tT$.
\section{Coordinate-space representation of the potential $v^{\rm S}_{12}$}
\label{app:a2}
The coordinate-space representation of a (regularized) term
$O({\bf K},{\bf{k}})$ in Eqs.~(\ref{eq:sci}) and~(\ref{eq:scib}) follows from
\begin{eqnarray}
O({\bf r})=\int \frac{d{\bf k}}{(2\pi)^3}
\int \frac{d{\bf K}}{(2\pi)^3} \, e^{i\,{\bf k}\cdot ({\bf r}^{\prime}+{\bf r})/2}
\,O({\bf K},{\bf{k}})\,e^{i\,{\bf K}\cdot ({\bf r}^{\prime}-{\bf r})}\ ,
\label{eq:ft}
\end{eqnarray}
where ${\bf r}$ is the relative position and ${\bf K}
\longrightarrow {\bf p} =-i\, {\bm \nabla}^\prime\delta({\bf r}^\prime-{\bf r})$,
the relative momentum operator.  For the momentum-space operator structures
present in Eqs.~(\ref{eq:sci}) and~(\ref{eq:scib}) one finds:
\begin{eqnarray}
1 &\longrightarrow& C_{R_{\rm S}}(r) \ , \\
k^2&\longrightarrow& - C^{(2)}_{R_{\rm S}}(r)-\frac{2}{r}\, C^{(1)}_{R_{\rm S}}(r)
\ ,\\
k^4 &\longrightarrow& C^{(4)}_{R_{\rm S}}(r)+\frac{4}{r}\,C^{(3)}_{R_{\rm S}}(r) \ ,\\
S_{12}({\bf k})&\longrightarrow&-\left[ C^{(2)}_{R_{\rm S}}(r)-\frac{1}{r}\, C^{(1)}_{R_{\rm S}}(r)\right] S_{12} \ , \\
i\, {\bf S} \cdot \left( {\bf K}\times{\bf k}\right) &\longrightarrow& -\frac{1}{r} \,
C^{(1)}_{R_{\rm S}}(r)\,{\bf L}\cdot{\bf S} \ , \\
{\bf K}^2 &\longrightarrow&  \left\{ {\bf p}^2 \, ,\, C_{R_{\rm S}}(r)\right\} \ , \\
 \label{eq:eb8}
 \left( {\bf K}\times{\bf k}\right)^2 &\longrightarrow&-\frac{1}{r^2}
 \left[ C^{(2)}_{R_{\rm S}}(r)-\frac{1}{r}\, C^{(1)}_{R_{\rm S}}(r)\right] {\bf L}^2
 -\left\{ {\bf p}^2\, ,\, \frac{1}{r}\, C^{(1)}_{R_{\rm S}}(r) \right\}-\frac{1}{r}\, C^{(3)}_{R_{\rm S}}(r) \ , \\
 \label{eq:eb9}
  \left[{\bf S}\cdot\left( {\bf K}\times{\bf k}\right)\right]^2&\longrightarrow&
  -\frac{1}{r^2}
 \left[ C^{(2)}_{R_{\rm S}}(r)-\frac{1}{r}\, C^{(1)}_{R_{\rm S}}(r)\right] \left({\bf L}\cdot {\bf S}\right)^2
 -\left\{ {\bf p}^2\frac{\left( 1+{\bm \sigma}_1\cdot{\bm \sigma}_2 \right)}{2}\!-\! {\bm \sigma}_1\cdot {\bf p}\,\, {\bm \sigma}_2\cdot {\bf p}\,\,
  ,\, \frac{1}{r}\, C^{(1)}_{R_{\rm S}}(r) \right\}.
\end{eqnarray}
where
\begin{equation}
C^{(n)}_{R_{\rm S}}(r) =\frac{d^n C_{R_{\rm S}}(r)}{dr^n} \ .
\end{equation}
Using the above expressions, the functions $v^{l}_{\rm S}(r)$ are obtained as
\begin{eqnarray}
v^c_{S}(r)&=& C_S\,C_{R_{\rm S}}(r)+C_1\left[- C^{(2)}_{R_{\rm S}}(r)-\frac{2}{r}\, C^{(1)}_{R_{\rm S}}(r)\right]+
D_1\left[C^{(4)}_{R_{\rm S}}(r)+\frac{4}{r}\,C^{(3)}_{R_{\rm S}}(r)\right]\ , \\
v^\tau_{S}(r)&=& C_2\left[- C^{(2)}_{R_{\rm S}}(r)-\frac{2}{r}\, C^{(1)}_{R_{\rm S}}(r)\right]+
D_2\left[C^{(4)}_{R_{\rm S}}(r)+\frac{4}{r}\,C^{(3)}_{R_{\rm S}}(r)\right]\ , \\
v^\sigma_{S}(r)&=& C_T\,C_{R_{\rm S}}(r)+C_3\left[- C^{(2)}_{R_{\rm S}}(r)-\frac{2}{r}\, C^{(1)}_{R_{\rm S}}(r)\right]+
D_3\left[C^{(4)}_{R_{\rm S}}(r)+\frac{4}{r}\,C^{(3)}_{R_{\rm S}}(r)\right]\ , \\
v^{\sigma\tau}_{S}(r)&=& C_4\left[- C^{(2)}_{R_{\rm S}}(r)-\frac{2}{r}\, C^{(1)}_{R_{\rm S}}(r)\right]+
D_4\left[C^{(4)}_{R_{\rm S}}(r)+\frac{4}{r}\,C^{(3)}_{R_{\rm S}}(r)\right]\ , \\
v^{t}_{S}(r)&=& -C_5\left[C^{(2)}_{R_{\rm S}}(r)-\frac{1}{r}\, C^{(1)}_{R_{\rm S}}(r)\right]+D_5\left[
C^{(4)}_{R_{\rm S}}(r)+\frac{1}{r}C^{(3)}_{R_{\rm S}}(r)-\frac{6}{r^2}C^{(2)}_{R_{\rm S}}(r)+\frac{6}{r^3}
C^{(1)}_{R_{\rm S}}(r)\right]\ , \\
v^{t\tau}_{S}(r)&=& -C_6\left[C^{(2)}_{R_{\rm S}}(r)-\frac{1}{r}\, C^{(1)}_{R_{\rm S}}(r)\right]+D_6\left[
C^{(4)}_{R_{\rm S}}(r)+\frac{1}{r}C^{(3)}_{R_{\rm S}}(r)-\frac{6}{r^2}C^{(2)}_{R_{\rm S}}(r)+\frac{6}{r^3}
C^{(1)}_{R_{\rm S}}(r)\right]\ , \\
v^{b}_{S}(r)&=& -C_7\frac{1}{r}C^{(1)}_{R_{\rm S}}(r)+D_7\left[\frac{1}{r}C^{(3)}_{R_{\rm S}}(r)+2\,\frac{1}{r^2}C^{(2)}_{R_{\rm S}}(r)-\frac{2}{r^3}C^{(1)}_{R_{\rm S}}(r)\right]\ , \\
v^{b\tau}_{S}(r)&=& D_8\left[\frac{1}{r}C^{(3)}_{R_{\rm S}}(r)+2\,\frac{1}{r^2}C^{(2)}_{R_{\rm S}}(r)-\frac{2}{r^3}C^{(1)}_{R_{\rm S}}(r)\right]\ , \\
v^{bb}_{S}(r)&=&-D_9\frac{1}{r^2}
 \left[ C^{(2)}_{R_{\rm S}}(r)-\frac{1}{r}\, C^{(1)}_{R_{\rm S}}(r)\right] \ , \\
v^{q}_{S}(r)&=&-D_{10}\frac{1}{r^2}
 \left[ C^{(2)}_{R_{\rm S}}(r)-\frac{1}{r}\, C^{(1)}_{R_{\rm S}}(r)\right] \ , \\
v^{q\sigma}_{S}(r)&=& -D_{11}\frac{1}{r^2}
 \left[ C^{(2)}_{R_{\rm S}}(r)-\frac{1}{r}\, C^{(1)}_{R_{\rm S}}(r)\right]\ , \\
v^{p}_{S}(r)&=& D_{12}\left[- C^{(2)}_{R_{\rm S}}(r)-\frac{2}{r}\, C^{(1)}_{R_{\rm S}}(r)\right]\ , \\
v^{p\sigma}_{S}(r)&=& D_{13}\left[- C^{(2)}_{R_{\rm S}}(r)-\frac{2}{r}\, C^{(1)}_{R_{\rm S}}(r)\right]\ , \\
v^{pt}_{S}(r)&=&-D_{14}\left[C^{(2)}_{R_{\rm S}}(r)-\frac{1}{r}\, C^{(1)}_{R_{\rm S}}(r)\right]\ , \\
v^{pt\tau}_{S}(r)&=&-D_{15}\left[C^{(2)}_{R_{\rm S}}(r)-\frac{1}{r}\, C^{(1)}_{R_{\rm S}}(r)\right] \ , \\
v^{T}_{S}(r)&=& C_0^{\rm IT}\,C_{R_{\rm S}}(r)+C_1^{\rm IT}\left[- C^{(2)}_{R_{\rm S}}(r)-\frac{2}{r}\, C^{(1)}_{R_{\rm S}}(r)\right]\,\ , \\
v^{\tau z}_{S}(r)&=& C_0^{\rm IV}\,C_{R_{\rm S}}(r)+C_1^{\rm IV}\left[- C^{(2)}_{R_{\rm S}}(r)-\frac{2}{r}\, C^{(1)}_{R_{\rm S}}(r)\right]\ , \\
v^{\sigma T}_{S}(r)&=& C_2^{\rm IT}\left[- C^{(2)}_{R_{\rm S}}(r)-\frac{2}{r}\, C^{(1)}_{R_{\rm S}}(r)\right]\ , \\
v^{\sigma\tau z}_{S}(r)&=& C_2^{\rm IV}\left[- C^{(2)}_{R_{\rm S}}(r)-\frac{2}{r}\, C^{(1)}_{R_{\rm S}}(r)\right]\ , \\
v^{t T}_{S}(r)&=&-C_3^{\rm IT}\left[C^{(2)}_{R_{\rm S}}(r)-\frac{1}{r}\, C^{(1)}_{R_{\rm S}}(r)\right] \ , \\
v^{t \tau z}_{S}(r)&=& -C_3^{\rm IV}\left[C^{(2)}_{R_{\rm S}}(r)-\frac{1}{r}\, C^{(1)}_{R_{\rm S}}(r)\right] \ , \\
v^{b T}_{S}(r)&=& -C_4^{\rm IT}\frac{1}{r}C^{(1)}_{R_{\rm S}}(r)\ , \\
v^{b \tau z}_{S}(r)&=& -C_4^{\rm IV}\frac{1}{r}C^{(1)}_{R_{\rm S}}(r)\ .
\end{eqnarray}
Note that in Eqs.~(\ref{eq:eb8}) and~(\ref{eq:eb9}) only the terms proportional
to ${\bf L}^2$ and $({\bf L}\cdot{\bf S})^2$ are retained.

\section{Solution of the Schr\"odinger equation with $v_{12}$}
\label{app:a3}
In this appendix, we discuss the solution of the Schr\"odinger equation
with $v_{12}$, which contains ${\bf p}^2$-dependent central and tensor terms.
For simplicity, we ignore the electromagnetic and charge-dependent parts of $v_{12}$---the
treatment in the presence of $v_{12}^{\rm EM}$ is discussed in the following appendix.
In spin $S$ and isospin $T$ channel, the potential reads
\begin{equation}
v^{TS}_{12}=v^c_{TS}(r)+v^t_T(r)\, S_{12}+ v^b_T(r)\, {\bf L}\cdot {\bf S}
+v^q_{TS}(r)\, {\bf L}^2+v^{bb}_T(r) \left({\bf L}\cdot {\bf S}\right)^2
+\left\{ v^p_{TS}(r)+v^{pt}_T(r)\, S_{12}\, ,\, {\bf p}^2\right\} \ ,
\end{equation}
with
\begin{equation}
{\bf p}^2=\frac{{\bf L}^2}{r^2}-\frac{2}{r} \frac{d }{dr}- \frac{d^2 }{dr^2}\ .
\end{equation}
For single channels ($J=L$, where $L$ and $J$ are the orbital and total
angular momenta), the Schr\"odinger equation for the reduced radial
function $u_{TSJ}(r)$ reads
\begin{equation}
-\left( 1+ \overline{v} \right) u^{\prime\prime} -\overline{v}^{\,\prime} u^\prime
+\left[ v-\frac{\overline{v}^{\,\prime\prime}}{2}-k^2\right]u=0 \ ,
\label{eq:ssgl}
\end{equation}
where
\begin{eqnarray}
v_{TSJ}&=&2\, \mu \left[v^c_{TS}+\delta_{S,1}\left( 2\, v^t_{T}-v^b_T\right) +J(J+1)\left( v^q_{TS}
+2\, \frac{v^p_{TS}}{r^2}+\delta_{S,1}\, 4 \,\frac{v^{pt}_T}{r^2}\right) +\delta_{S,1} v^{bb}_{T} \right]
+\frac{J(J+1)}{r^2} \ , \\
\overline{v}_{TS}&=&4\, \mu\left( v^p_{TS}+\delta_{S,1}\, 2\, v^{pt}_{T}\right) \ ,
\end{eqnarray}
$\mu$ is the reduced mass, and the subscripts have been dropped for brevity.  The dependence
on the first derivative $u^\prime$ is removed by setting
\begin{equation}
u=\lambda\, w\ ,
\label{eq:ec6}
\end{equation}
and by requiring that terms proportional to $w^\prime$ vanish.  One finds that $\lambda$
must satisfy
\begin{equation}
2\left( 1+ \overline{v}\right)\lambda^\prime + \overline{v}^{\, \prime} \lambda=0 \ ,
\end{equation}
which has the solution
\begin{equation}
\lambda=\left(1+\overline{v}\right)^{-1/2} \ .
\label{eq:ec8}
\end{equation}
The function $w$ then satisfies
\begin{equation}
w^{\prime\prime}=f\, w \ , \qquad \left(1+\overline{v}\right) f= v-
\frac{(\overline{v}^{\,\prime}/2)^2}{1+\overline{v}}-k^2 \ ,
\label{eq:ec9}
\end{equation}
with the boundary condition (reinstating the appropriate superscripts and subscripts for the case
under consideration)
\begin{equation}
\frac{w_{TSJ}(r)}{r} \simeq \frac{1}{2} \left[ h^{(2)}_J(kr) + S^{JST}_{JJ}(k)\, h^{(1)}_J(kr) \right] \ ,
\end{equation}
where the Hankel functions are defined as
$h^{(1,2)}_L(kr) = j_L(kr) \pm i\, n_L(kr)$,
$j_L(kr)$ and $n_L(kr)$ being the regular and irregular spherical Bessel functions, respectively.
The differential equation above is solved with the standard Numerov method.

In coupled channels ($L=J \pm 1$) it is convenient to introduce the $2\times2$ matrices
$V$ and $\overline{V}$ with matrix elements given respectively by
\begin{eqnarray}
\!\!\!\!v^{TJ}_{--}\!\!&=&\!\! 2\, \mu \left[ v^c_{T1}-2\,\frac{J-1}{2J+1}\, v^t_T
+(J-1) v^b_T +J(J-1)\left( v^q_{T1}
+2\, \frac{v^p_{T1}}{r^2}- 4\,\frac{J-1}{2J+1}\,\frac{v^{pt}_T}{r^2}\right) +(J-1)^2 v^{bb}_{T} \right] \nonumber \\
&&+\frac{J(J-1)}{r^2} \ , \\
\!\!\!\!v^{TJ}_{++}\!\!&=&\!\! 2\, \mu \left[ v^c_{T1}-2\,\frac{J+2}{2J+1}\, v^t_T
-(J+2) v^b_T +(J+1)(J+2)\left( v^q_{T1}
+2\, \frac{v^p_{T1}}{r^2}- 4\,\frac{J+2}{2J+1}\,\frac{v^{pt}_T}{r^2}\right) +(J+2)^2 v^{bb}_{T} \right] \nonumber\\
&&+\frac{(J+1)(J+2)}{r^2}\ , \\
\!\!\!\!v^{TJ}_{-+}\!\!&=&\!\!1 2\, \mu \frac{\sqrt{J(J+1)}}{2J+1}
\left( v^t_T+ 2\,\frac{J^2+J+1}{r^2} \,v^{pt}_T\right)  , \qquad v^{TJ}_{+-}=v^{TJ}_{-+}\ ,
\end{eqnarray}
and
\begin{eqnarray}
\overline{v}^{TJ}_{--}&=&4\, \mu\left( v^p_{T1}-2\, \frac{J-1}{2J+1}\, v^{pt}_{T}\right) \ , \\
\overline{v}^{TJ}_{++}&=&4\, \mu\left( v^p_{T1}-2\, \frac{J+2}{2J+1}\, v^{pt}_{T}\right) \ , \\
\overline{v}^{TJ}_{-+}&=&24\, \mu \,  \frac{\sqrt{J(J+1)}}{2J+1}\, v^{pt}_{T} \ , \qquad
\overline{v}^{TJ}_{+-}=\overline{v}^{TJ}_{-+} \ .\\
\end{eqnarray}
where the subscript $-$ or $+$ specifies the orbital angular momentum $L=J-1$ or $L=J+1$.
With these definitions, the coupled-channel Schr\"odinger equation can be
written as
\begin{equation}
-\left( 1+ \overline{V} \right) U^{\prime\prime} -\overline{V}^{\,\prime} U^\prime
+\left[ V-\frac{\overline{V}^{\,\prime\prime}}{2}-k^2\right]U=0 \ ,
\label{eq:ssglv}
\end{equation}
where the transpose of the $U$ vector is given by $U^T=\left(u_{--},u_{+-}\right)$
or $U^T=\left(u_{-+},u_{++}\right)$, depending on whether the incoming wave
has $L=J-1$ or $L=J+1$.
Introducing the $2\times2$ matrix $\Lambda$ with
\begin{equation}
U=\Lambda\, W \ ,
\end{equation}
and requiring that terms proportional to $W^\prime$ vanish lead to
\begin{equation}
2\left( 1+ \overline{V}\right)\Lambda^\prime + \overline{V}^{\, \prime} \Lambda=0 \ .
\end{equation}
The set of first order differential equations above is solved with the Runge-Kutta method
by integrating out $\longrightarrow$ in.  Note that in the limit $r \rightarrow \infty$, $\Lambda$
reduces to the identity matrix (and hence the asymptotic behavior of $w_\mp$ is the same
as that of $u_\mp$).  Straightforward manipulations allow one to cast the Schr\"odinger
equation for $W$ in the standard form
\begin{equation}
W^{\prime\prime}=F\, W\ , \qquad \left(1+\overline{V}\right) \Lambda\, F\, \Lambda^{-1}=  V-
\frac{1}{4}\, \overline{V}^{\,\prime} (1+\overline{V})^{-1}  \,\overline{V}^{\,\prime}-k^2 \ ,
\end{equation}
with the boundary conditions (again, reinstating superscripts and subscripts)
\begin{equation}
\frac{w^{TSJ}_{L^\prime L}(r)}{r} \simeq \frac{1}{2} \left[ \delta_{L^\prime L}\,
h^{(2)}_{L^\prime}(kr) + S^{JST}_{L^\prime L}(k)\, h^{(1)}_{L^\prime}(kr) \right] \ ,
\end{equation}
where $L=J\mp 1$ is the orbital angular momentum of the incoming wave.
\section{$pp$ phase shifts and effective range expansion}
\label{app:a4}
We discuss briefly the calculation of the $pp$ phase shifts and
effective range expansion with inclusion of the full electromagnetic potential
$v^{\rm EM}_{12}$~\cite{Wiringa95}. Radial wave functions
behave in the asymptotic region ($r\gtrsim 30$ fm) as
\begin{equation}
\frac{u_L(r)}{r}  \simeq \frac{1}{2} \left[ \overline{h}^{\,(2)}_L(kr;\eta^\prime) 
+ e^{2i\delta^{\rm EM}_L}\, \overline{h}^{\, (1)}_L(kr;\eta^\prime) \right] \ ,
\end{equation}
where $L=J$ for single channels or $L=L^\prime=J\mp 1$
for coupled channels (the pair isospin and spin subscripts $T$ and $S$ have been dropped
for simplicity), $\overline{h}^{\, (1,2)}_L(kr;\eta^\prime)$ are defined in terms
of regular, $\overline{F}_L(kr;\eta^\prime)$, and irregular,
$\overline{G}_L(kr;\eta^\prime)$, electromagnetic (EM) functions as
\begin{equation}
\overline{h}^{\,(1,2)}_L(kr) =\frac{\overline{F}_L(kr;\eta^\prime)}{kr}
 \mp i\, \frac{\overline{G}_L(kr;\eta^\prime)}{kr} \ ,
\end{equation}
$\delta_L^{\rm EM}$ are the EM phase shifts shown in Sec.~\ref{sec:res},
and the Coulomb parameter $\eta^\prime$ is defined~\cite{Berg88} as
\begin{equation}
\eta^\prime=\frac{\alpha M_p}{2\, k}\,\frac{1+2\, k^2/M_p^2}{\sqrt{1+k^2/M_p^2}} \ .
\end{equation}
The EM functions, generically denoted as $X_L(kr;\eta^\prime)$, are solutions of the radial equation
\begin{equation}
\left[\frac{d^2}{dr^2}+k^2-\frac{L(L+1)}{r^2}-M_p \left[V_{C1}(r)+V_{C2}(r)+V_{VP}(r)\right] \right]
X_L(kr;\eta^\prime)=0 \ ,
\end{equation}
where $V_{C1}$ ($V_{C2}$) and $V_{VP}$ are respectively
the first-order (second-order) Coulomb and vacuum polarization terms.
These terms include form factors to remove singularities in the $r=0$
limit~\cite{Wiringa95}.  Note that the Darwin-Foldy and magnetic moment
corrections are not included above, since at large $r$ the former falls off
exponentially and the latter behaves as $1/r^3$. 

Following Ref.~\cite{Heller60} and treating the $V_{C2}(r)$ and $V_{VP}(r)$ corrections
in first order perturbation theory, one finds that $\overline{F}_L(kr;\eta^\prime$ and
$\overline{G}_L(kr;\eta^\prime)$ can be expressed as
\begin{eqnarray}
\label{eq:fem}
\overline{F}_L(kr;\eta^\prime)&=&F_L(kr;\eta^\prime)\left[1-\int_r^\infty{\rm d}r^\prime\, G_L(kr^\prime;\eta^\prime)
\, V(r^\prime)\, F_L(kr^\prime;\eta^\prime)\right] \nonumber \\
&&+G_L(kr;\eta^\prime)\left[{\rm tan}(\rho_L+\tau_L)+\int_r^\infty{\rm d}r^\prime\, F_L(kr^\prime;\eta^\prime)
\, V(r^\prime)\, F_L(kr^\prime;\eta^\prime)\right] \\
\label{eq:gem}
\overline{G}_L(kr;\eta^\prime)&=&G_L(kr;\eta^\prime)\left[1+\int_r^\infty{\rm d}r^\prime\, G_L(kr^\prime;\eta^\prime)
\,V(r^\prime)\, F_L(kr^\prime;\eta^\prime)\right] \nonumber \\
&&-F_L(kr;\eta^\prime)\left[{\rm tan}(\rho_L+\tau_L)+\int_r^\infty{\rm d}r^\prime\, G_L(kr^\prime;\eta^\prime)
\,V(r^\prime)\, G_L(kr^\prime;\eta^\prime)\right] \ ,
\end{eqnarray}
where the $F_L$ and $G_L$ are standard Coulomb functions, the function $V(r)$ is proportional to
$V_{C2}(r)$ and $V_{VP}(r)$,
\begin{equation}
V(r)=\frac{M_p}{k}\, \left[ V_{C2}(r)+V_{VP}(r)\right] \ ,
\end{equation}
and the phase shifts $\rho_L$ and $\tau_L$ corresponding, respectively, to $V_{C2}$ and
$V_{VP}$ are given (in first order perturbation theory) by
\begin{equation}
\label{eq:phem}
{\rm tan} (\rho_L+\tau_L) \simeq \rho_L+\tau_L =-\int_0^\infty {\rm d}r\,  F_L(kr;\eta^\prime)
\, V(r)\, F_L(kr;\eta^\prime) \ .
\end{equation}
In the absence of $V_{C2}$ and $V_{VP}$, the solutions
$\overline{F}_L$ and $\overline{G}_L$ reduce to the regular
and irregular Coulomb functions. 
In the computer programs Eqs.~(\ref{eq:fem})--(\ref{eq:gem}) are used to
construct the EM functions and Eq.~(\ref{eq:phem}) to obtain the phase
shifts $\rho_L$ and $\tau_L$.  

The effective range expansion in the $^1$S$_0$ channel is
obtained as~\cite{Heller60,Berg88,Vands83}
\begin{equation}
F_{\rm EM}(k^2)=-\frac{1}{a_{\rm EM}}+\frac{1}{2}\, r_{\rm EM}\, k^2+\dots \ ,
\end{equation}
with
\begin{eqnarray}
F_{\rm EM}(k^2)=k\, C^2_0(\eta^\prime)\, \frac{ (1+\chi_0)\, {\rm cot}\,\delta^{\rm EM}_0-{\rm tan}\,\tau_0}
{(1+A_1)(1-\chi_0)}+2\,k\,\eta^\prime\, h(\eta^\prime)\,(1-A_2) 
+\, k^2\, d \left[C_0^4(\eta^\prime)-1\right]+k\, {\widetilde l}_0 \ ,
\label{eq:erf}
\end{eqnarray}
where
\begin{eqnarray}
&&C_0^2(\eta^\prime)=\frac{2\pi\,\eta^\prime}{e^{2\pi\eta^\prime}-1} \ , \qquad
h(\eta^\prime)=-\gamma-{\rm ln}\, \eta^\prime+\sum_{n=1}^\infty\frac{\eta^{\prime 2} }{n\,(n^2+\eta^{\prime 2})} \ , \\
&& \chi_o=- \frac{4\alpha}{3\pi} \, \eta^\prime\int_0^\infty
dr\,  \frac{I(r)}{r}\, F_0(kr;\eta^\prime)\, G_0(kr;\eta^\prime)
\ , \qquad \widetilde{l}_0=-\frac{4\, \alpha}{3\pi}\, \eta^\prime \int_0^\infty
dr\, \frac{I(r)}{r}\Big[ C^2_0(\eta^\prime)\, G^2_0(kr;\eta^\prime)-1\Big] \ ,\\
&&d=\frac{\alpha}{M_p} \ , \qquad A_1=4\,d\, k\,\eta^\prime\left[ {\rm ln}\left( 2\,d\, k\,\eta^\prime\right) 
+h(\eta^\prime)+2\,\gamma-1\right]\ ,\qquad
A_2=2\,d\, k\,\eta^\prime\left( 2\,{\rm ln}\, \alpha +2\,\gamma-1\right)+\frac{A_1}{2} \ , 
\end{eqnarray}
$\gamma$ is Euler's constant, and the function $I(r)$ entering the
vacuum polarization potential $V_{VP}(r)$ is defined as in Ref.~\cite{Heller60},
\begin{equation}
I(r)=\int_1^\infty dx\, e^{2 m_erx}\left(1+\frac{1}{2\, x^2}\right) \frac{\sqrt{x^2-1}}{x^2} \ .
\end{equation}
\begin{center}
\begin{figure*}[bth]
\includegraphics[width=5in]{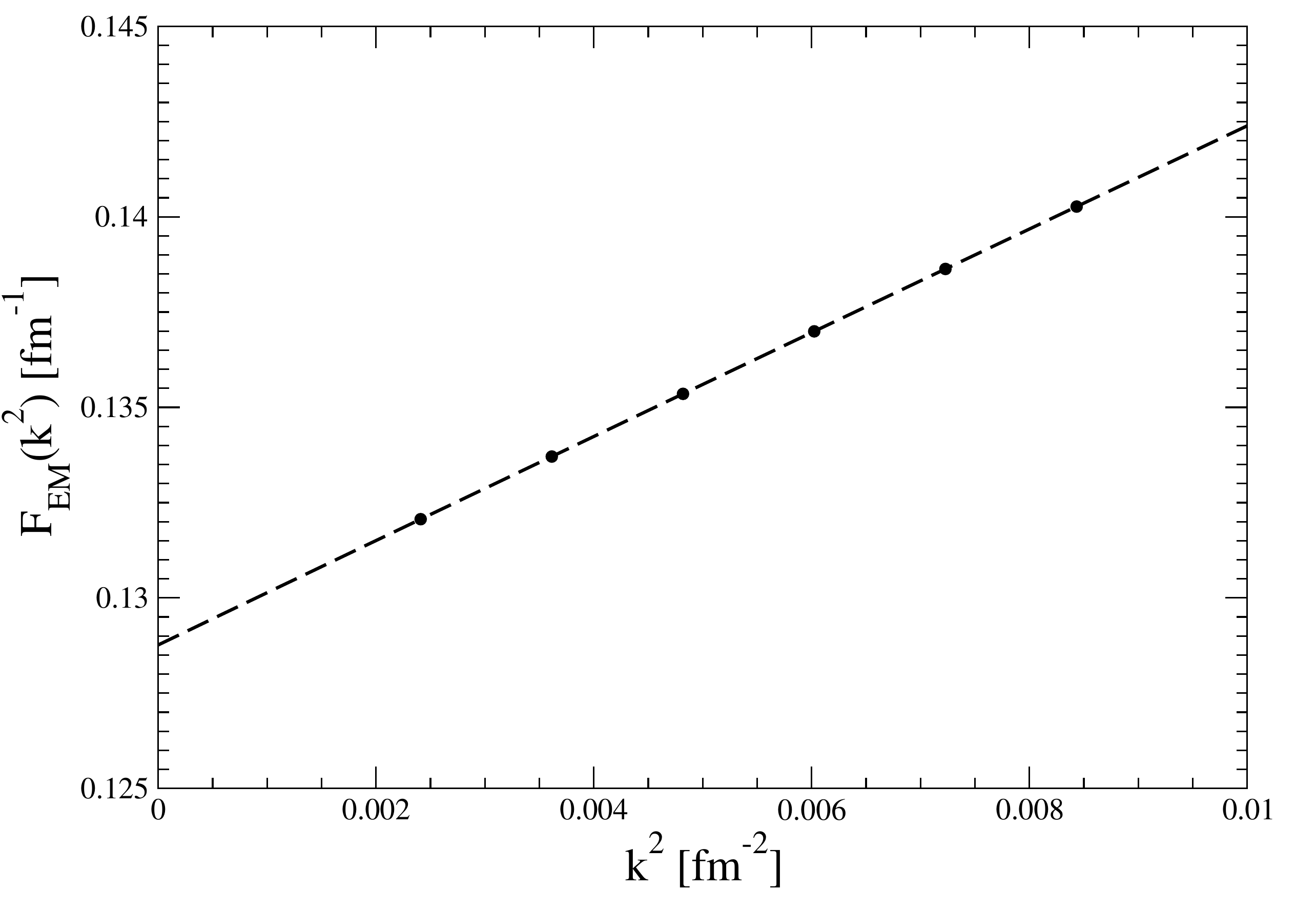}
\caption{The effective range function of Eq.~(\ref{eq:erf})
for the potential model b with $(R_{\rm L},R_{\rm S})=(1.0,0.7)$ fm.
The dashed line is a straight line fit.}
\label{fig:erfnt}
\end{figure*}
\end{center}
The effective range function $F_{\rm EM}(k^2)$ corresponding
to model b is shown in Fig.~\ref{fig:erfnt}.
The numerical methods are stable down to lab energies of 1 keV.

\section{Tables of phase shifts and figures of potential components}
\label{app:a5}
\begin{center}
\begin{figure}[bth]
\includegraphics[width=7in]{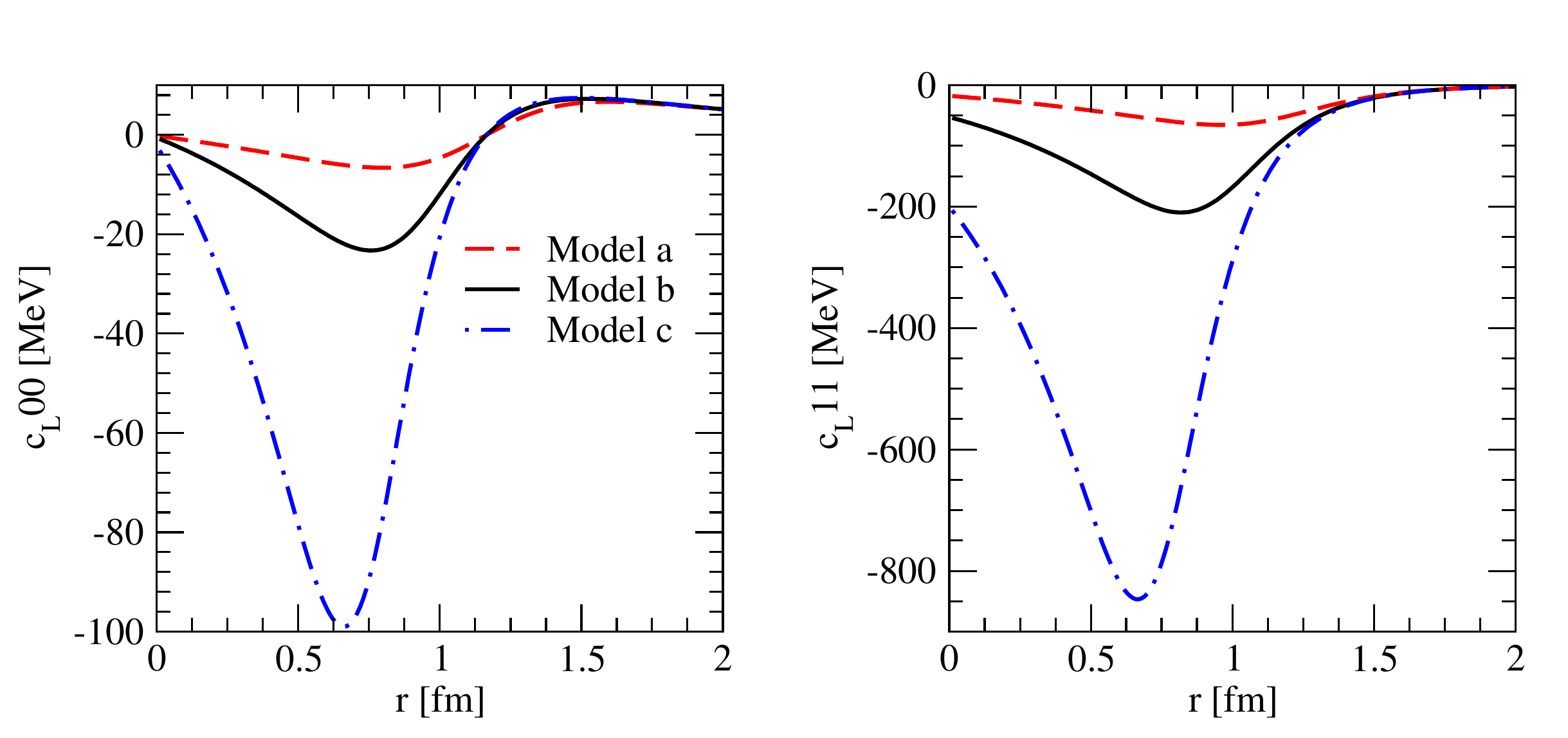}
\caption{(Color online) Central components of the long-range potential $v^{\rm L}_{12}$
in pair spin-isospin channels  $ST=00$ and $11$.}
\label{fig:f00l}
\end{figure}
\end{center}

\begin{center}
\begin{figure}[bth]
\includegraphics[width=7in]{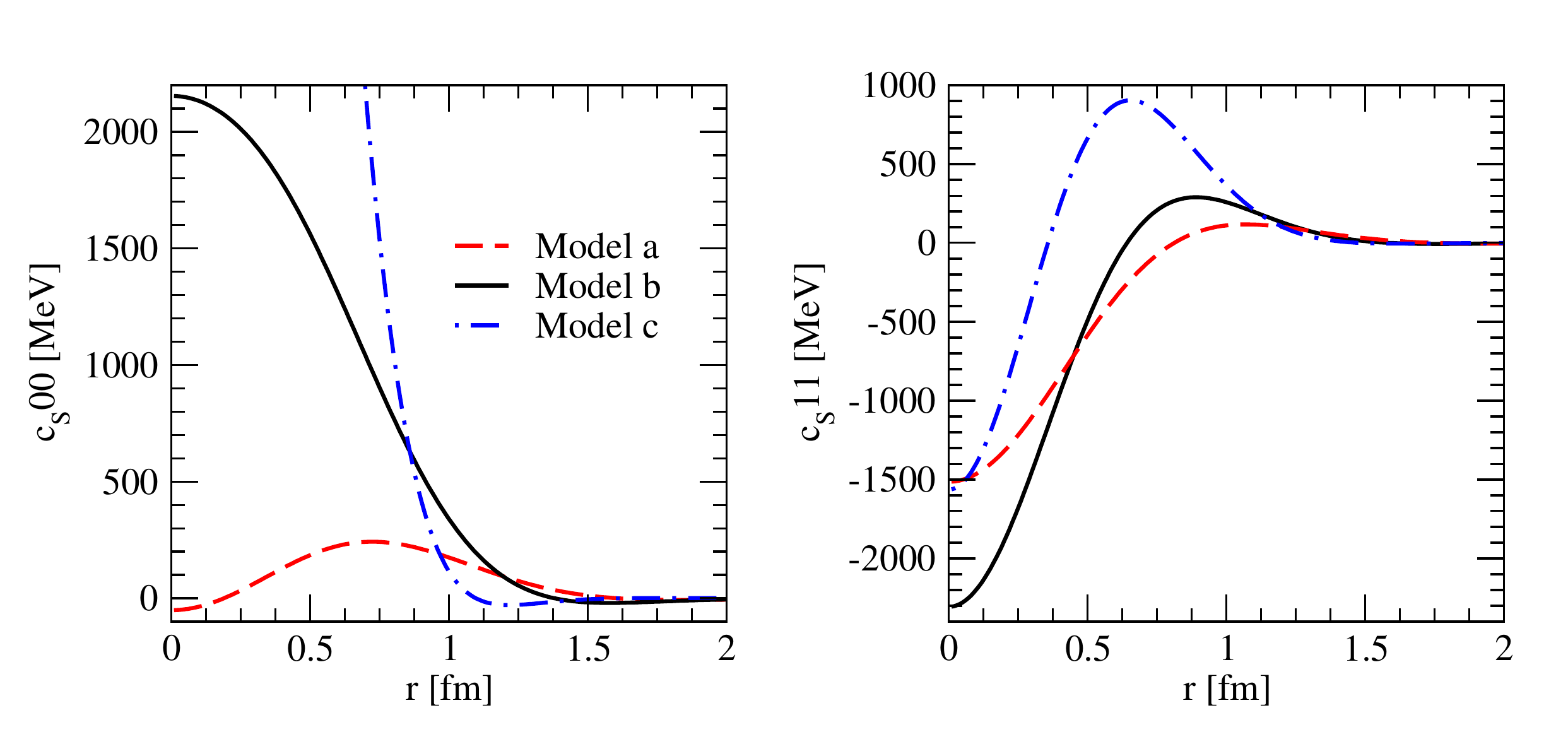}
\caption{(Color online) Same as in Fig.~\ref{fig:f00l}, but for the short-range charge-independent potential
$v^{\rm S,CI}_{12}$.}
\label{fig:f00s}
\end{figure}
\end{center}

\begin{center}
\begin{figure}[bth]
\includegraphics[width=7in]{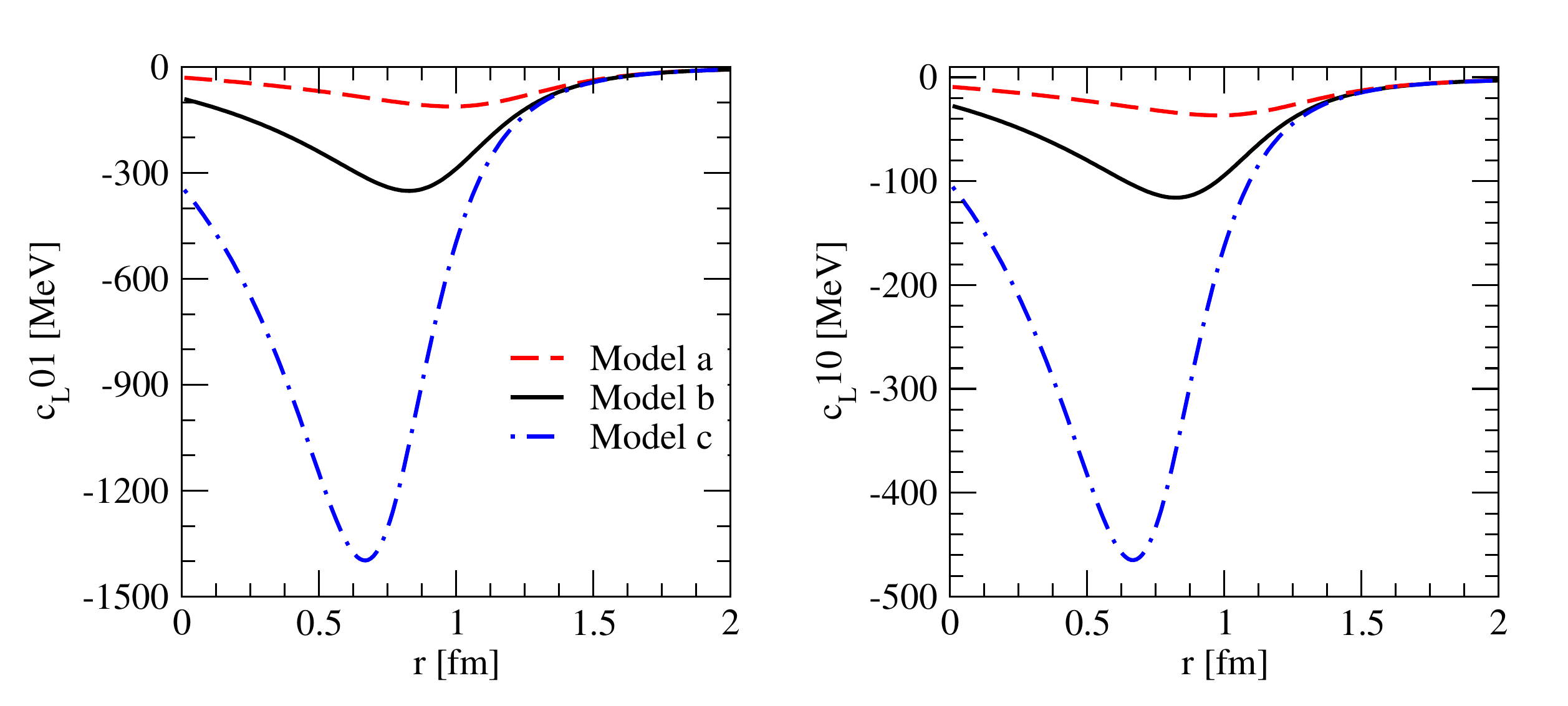}
\caption{(Color online) Same as in Fig.~\ref{fig:f00l} but in pair spin-isospin channels $ST=01$ and $10$.}
\label{fig:f01l}
\end{figure}
\end{center}

The $pp$ and $np$ phase shifts calculated with model b are listed in
Tables~\ref{phspp}--\ref{phsnp0}, while the various components of the
long-range ($v^{\rm L}_{12}$) and short-range ($v^{\rm S,CI}_{12}$)
potentials corresponding to models a, b, and c and projected
out in pair spin and isospin $S=0,1$ and $T=01$, are shown in
Figs.~\ref{fig:f00l}--\ref{fig:ftp0s}.

\mediumtext
\begin{table}
\caption{$pp$ phase shifts in degrees for potential model b with $(R_{\rm L},R_{\rm S})=(1.0,0.7)$ fm.
The phases are relative to electromagnetic functions.}
\begin{ruledtabular}
\begin{tabular}{rdddddddddd}
\multicolumn{1}{c}{$E_{\rm lab}$} &
\multicolumn{1}{c}{$^{1}S_{0}$}  & 
\multicolumn{1}{c}{$^{1}D_{2}$} & 
\multicolumn{1}{c}{$^{1}G_{4}$}& 
\multicolumn{1}{c}{$^{3}P_{0}$} & 
\multicolumn{1}{c}{$^{3}P_{1}$} &
\multicolumn{1}{c}{$^{3}F_{3}$} & 
\multicolumn{1}{c}{$^{3}P_{2}$ }& 
\multicolumn{1}{c}{$\epsilon_{2}$} & 
\multicolumn{1}{c}{$^{3}F_{2}$} & 
\multicolumn{1}{c}{$^{3}F_{4}$} \\
   \tableline
  1 &  32.69 &   0.00  & 0.00   & 0.14 &  -0.08 &  -0.00  &  0.02 &  -0.00&   -0.00 &   0.00  \\
  5  & 55.00 &   0.04  &  0.00 &   1.64 &  -0.90 &  -0.00 &   0.22 &  -0.05 &  -0.01 &   0.01 \\
 10 & 55.49  &  0.17   & 0.00 &   3.90  & -2.06 &  -0.03  &  0.64  & -0.19  & -0.01  &  0.02  \\
 25 &  49.13 &   0.69   & 0.04  &  9.21  & -4.95 &  -0.23   & 2.42  & -0.80   & 0.06   & 0.04  \\
 50  & 39.52 &   1.68   & 0.16 &  12.77 &  -8.38  & -0.70  &  5.73  & -1.71&    0.27   & 0.14 \\
100 &  25.66 &   3.77 &   0.43 &  11.21 & -13.42 &  -1.58  & 11.02 &  -2.73 &   0.73 &   0.47 \\
150  & 15.44 &   5.75  &  0.71 &   6.21 & -17.63  & -2.28  & 14.16  & -3.05  &  1.10   & 0.97  \\
200 &   7.20   & 7.38  &  1.01  &  0.50 & -21.38  & -2.90  & 15.90  & -2.97   & 1.30   & 1.55   \\
250 &   0.22 &   8.59  &  1.33  & -5.18 & -24.68  & -3.52  & 16.89  & -2.65   & 1.27   & 2.16 \\
300  & -5.88 &   9.36  &  1.66 & -10.62 & -27.55  & -4.20  & 17.45  & -2.19   & 0.98   & 2.76  
\end{tabular}
\end{ruledtabular}
\label{phspp}
\end{table}

\mediumtext
\begin{table}
\caption{$T=1$ $np$ phase shifts in degrees for potential model b with $(R_{\rm L},R_{\rm S})=(1.0,0.7)$ fm.
The phases are relative to spherical Bessel functions.}
\begin{ruledtabular}
\begin{tabular}{rdddddddddd}
\multicolumn{1}{c}{$E_{\rm lab}$} &
\multicolumn{1}{c}{$^{1}S_{0}$}  & 
\multicolumn{1}{c}{$^{1}D_{2}$} & 
\multicolumn{1}{c}{$^{1}G_{4}$}& 
\multicolumn{1}{c}{$^{3}P_{0}$} & 
\multicolumn{1}{c}{$^{3}P_{1}$} &
\multicolumn{1}{c}{$^{3}F_{3}$} & 
\multicolumn{1}{c}{$^{3}P_{2}$ }& 
\multicolumn{1}{c}{$\epsilon_{2}$} & 
\multicolumn{1}{c}{$^{3}F_{2}$} & 
\multicolumn{1}{c}{$^{3}F_{4}$} \\
   \tableline
  1 &  62.10 &   0.00 &   0.00  &  0.18 &  -0.11 &  -0.00 &   0.02  & -0.00 &   0.00 &   0.00\\
  5  & 63.65  &  0.04  &  0.00  &  1.67  & -0.92  & -0.00  &  0.24  & -0.05  &  0.01  &  0.00 \\  
 10 &  60.00 &   0.16 &   0.00  &  3.80 &  -2.02  & -0.03  &  0.68 &  -0.19 &   0.02  &  0.00  \\ 
 25 &  50.83 &   0.67 &   0.03 &   8.71  & -4.72  & -0.20   & 2.53  & -0.76 &   0.11 &   0.01  \\ 
 50 &  40.22  &  1.69 &   0.14 &  11.90&   -7.88 &  -0.63 &   5.95 &  -1.63 &   0.33 &   0.08   \\
100  & 25.84 &   3.86 &   0.40  & 10.06&  -12.42&   -1.46 &  11.35&   -2.58&    0.81 &   0.38  \\ 
150  & 15.46 &   5.90  &  0.69  &  4.97 & -16.17 &  -2.12 &  14.49 &  -2.81 &   1.20 &   0.84   \\
200 &   7.13 &   7.58  &  1.00 &  -0.77 & -19.50 &  -2.70  & 16.17  & -2.64 &   1.44  &  1.41   \\
250  &  0.09 &   8.81  &  1.33 &  -6.48 & -22.43 &  -3.27 &  17.05  & -2.24  &  1.45 &   2.01   \\
300  & -6.04 &   9.59  &  1.67 & -11.93 & -24.96  & -3.89 &  17.49&   -1.72 &   1.21 &   2.60  
\end{tabular}
\end{ruledtabular}
\label{phsnp1}
\end{table}

\mediumtext
\begin{table}
\caption{Same as in Table~\ref{phsnp1} but for $T=0$ $np$ phase shifts.}
\begin{ruledtabular}
\begin{tabular}{rdddddddddd}
\multicolumn{1}{c}{$E_{\rm lab}$} &
\multicolumn{1}{c}{$^{1}P_{1}$}  & 
\multicolumn{1}{c}{$^{1}F_{3}$} & 
\multicolumn{1}{c}{$^{3}D_{2}$}& 
\multicolumn{1}{c}{$^{3}G_{4}$} & 
\multicolumn{1}{c}{$^{3}S_{1}$} &
\multicolumn{1}{c}{$\epsilon_{1}$} & 
\multicolumn{1}{c}{$^{3}D_{1}$ }& 
\multicolumn{1}{c}{$^{3}D_{3}$} & 
\multicolumn{1}{c}{$\epsilon_{3}$} & 
\multicolumn{1}{c}{$^{3}G_{3}$} \\
   \tableline
  1  & -0.19 &  -0.00 &   0.01 &   0.00 &  147.81&    0.10 &  -0.00  &  0.00  &  0.00 &  -0.00\\
  5  & -1.53 &  -0.01 &   0.22  &  0.00 &  118.32  &  0.63 &  -0.17 &   0.00 &   0.01  & -0.00\\
 10  & -3.15 &  -0.07 &   0.85 &   0.01 &  102.80 &   1.06 &  -0.65 &   0.00 &   0.08 &  -0.00\\    
 25  & -6.55 &  -0.43 &   3.70  &  0.17  &  80.86  &  1.53 &  -2.77  &  0.00   & 0.55  & -0.04\\
 50  & -9.87  & -1.16 &   8.89  &  0.73  &  63.00  &  1.62 &  -6.42 &   0.18  &  1.62 &  -0.25\\ 
100&  -14.05 &  -2.33 &  17.21 &   2.20 &   43.53 &   1.67 & -12.31 &   1.16 &   3.54 &  -0.97\\
150 & -17.48 &  -3.12 &  22.33 &   3.71 &   31.32 &   1.92 & -16.61 &   2.34 &   4.87 &  -1.88\\ 
200 & -20.78  & -3.69  & 25.02  &  5.10  &  22.35  &  2.34 & -19.83  &  3.17  &  5.72  & -2.83\\
250 & -24.04 &  -4.14 &  26.09  &  6.36  &  15.26 &   2.84 & -22.27  &  3.40 &   6.23&   -3.76\\
300 & -27.23 &  -4.56 &  26.10 &   7.46   &  9.40  &  3.39 & -24.11  &  3.01 &   6.52 &  -4.62\\
\end{tabular}
\end{ruledtabular}
\label{phsnp0}
\end{table}
\begin{center}
\begin{figure}[bth]
\includegraphics[width=7in]{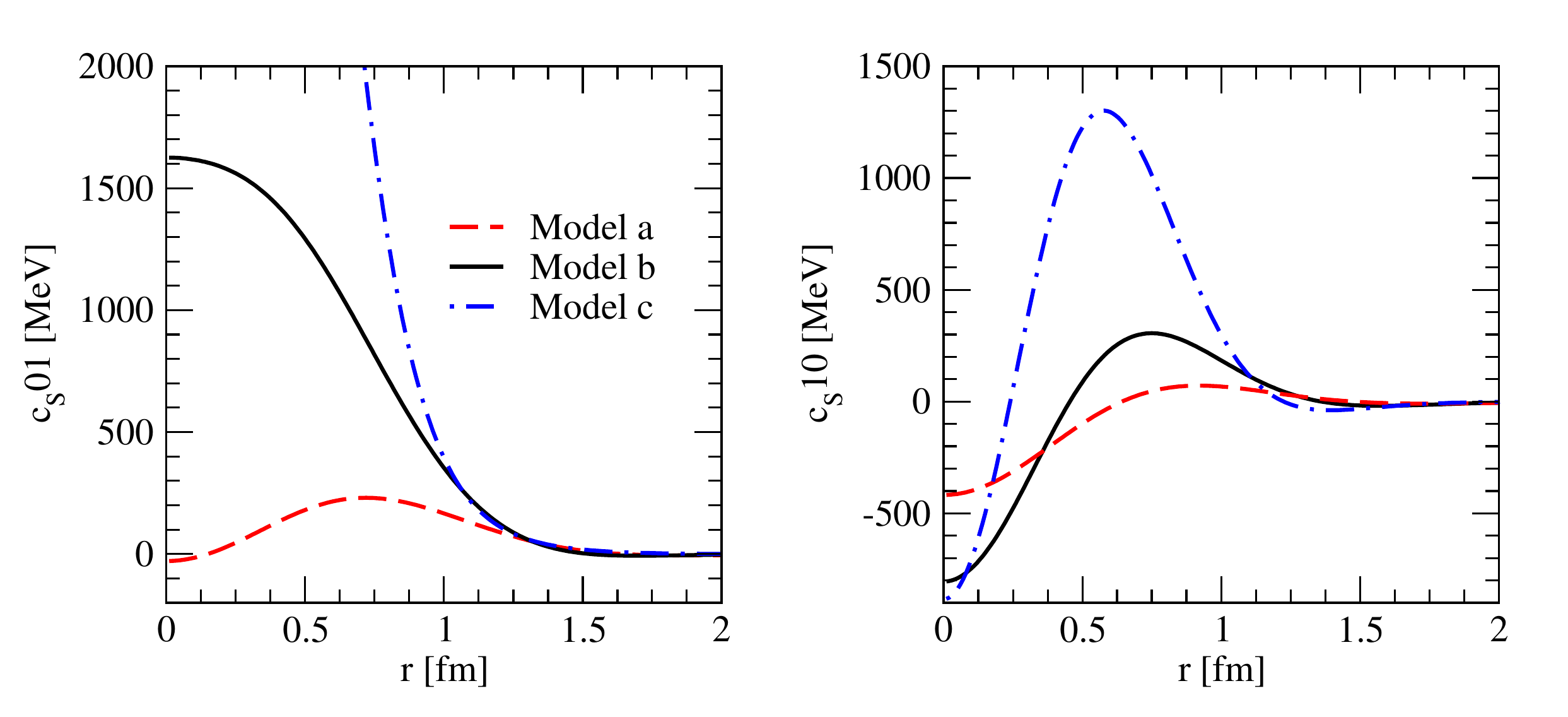}
\caption{(Color online) Same as in Fig.~\ref{fig:f00s} but in pair spin-isospin channels $ST=01$ and $10$.}
\end{figure}
\end{center}
\begin{center}
\begin{figure}[bth]
\includegraphics[width=7in]{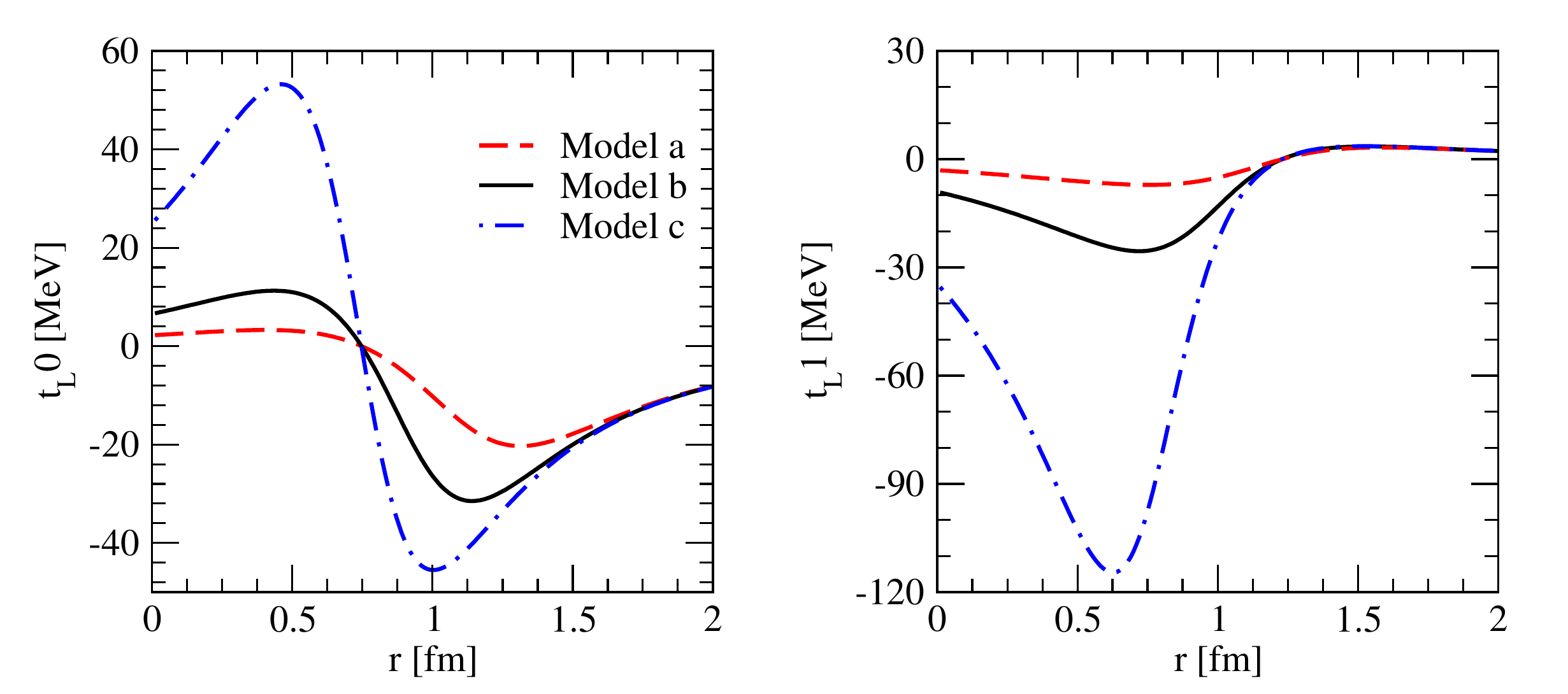}
\caption{(Color online) Tensor components of the long-range potential $v^{\rm L}_{12}$
in pair isospin channels  $T=0$ and $1$.}
\label{fig:ft0l}
\end{figure}
\end{center}
\begin{center}
\begin{figure}[bth]
\includegraphics[width=7in]{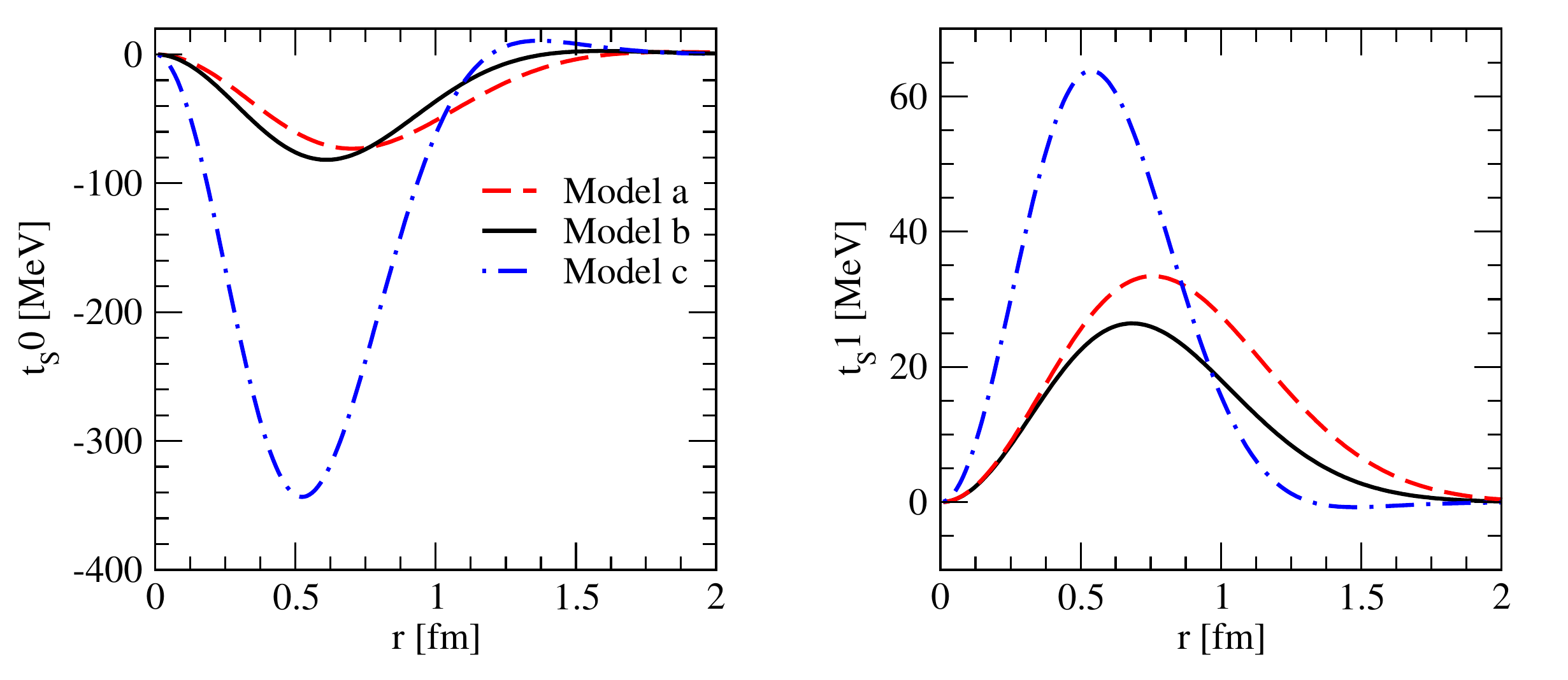}
\caption{(Color online) Same as in Fig.~\ref{fig:ft0l} but for the short-range charge-independent potential
$v^{\rm S,CI}_{12}$. }
\label{fig:ft0s}
\end{figure}
\end{center}
\begin{center}
\begin{figure}[bth]
\includegraphics[width=7in]{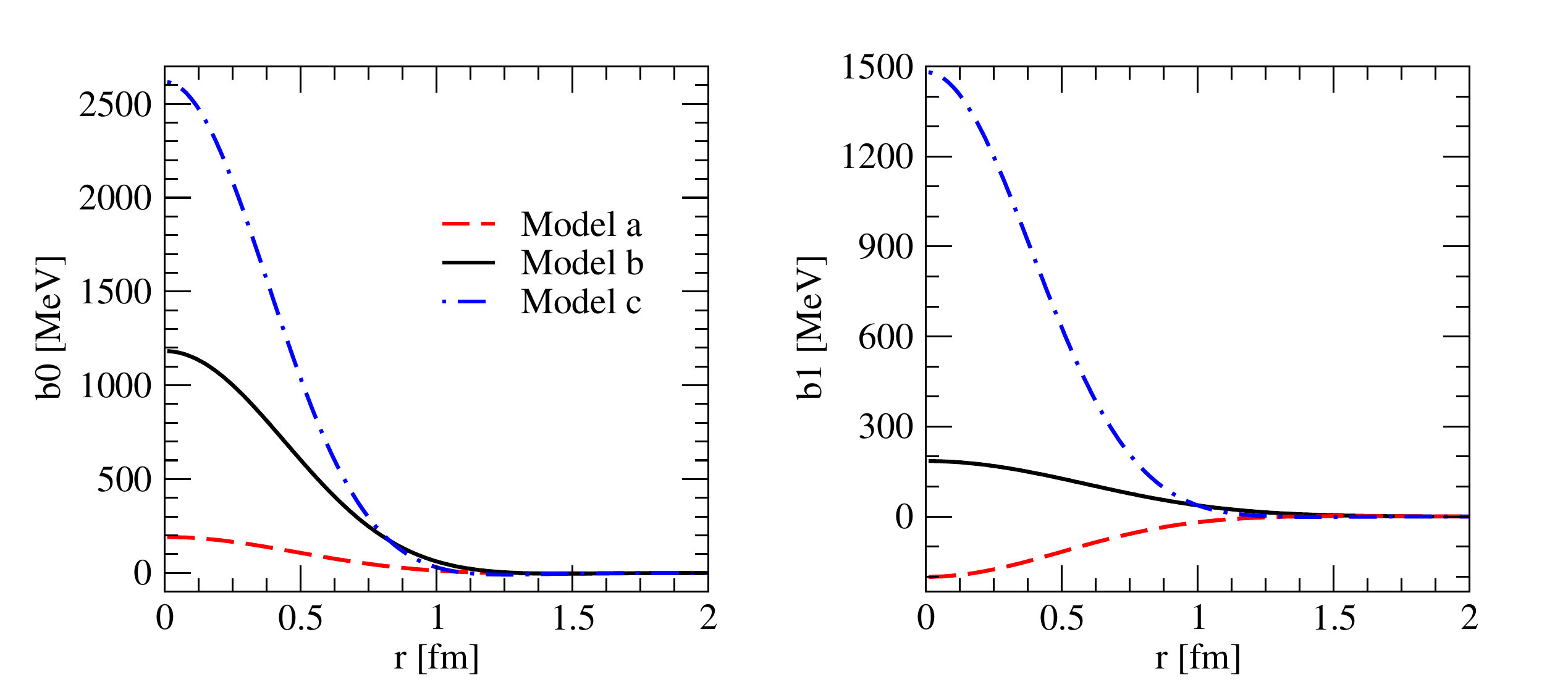}
\caption{(Color online) Spin-orbit components of the short-range charge-independent potential
$v^{\rm S,CI}_{12}$ in pair isospin channels  $T=0$ and $1$.}
\label{fig:fb0s}
\end{figure}
\end{center}
\begin{center}
\begin{figure}[bth]
\includegraphics[width=3.5in]{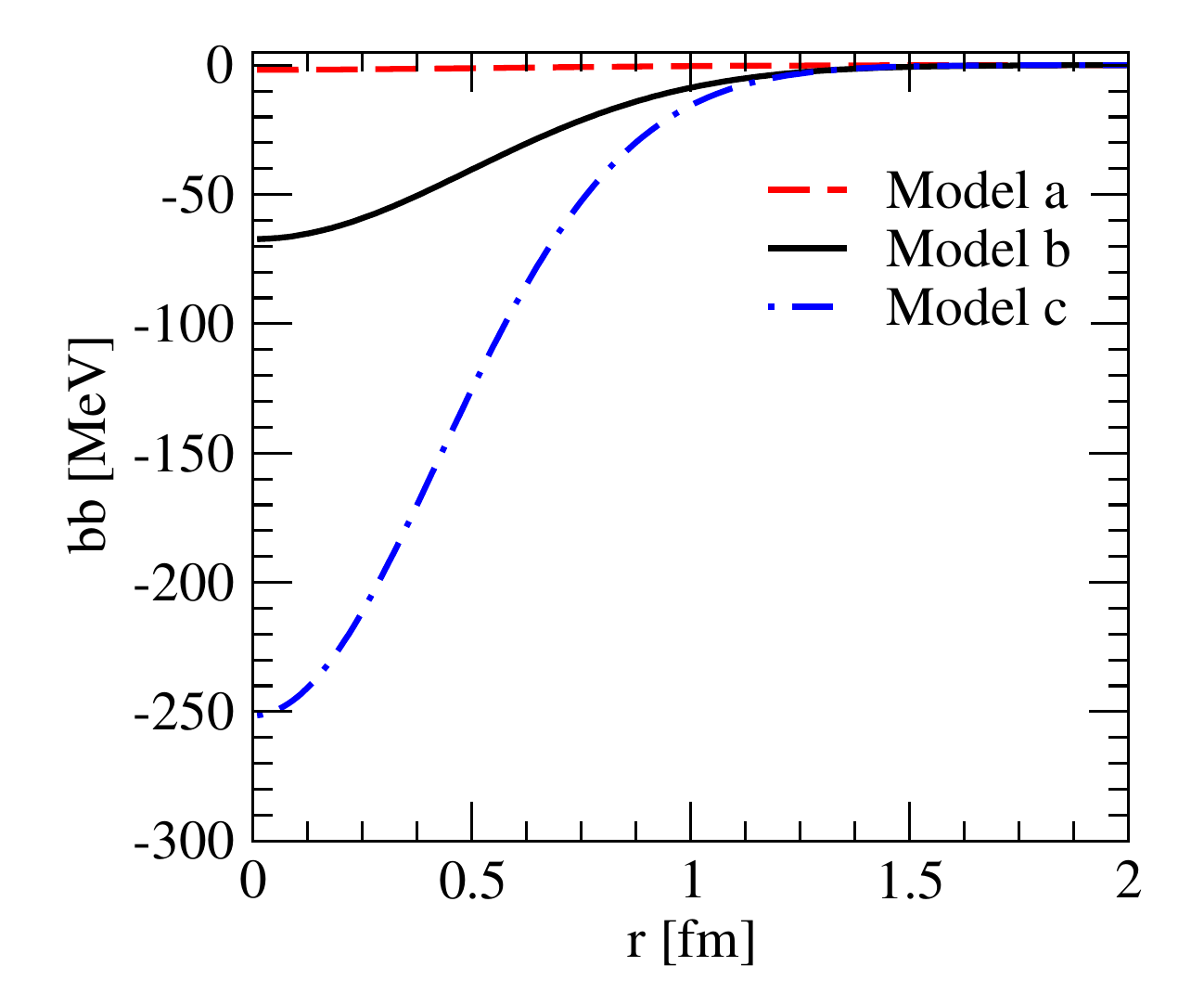}
\caption{(Color online) Spin and isospin independent quadratic spin-orbit components of the short-range charge-independent potential
$v^{\rm S,CI}_{12}$.}
\label{fig:fbbs}
\end{figure}
\end{center}
\begin{center}
\begin{figure}[bth]
\includegraphics[width=7in]{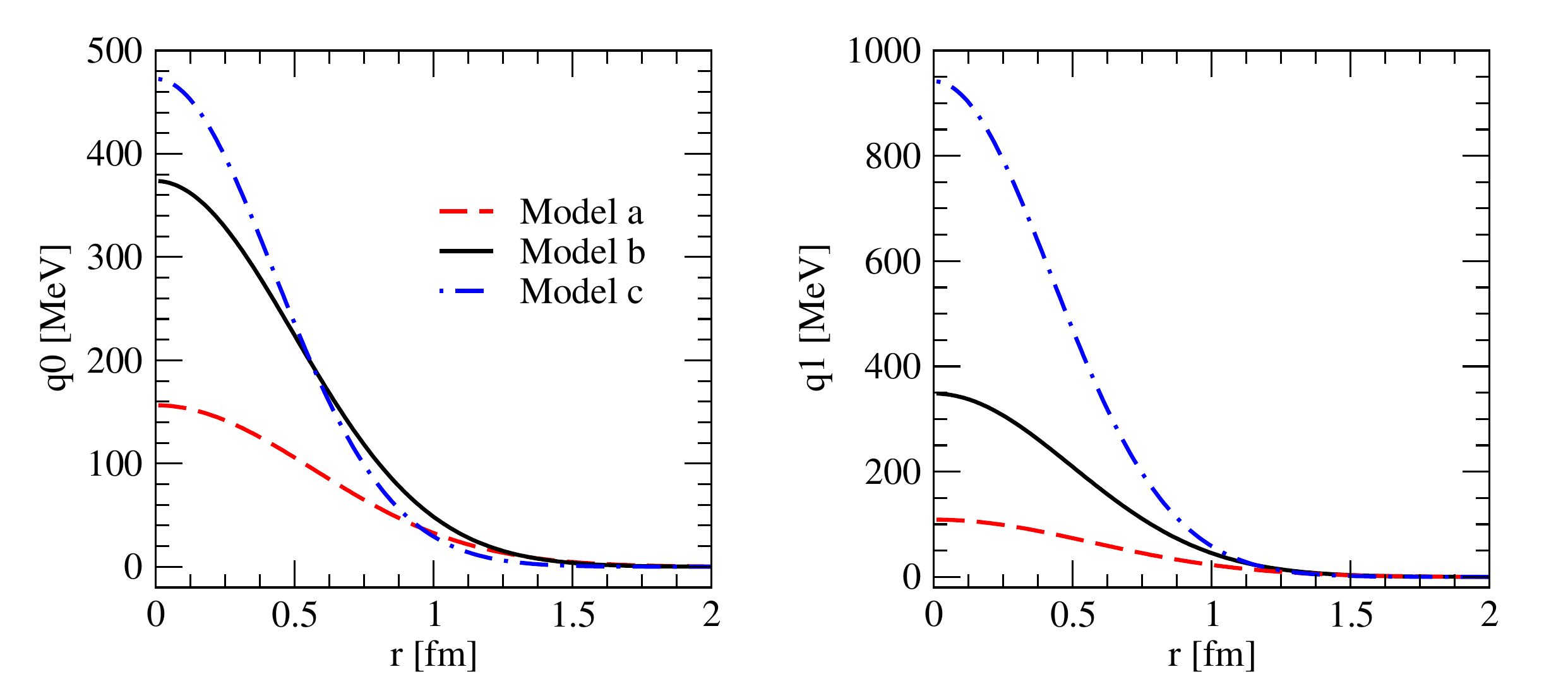}
\caption{(Color online) Quadratic orbital angular momentum components of the short-range charge-independent potential
$v^{\rm S,CI}_{12}$ in pair spin channels $S=0$ and $1$.}
\label{fig:fq0s}
\end{figure}
\end{center}
\begin{center}
\begin{figure}[bth]
\includegraphics[width=7in]{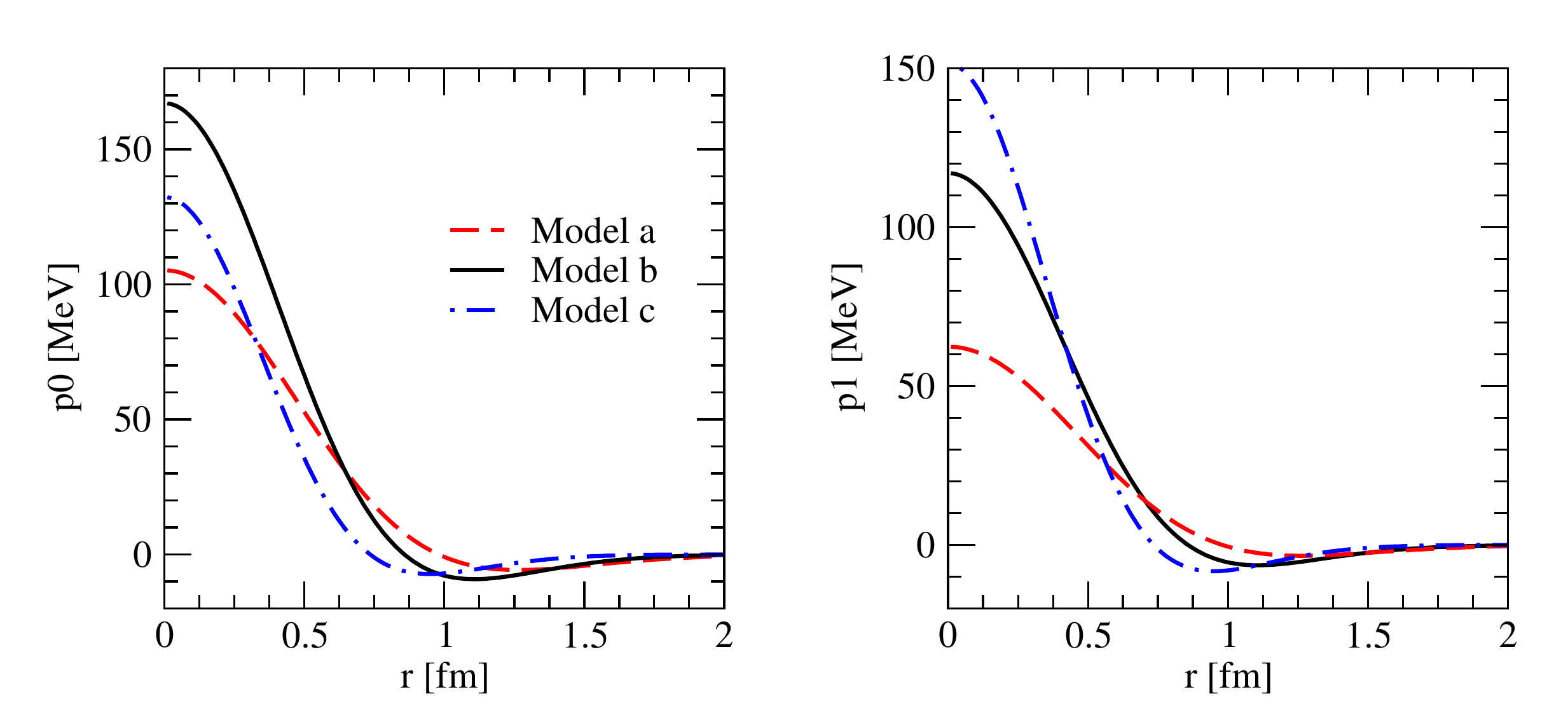}
\caption{(Color online) Quadratic relative momentum components of the short-range charge-independent potential
$v^{\rm S,CI}_{12}$ in pair spin channels $S=0$ and $1$.}
\label{fig:fp0s}
\end{figure}
\end{center}
\begin{center}
\begin{figure}[bth]
\includegraphics[width=7in]{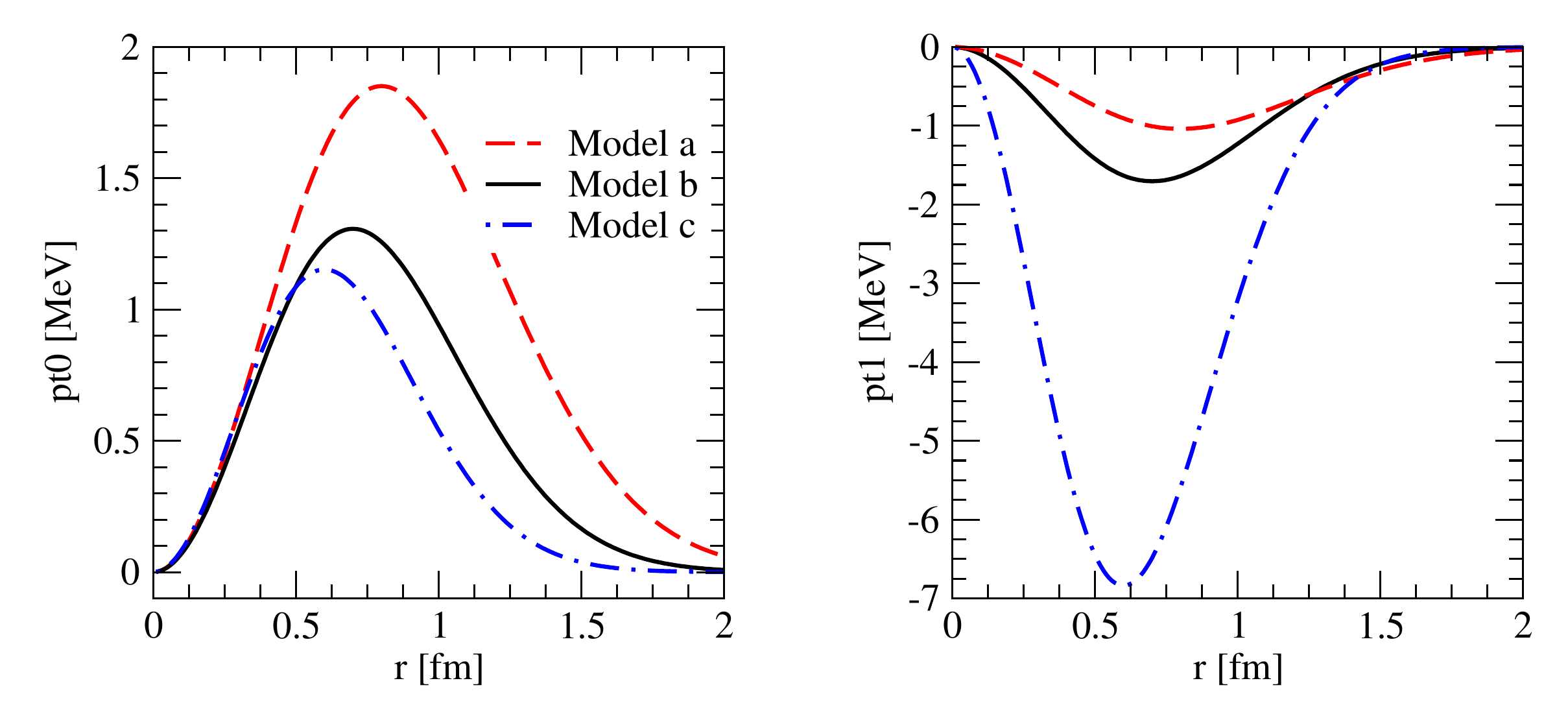}
\caption{(Color online) Quadratic-relative-momentum-tensor components of the short-range charge-independent potential
$v^{\rm S,CI}_{12}$ in pair isospin channels $T=0$ and $1$.}
\label{fig:ftp0s}
\end{figure}
\end{center}

\begin{thebibliography}{100}
%
%
\bibitem{Stapp:1956mz} 
H.P.\ Stapp, T.J.\ Ypsilantis, and N.\ Metropolis,
Phys.\ Rev.\  {\bf 105}, 302 (1957).
%
\bibitem{arndt1966chi}
R.A.\ Arndt and M.H.\ MacGregor,
Methods in Computational Physics {\bf 6}, 253 (1966).
%
\bibitem{Stoks93}
V.G.J.\ Stoks, R.A.M.\ Klomp, M.C.M.\ Rentmeester, and J.J.\ de Swart,
Phys.\ Rev.\ C {\bf 48}, 792 (1993).
%
\bibitem{Stoks94}
V.G.J.\ Stoks, R.A.M.\ Klomp, C.P.F.\ Terheggen, and J.J.\ de Swart,
Phys.\ Rev.\ C {\bf 49}, 2950 (1994).
%
\bibitem{Wiringa95}
R.\ B.\ Wiringa, V.\ G.\ J.\ Stoks, and R.\ Schiavilla,
Phys.\ Rev.\ C {\bf 51}, 38 (1995).
%
\bibitem{Rentmeester:1999vw} 
M.C.M.\ Rentmeester, R.G.E.\ Timmermans, J.L.\ Friar, and J.J.\ de Swart,
Phys.\ Rev.\ Lett.\  {\bf 82}, 4992 (1999).
%
\bibitem{Machleidt:2000ge} 
R.\ Machleidt,
Phys.\ Rev.\ C {\bf 63}, 024001 (2001).
%
\bibitem{Rentmeester:2003mf} 
M.C.M.\ Rentmeester, R.G.E.\ Timmermans, and J.J.\ de Swart,
Phys.\ Rev.\ C {\bf 67}, 044001 (2003).
%
\bibitem{Gross08}
F.L.\ Gross and A.\ Stadler,
Phys.\ Rev.\ C {\bf 78}, 014005 (2008).
%
\bibitem{Navarro13}
R.\ Navarro P\'erez, J.E.\ Amaro, and E.\ Ruiz Arriola,
Phys.\ Rev.\ C {\bf 88}, 064002 (2013).
%
\bibitem{Perez:2013oba}
R.\ Navarro P\'erez, J.E.\ Amaro, and E.\ Ruiz Arriola,
Phys.\ Rev.\ C {\bf 89},  024004 (2014).
%
\bibitem{Navarro14} 
R.\ Navarro P\'erez, J.E.\ Amaro, and E.\ Ruiz Arriola,
Phys.\ Rev.\ C {\bf 89}, 064006 (2014).
%
\bibitem{Inverse}
K.\ Chadan and P.C.\ Sabatier, 
{\it Inverse Problems in Quantum Scattering Theory}
(Springer-Verlag, New York, 1989).
%
\bibitem{Baye:2014oea} 
D.\ Baye, J.M.\ Sparenberg, A.M.\ Pupasov-Maksimov, and B.F.\ Samsonov,
J.\ Phys.\ A {\bf 47}, 243001 (2014).
%

\bibitem{Bogner:2001gq} 
  S.~K.~Bogner, T.~T.~S.~Kuo, A.~Schwenk, D.~R.~Entem and R.~Machleidt,
  Phys.\ Lett.\ B {\bf 576}, 265 (2003)
%
\bibitem{Bogner:2003wn} 
  S.~K.~Bogner, T.~T.~S.~Kuo and A.~Schwenk,
  Phys.\ Rept.\  {\bf 386}, 1 (2003)
\bibitem{Arriola:2014fqa} 
  E.~Ruiz~Arriola, S.~Szpigel and V.~S.~Timoteo,
  Annals Phys.\  {\bf 353}, 129 (2014)
  %
\bibitem{Carlson2014}
J.\ Carlson, S.\ Gandolfi, F.\ Pederiva, S.C.\ Pieper, R.\ Schiavilla, K.E.\ Schmidt, and R.B.\ Wiringa,
arXiv:1412.3081[nucl-th].
%
\bibitem{Hatsuda:2012hw} 
T.\ Hatsuda,
J.\ Phys.\ Conf.\ Ser.\  {\bf 381}, 012020 (2012).
%
\bibitem{Detmold:2015jda} 
W.\ Detmold,
Lect.\ Notes Phys.\  {\bf 889}, 153 (2015).
%
\bibitem{Klomp:1991vz}
R.A.M.\ Klomp, V.G.J.\ Stoks. and J.J.\ de Swart,
Phys.\ Rev.\ C {\bf 44}, 1258 (1991).
%
\bibitem{Weinberg:1990rz} 
S.\ Weinberg,
Phys.\ Lett.\ B {\bf 251}, 288 (1990).
%
\bibitem{Entem11} 
R.\ Machleidt and D.R.\ Entem,
Phys.\ Rep.\ {\bf 503}, 1 (2011).
%
\bibitem{Kaiser:1997mw} 
N.\ Kaiser, R.\ Brockmann, and W.\ Weise,
Nucl.\ Phys.\ A {\bf 625}, 758 (1997).
%
\bibitem{Pastore2008} 
S.\ Pastore, R.\ Schiavilla, and J.L.\ Goity
Phys.\ Rev.\ C {\bf 78}, 064002 (2008). 
%
\bibitem{Schiavilla92}
R.\ Schiavilla, R.B.\ Wiringa, V.R.\ Pandharipande, and J.\ Carlson,
Phys.\ Rev.\ C {\bf 45}, 2628 (1992).
%
\bibitem{Viviani96}
M.\ Viviani, R.\ Schiavilla, and A.\ Kievsky,
Phys.\ Rev.\ C {\bf 54}, 534 (1996).
%
\bibitem{Girlanda10}
L.\ Girlanda, A.\ Kievsky, L.E.\ Marcucci, S.\ Pastore, R.\ Schiavilla, and M.\ Viviani,
Phys.\ Rev.\ Lett.\ {\bf 105}, 232502 (2010).
%
\bibitem{Marcucci01}
L.E.\ Marcucci, R.\ Schiavilla, M.\ Viviani, A.\ Kievsky, S.\ Rosati, and J.F.\ Beacom,
Phys.\ Rev.\ C {\bf 63}, 015801 (2001).
%
 \bibitem{Bystricky1}
J.\ Bystricky, F.\ Lehar, and P,\ Winternitz, 
J.\ Phys.\ {\bf 39}, 1 (1978).
%
\bibitem{Bystricky2}
J.\ Bystricky, C.\ Lechanoine-Leluc, and F.\ Lehar, 
J.\ Phys. {\bf 48}, 199 (1987).
%
\bibitem{Entem03}
D.R.\ Entem and R.\ Machleidt,
Phys. Rev. C 68, 041001(R) (2003).
%
\bibitem{Epelbaum:2004fk} 
E.\ Epelbaum, W.\ Gl\"ockle, and U.-G.\ Meissner,
Nucl.\ Phys.\ A {\bf 747}, 362 (2005).
%
\bibitem{Entem2015}
D.R.\ Entem, N.\ Kaiser, R.\ Machleidt, and Y.\ Nosyk,
Phys.\ Rev.\ C {\bf 91}, 014002 (2015). 
%
\bibitem{Epelbaum:2014efa} 
E.\ Epelbaum, H.\ Krebs, and U.-G.\ Meissner,
arXiv:1412.0142 [nucl-th].
%
\bibitem{NavarroPerez:2012vr} 
R.\ Navarro Perez, J.E.\ Amaro, and E.\ Ruiz Arriola,
arXiv:1202.6624 [nucl-th].
%
\bibitem{Perez:2014jsa} 
R.\ Navarro Perez, J.E.\ Amaro, and E.\ Ruiz Arriola,
Phys.\ Lett.\ B {\bf 738}, 155 (2014).
%
\bibitem{Perez:2013za} 
R.\ Navarro Perez, J.E.\ Amaro, and E.\ Ruiz Arriola,
PoS CD {\bf 12}, 104 (2013).
%
\bibitem{Epelbaum98}
E.\ Epelbaum, W.\ Gl\"ockle, and U.-G. Meissner,
Nucl.\ Phys.\ {\bf A637}, 107 (1998).
%
\bibitem{Pastore09}
S.\ Pastore, L.\ Girlanda, R.\ Schiavilla, M.\ Viviani, and R.B.\ Wiringa,
Phys.\ Rev.\ C {\bf 80}, 034004 (2009).
%
\bibitem{Viviani14}
M.\ Viviani, A.\ Baroni, L.\ Girlanda, A.\ Kievsky, L.E.\ Marcucci, and R.\ Schiavilla,
Phys.\ Rev.\ C {\bf 89}, 064004 (2014).
%
\bibitem{Kaiser1998}
N.\ Kaiser, S. Gerstend\"orfer, and W. Weise
Nucl.\ Phys.\ {\bf A 637}, 395 (1998).
%
\bibitem{Krebs07}
H.\ Krebs, E.\ Epelbaum, and Ulf.-G. Mei\ss{}ner,
Eur.\ Phys.\ J.\ A {\bf 32}, 127 (2007).
%
\bibitem{Epelbaum04}
E.\ Epelbaum, W.\ Gl\"{o}ckle, and U.-G. Mei\ss{}ner,
Eur.\ Phys.\ J.\ A {\bf 19}, 125 (2004).
%
\bibitem{Green76}
A.M.\ Green,
Rep.\ Prog.\ Phys.\ {\bf 39}, 1109 (1976).
%
\bibitem{Fettes00}
N.\ Fettes, U.-G.\ Meissner, M.\ Moj\u{z}i\u{s}, S.\ Steininger,
Ann.\ Phys.\ {\bf 283}, 273 (2000).
%
\bibitem{Valderrama:2008kj} 
  M.\ Pav\'on Valderrama and E.\ Ruiz Arriola,
  Phys.\ Rev.\ C {\bf 79}, 044001 (2009).
%
\bibitem{PavonValderrama:2010fb} 
  M.\ Pav\'on Valderrama and E.\ Ruiz Arriola,
  Phys.\ Rev.\ C {\bf 83}, 044002 (2011).
%
\bibitem{Friar04}
J.L.\ Friar, U.\ van Kolck, M.C.M.\ Rentmeester, and R.G.E.\ Timmermans,
Phys.\ Rev.\ C {\bf 70}, 044001 (2004).
%
\bibitem{Epelbaum05}
E.\ Epelbaum and U.-G.\ Meissner,
Phys.\ Rev.\ C {\bf 72}, 044001 (2005).
%
\bibitem{Kolck98}
U.\ van Kolck, M.C.M.\ Rentmeester, J.L.\ Friar, T.\ Goldman, and J.J.\ de Swart,
Phys.\ Rev.\ Lett.\ {\bf 80}, 4386 (1998).
%
\bibitem{Kaiser06}
N.\ Kaiser,
Phys.\ Rev.\ C {\bf 73}, 044001 (2006).
%
\bibitem{Gezerlis14}
A.\ Gezerlis, I.\ Tews, E.\ Epelbaum, S.\ Gandolfi, K.\ Hebeler, A.\ Nogga, and A.\ Schwenk,
Phys.\ Rev.\ Lett.\ {\bf 111}, 032501 (2013);
A.\ Gezerlis, I.\ Tews, E.\ Epelbaum, M.\ Freunek, S.\ Gandolfi, K.\ Hebeler, A.\ Nogga, and A.\ Schwenk,
Phys.\ Rev.\ C {\bf 90}, 054323 (2014).
%
\bibitem{Stoks:1990us} 
V.G.J.\ Stoks and J.J.\ de Swart,
Phys.\ Rev.\ C {\bf 42}, 1235 (1990).
%
\bibitem{POUNDerS} 
M.\ Kortelainen, T.\ Lesinski, J.\ More, W.\ Nazarewicz, J.\ Sarich, N.\ Schunck, M.V.\ Stoitsov, and S.\ Wild,
Phys.\ Rev.\ C {\bf 82}, 024313 (2010).
%
\bibitem{Miller90}
G.A.\ Miller, M.K.\ Nefkens, and I.\ Slaus,
Phys.\ Rep.\ {\bf 194}, 1 (1990).
%
\bibitem{Machleidt01}
R.\ Machleidt,
Phys.\ Rev.\ C {\bf 63}, 024001 (2001).
%
 \bibitem{Gonzalez06}
D.E.\ Gonz\'alez Trotter {\it et al.},
Phys.\ Rev.\ C {\bf 73}, 034001 (2006).
%
\bibitem{Chen08}
Q.\ Chen {\it et al.},
Phys.\ Rev.\ C {\bf 77}, 054002 (2008).
%
\bibitem{Vandl82}
C.\ van der Leun and C.\ Alderlisten, 
Nucl.\ Phys.\ A {\bf 380}, 261 (1982).
%
\bibitem{Ericson83}
T.E.O.\ Ericson and M.\ Rosa-Clot,
Nucl.\ Phys.\ A {\bf 405}, 497 (1983).
%
\bibitem{Rodning90}
N.L.\ Rodning and L.D.\ Knutson, 
Phys.\ Rev.\ C {\bf 41}, 898 (1990).
%
\bibitem{Huber98}
A.\ Huber \textit{et al.},
Phys.\ Rev.\ Lett.\ {\bf 80}, 468 (1998).
%
\bibitem{Bishop79}
D.M.\ Bishop and L.M.\ Cheung, 
Phys.\ Rev.\ A {\bf 20}, 381 (1979).
%
\bibitem{Piarulli13}
M.\ Piarulli, L.\ Girlanda, L.E.\ Marcucci, S.\ Pastore, R.\ Schiavilla, and M.\ Viviani,
Phys.\ Rev.\ C {\bf 87}, 014006 (2013).
%
\bibitem{Kievsky08}
A.\ Kievsky, S.\ Rosati, M.\ Viviani, L.E.\ Marcucci, and L.\ Girlanda,
J.\ Phys.\ G {\bf 35}, 063101 (2008).
%
\bibitem{Perez:2014bua} 
  R.~N.~Perez, J.~E.~Amaro and E.~R.~Arriola,
  arXiv:1411.1212 [nucl-th].
%
%
\bibitem{Berg88}
J.R.\ Bergervoet, P.C.\ van Campen, W.A. van de Sanden, and J.J.\ de Swart,
Phys.\ Rev.\ C {\bf 38}, 15 (1988).
%
\bibitem{Heller60}
L.\ Heller, Phys.\ Rev.\ {\bf 120}, 627 (1960).
%
\bibitem{Vands83}
W.A.\ van de Sanden, A.H.\ Emmen, J.J.\ de Swart, Report No. THEF-NYM-83.11, Nijmegen (1983), unpublished; quoted in~\cite{Berg88}.
%
%
%
\end{thebibliography}
\end{document}